\documentclass[preprint,twocolumn]{aastex63}
\usepackage{float, bm, graphicx, amsmath, morefloats}
\usepackage[caption=false]{subfig}
\shorttitle{Double-peaked Type Ibc SNe}
\shortauthors{Das et al.}

\usepackage{float,graphicx,amsmath,tabularx,booktabs,natbib,threeparttable}
\usepackage{float, bm, graphicx, amsmath, morefloats}
\usepackage[caption=false]{subfig}
\usepackage{float,graphicx,amsmath,tabularx,booktabs,natbib,threeparttable}
\usepackage{subfig}
\usepackage[para]{footmisc}
\usepackage{tablefootnote}

\usepackage{graphicx}
\usepackage[caption=false]{subfig}
\usepackage{lipsum} 
\usepackage{placeins}

\newcommand{\lpipe}{\textsc{lpipe}}
\newcommand{\emcee}{\textsc{emcee}}

\newcommand{\HEAsoft}{\textsc{HEAsoft}}

\usepackage{comment}
\definecolor{dark-red}{rgb}{0.4,0.15,0.15}
\definecolor{dark-blue}{rgb}{0.15,0.15,0.4}
\definecolor{medium-blue}{rgb}{0,0,0.5}
\hypersetup{
    colorlinks, linkcolor={dark-red},
    citecolor={dark-blue}, urlcolor={medium-blue}
}

\newcommand{\beqa}{\begin{eqnarray}} 
\newcommand{\eeqa}{\end{eqnarray}}

\newcommand{\bsub}{\begin{subequations}}
\newcommand{\esub}{\end{subequations}}
\newcommand{\beal}{\begin{align}}
\newcommand{\ealn}{\end{align}}
\newcommand{\Mni}{$M_\mathrm{Ni}$}
\newcommand{\Mej}{$M_\mathrm{ej}$}

\newcommand{\msun}{M$_{\sun}$}

\newcommand{\Ni}{\ensuremath{^{56}\mathrm{Ni}}}
\newcommand{\Msun}{{\ensuremath{\mathrm{M}_{\odot}}}}
\newcommand{\Rsun}{{\ensuremath{\mathrm{R}_{\odot}}}}

\graphicspath{{./}{figures/}}

\begin{document}

\title{\vspace{0.5cm} Probing pre-supernova mass loss in double-peaked Type Ibc SNe from the Zwicky Transient Facility}

\author[0000-0001-8372-997X]{Kaustav~K.~Das}\thanks{E-mail: kdas@astro.caltech.edu}
\affiliation{Cahill Center for Astrophysics, California Institute of Technology, MC 249-17, 
1200 E California Boulevard, Pasadena, CA, 91125, USA}

\author[0000-0002-5619-4938]{Mansi~M.~Kasliwal}
\affiliation{Cahill Center for Astrophysics, 
California Institute of Technology, MC 249-17, 
1200 E California Boulevard, Pasadena, CA, 91125, USA}

\author[0000-0003-1546-6615]{Jesper Sollerman}
\affiliation{The Oskar Klein Centre, Department of Astronomy, Stockholm University, AlbaNova, SE-10691 Stockholm, Sweden}

\author[0000-0002-4223-103X]{Christoffer Fremling}
\affiliation{Cahill Center for Astrophysics, 
California Institute of Technology, MC 249-17, 
1200 E California Boulevard, Pasadena, CA, 91125, USA}

\author{I. Irani}
\affiliation{Department of Particle Physics and Astrophysics, Weizmann Institute of Science, 234 Herzl St, 76100 Rehovot, Israel}

\author[0000-0002-4972-3803]{Shing-Chi Leung}
\affiliation{Department of Mathematics and Physics, SUNY Polytechnic Institute, 100 Seymour Road, Utica, NY 13502, USA
}

\author[0000-0003-1546-6615]{Sheng Yang}
\affiliation{The Oskar Klein Centre, Department of Astronomy, Stockholm University, AlbaNova, SE-10691 Stockholm, Sweden}

\author{Samantha Wu}
\affiliation{Cahill Center for Astrophysics, California Institute of Technology, MC 249-17, 
1200 E California Boulevard, Pasadena, CA, 91125, USA}

\author{Jim Fuller}
\affiliation{Cahill Center for Astrophysics, California Institute of Technology, MC 249-17, 
1200 E California Boulevard, Pasadena, CA, 91125, USA}



\author{Shreya Anand}
\affiliation{Cahill Center for Astrophysics, 
California Institute of Technology, MC 249-17, 
1200 E California Boulevard, Pasadena, CA, 91125, USA}

\author{Igor Andreoni}
\affiliation{Joint Space-Science Institute, University of Maryland, College Park, MD 20742, USA}
\affiliation{Department of Astronomy, University of Maryland, College Park, MD 20742, USA}
\affiliation{Astrophysics Science Division, NASA Goddard Space Flight Center, Mail Code 661, Greenbelt, MD 20771, USA}

\author{C. Barbarino}
\affiliation{The Oskar Klein Centre, Department of Astronomy, Stockholm University, AlbaNova, SE-10691 Stockholm, Sweden}

\author{Thomas G. Brink}
\affiliation{Department of Astronomy, University of California, Berkeley, CA 94720-3411, USA}

\author[0000-0002-8989-0542]{Kishalay De}
\affiliation{MIT-Kavli Institute for Astrophysics and Space Research
77 Massachusetts Ave. Cambridge, MA 02139, USA}

\author{Alison Dugas}
\affiliation{Department of Physics and Astronomy, Watanabe 416, 2505 Correa Road, Honolulu, HI 96822, USA}

\author[0000-0001-5668-3507]{Steven L. Groom}
\affiliation{IPAC, California Institute of Technology, 1200 E. California Blvd, Pasadena, CA 91125, USA}

\author[0000-0003-3367-3415]{George~Helou}
\affiliation{IPAC, California Institute of Technology, 1200 E. California Blvd, Pasadena, CA 91125, USA}

\author{K-Ryan Hinds}
\affiliation{Astrophysics Research Institute, Liverpool John Moores University, IC2,  Liverpool L3 5RF, UK}

\author[0000-0002-9017-3567]{Anna Y. Q.~Ho}
\affiliation{Miller Institute for Basic Research in Science, 468 Donner Lab, Berkeley, CA 94720, USA}
\affiliation{Department of Astronomy, University of California, Berkeley, 501 Campbell Hall, Berkeley, CA, 94720, USA}
\affiliation{Lawrence Berkeley National Laboratory, 1 Cyclotron Road, MS 50B-4206, Berkeley, CA 94720, USA}
\affiliation{Department of Astronomy, Cornell University, Ithaca, NY 14853, USA}

\author{Viraj Karambelkar}
\affiliation{Cahill Center for Astrophysics, 
California Institute of Technology, MC 249-17, 
1200 E California Boulevard, Pasadena, CA, 91125, USA}

\author[0000-0001-5390-8563]{S.~R.~Kulkarni}
\affiliation{Cahill Center for Astrophysics, 
California Institute of Technology, MC 249-17, 
1200 E California Boulevard, Pasadena, CA, 91125, USA}

\author[0000-0001-8472-1996]{Daniel A.~Perley}
\affiliation{Astrophysics Research Institute, Liverpool John Moores University, IC2,  Liverpool L3 5RF, UK}

\author{Josiah Purdum}
\affiliation{Caltech Optical Observatories, California Institute of Technology, Pasadena, CA 91125, USA}

\author{Nicolas Regnault}
\affiliation{LPNHE, CNRS/IN2P3 \& Sorbonne Universit`e, 4 place Jussieu, 75005 Paris, France}

\author[0000-0001-6797-1889]{Steve Schulze}
\affiliation{The Oskar Klein Centre, Department of Astronomy, Stockholm University, AlbaNova, SE-10691 Stockholm, Sweden}

\author{Yashvi Sharma}
\affiliation{Cahill Center for Astrophysics, 
California Institute of Technology, MC 249-17, 
1200 E California Boulevard, Pasadena, CA, 91125, USA}

\author{Tawny Sit}
\affiliation{Department of Astronomy, The Ohio State University, Columbus, OH 43210, USA}

\author{Niharika Sravan}
\affiliation{Department of Physics, Drexel University, Philadelphia, PA 19104, USA}

\author{Gokul P. Srinivasaragavan}
\affiliation{Department of Astronomy, University of Maryland, College Park, MD 20742, USA}

\author{Robert Stein}
\affiliation{Cahill Center for Astrophysics, 
California Institute of Technology, MC 249-17, 
1200 E California Boulevard, Pasadena, CA, 91125, USA}

\author{Kirsty Taggart}
\affiliation{Department of Astronomy and Astrophysics, University of California, Santa
Cruz, CA 95064, USA}

\author[0000-0003-3433-1492]{Leonardo Tartaglia}
\affiliation{The Oskar Klein Centre, Department of Astronomy, AlbaNova, SE-106 91 Stockholm, Sweden}
\affiliation{INAF - Osservatorio Astronomico di Padova, Vicolo dell’Osservatorio 5, I-35122 Padova, Italy}

\author{Anastasios Tzanidakis}
\affiliation{Cahill Center for Astrophysics, 
California Institute of Technology, MC 249-17, 
1200 E California Boulevard, Pasadena, CA, 91125, USA}

\author[0000-0002-9998-6732]{Avery Wold}
\affiliation{IPAC, California Institute of Technology, 1200 E. California Blvd, Pasadena, CA 91125, USA}

\author[0000-0003-1710-9339]{Lin~Yan}
\affil{Caltech Optical Observatories, California Institute of Technology, Pasadena, CA 91125, USA}

\author[0000-0001-6747-8509]{Yuhan Yao}
\affiliation{Cahill Center for Astrophysics, 
California Institute of Technology, MC 249-17, 
1200 E California Boulevard, Pasadena, CA, 91125, USA}

\author{Jeffry Zolkower}
\affiliation{Caltech Optical Observatories, California Institute of Technology, Pasadena, CA 91125, USA}





\begin{abstract}

Eruptive mass loss of massive stars prior to supernova (SN) explosion is key to understanding their evolution and end fate.  An observational signature of pre-SN mass loss is the detection of an early, short-lived peak prior to the radioactive-powered peak in the lightcurve of the SN. This is usually attributed to the SN shock passing through an extended envelope or circumstellar medium (CSM). Such an early peak is common for double-peaked Type IIb SNe with an extended Hydrogen envelope but is uncommon for normal Type Ibc SNe with very compact progenitors. In this paper, we systematically study a sample of 14 double-peaked Type Ibc SNe out of 475 Type Ibc SNe detected by the Zwicky Transient Facility. The rate of these events is $\sim 3-9 \%$ of Type Ibc SNe.  A strong correlation is seen between the peak brightness of the first and the second peak. We perform a holistic analysis of this sample's photometric and spectroscopic properties. We find that six SNe have ejecta mass less than 1.5\ \Msun. Based on the nebular spectra and lightcurve properties, we estimate that the progenitor masses for these are less than $\sim$ 12 \Msun. The rest have an ejecta mass $>$ 2.4\ \Msun\ and a higher progenitor mass. This sample suggests that the SNe with low progenitor masses undergo late-time binary mass transfer. Meanwhile, the SNe with higher progenitor masses are consistent with wave-driven mass loss or pulsation-pair instability-driven mass loss simulations. 

\end{abstract}

\keywords{}

\section{Introduction} \label{sec:intro}

Most massive stars undergo mass loss during their lifetime. This can affect the star’s luminosity, burning lifetime, apparent temperature, Helium-core mass, and impact its end fate. The mass loss has a great influence on the late-time evolution of massive stars and the resultant supernova (SN) \citep[e.g.,][]{Smith2014}. Pre-SN mass loss also has an impact on other areas of astronomy since it affects predictions for ionizing radiation, wind feedback from stellar remnants, and the origin of compact stellar remnants. 

Early observations of SNe and theoretical models indicate that enhanced mass loss and pre-SN outbursts may occur in progenitors of many types of core-collapse SNe (CCSNe).  Different evidence include the direct detection of pre-cursor outbursts \citep{pastorello07-06jc, Nora2015, Nora2021, Strotjohann2023, Jacobson2022}, bright UV emission in Type IIP SNe at early times 
\citep[e.g.,][]{Morozova2018, Bostroem2019}; 
narrow spectral lines originating from a dense circumstellar medium ionized by the explosion’s
shock \citep[as in Type IIn, Type Ibn, Type Icn SNe, and Type II SNe e.g.,][]
{smith2017,pastorello2008,perley2022, Bruch2021}. 
Various mechanisms have been proposed to explain this mass loss, like standard nuclear burning instabilities and gravity waves \citep{Arnett2011, Quatert2012, Wu2021, Wu2022, Leung2021}, silicon deflagration \citep{Woosley2015}, radiation-driven steady winds \citep{Crowther2007}, pulsation-pair instability driven mass loss \citep{Leung2019} and binary interactions \citep{Wu2022b}.

The detection of the first peak in the lightcurve of a double-peaked SN constitutes an observational signature of circumstellar matter (CSM) or extended envelope around the progenitor. If strong mass loss occurred shortly before the SN explosion, it would create a layer of CSM around the SN progenitor. The shock cooling emission  \citep[i.e., bright post-breakout emission;][]{Rabinak2011, nakarpiro2014, piro2015, Waxman2017, Piro2021, Morag2023, Khatami2023} is seen as the SN shock passes through this ejected material. This should manifest as an early peak in the SN lightcurve. This is common for Type IIb SNe where the extended material is attributed to the outer He/H envelope. However, the progenitors of Type Ib and Ic SNe (SNe Ibc) are suggested to be very compact Wolf-Rayet (WR) stars or helium stars whose hydrogen envelopes have been stripped off via mass loss \citep[e.g.,][]{Yoon2015}. Eruptive mass loss prior to a supernova explosion could provide a medium for the shock to propagate through. 

\textcolor{black}{This early peak has been detected in a few Type Ibc SNe in the past. The presence of CSM is likely responsible for the first peak of several peculiar SNe Ic like SN 2006aj 
\cite[e.g.,][]{Modjaz2006}, SN 2010mb \citep{Ben-Ami2014}, iPTF15dtg \citep{Taddia2016}, SN 2020bvc \citep{Ho2020}, and double-peaked superluminous SNe Ic 
(e.g., PTF12dam; \citealt{Vreeswijk2017}, LSQ14bdq; \citealt{Nicholl2015, Nicholl2016}, DES14X3taz; \citealt{Smith2016}).  
The double-peak is also seen in a few ordinary Type Ibc SNe: SN LSQ14efd \citep{Barbarino2017}, iPTF 16hgs \citep{De2018a}, SN 2017ein \citep{Xiang2019}, SN 2018lqo \citep{De2020}, SN 2019ehk \citep{Jacobson2020a, Nakaoka2021, De2021}, SNe 2021gno and 2021inl \citep{Jacobson2022b}, SN 2022oqm \citep{Irani2022} and ultra-stripped SNe: SN 2019dge \citep{Yao2020}, iPTF14gqr \citep{De2021}. The low number of detections could be because of an observational bias as the detection of these sources requires fast cadence and early follow-up. }

Modern high-cadence surveys such as the Zwicky Transient Facility \citep[ZTF;][]{Graham2019, Bellm2019, Masci2019} act as a discovery engine for such events. In this paper, we present a sample of 17 double-peaked Type Ibc SNe detected by the ZTF. These detections are part of the Census of the Local Universe survey (CLU; \citealp{De2020}) and the  Bright Transient Survey (BTS; \citealp{Fremling2020, Perley2020}). The CLU survey is designed as a volume-limited survey with the objective of classifying all SNe within 200 Mpc, whose host galaxies are part of the CLU galaxy catalog \citep{Cook2019}. BTS is a magnitude-limited survey focused  on spectroscopically classifying SNe with a peak magnitude brighter than 18.5 mag. In this paper, we perform a holistic analysis of the lightcurves for both the shock-cooling and the radioactive peaks, as well as for early time, photospheric and nebular spectra of the sample. Based on the estimated CSM and progenitor properties, we provide constraints on the mass loss and progenitor channels.

The sample selection 
is described in Section~\ref{sec:sample}. We describe the photometric and spectroscopic data in Section \ref{sec:data}. We present our analysis and results from the spectra and the light curves in Section~\ref{sec:analysis}. We discuss the inferred progenitor masses in Section~\ref{sec:zams} and  the mass-loss scenarios in Section~\ref{sec:massloss}. We provide a brief summary of the results and future goals in Section \ref{sec:summary}.

\section{Sample Selection}\label{sec:sample}

In this paper, we use SNe observed by the ZTF. The ZTF camera \citep{Dekany20} is mounted on the Palomar 48-inch (P48) Oschin Schmidt telescope and has a field of view spanning 47 square degrees. ZTF images the entire Northern sky every $\sim$2 nights in $g$ and $r$ bands and achieves a median depth of approximately 20.5 mag \citep{Bellm2019b}. We use ZTF discoveries and follow-up spectra that are part of the BTS and CLU surveys.

We apply the following selection criteria on the ZTF SN sample obtained from the BTS and CLU surveys (1 April 2018 to 25 October 2022): 

1. The transient should be classified as a stripped-envelope SN (SESN) (Type Ib, Ib-pec, Ibn, Ic, Ic-pec, Icn, Ic-BL, but Type IIb are not included) based on photospheric spectral template matching and manual inspection. As per classification status on 25th October 2022, there are 185 SNe classified as Type Ib, 27 SNe classified as Type Ibn, 176 classified as Type Ic,  59 classified as Type Ic-BL, and 28 SNe classified as Type Ib/c with unclear type 
(Yang et al. in prep).


2. In our analysis, we utilize the ZTF forced-photometry service developed by \citet{Masci2019} to perform forced photometry in $g$, $r$ and $i$ bands on the ZTF difference images. We consider 3$\sigma$ measurements as a threshold. 59 Type Ib/Ibn SNe, 86 Type Ic/Icn/Ic-BL SNe and 47 Type Ib/c SNe (classification not distinguishable between Ib and Ic) have good quality early-time lightcurves, where the gap between the first and second detection as well as the gap between the last non-detection and the first detection is less than 5 days. Hence, we did not miss the first peak for these SNe. 

3. We manually inspect the lightcurves of these transients to look for an early bump with a rise and decline or just the decline in either of the ZTF $g$-band or $r$-band photometry followed by a second peak. We find 19 such SNe with 10 Type Ib, 4 Type Ic, 3 Type Ic-BL, 2 Type Ib/c. We list the details of the sample in Table~\ref{tablesample}.


4. The early lightcurve decline or rise should be present in at least two filters. There were 2 SNe\footnote{ZTF18acsodbf, ZTF19abzzuhj} where we could see an early decline which could correspond to a first peak, but since they were seen only in the $r$ or $g$ band we do not include them in our sample. The summary of the sources in the sample is provided in Table \ref{tablesample}. \\

Thus, the lower and upper limit on the rate of these events is $\sim 17/475 = 3.5\%$ $\sim 17/192 = 8.8\%$ of Type Ibc(BL) SNe respectively. 



\begin{figure*}
    \centering
    \includegraphics[width=0.49\textwidth]{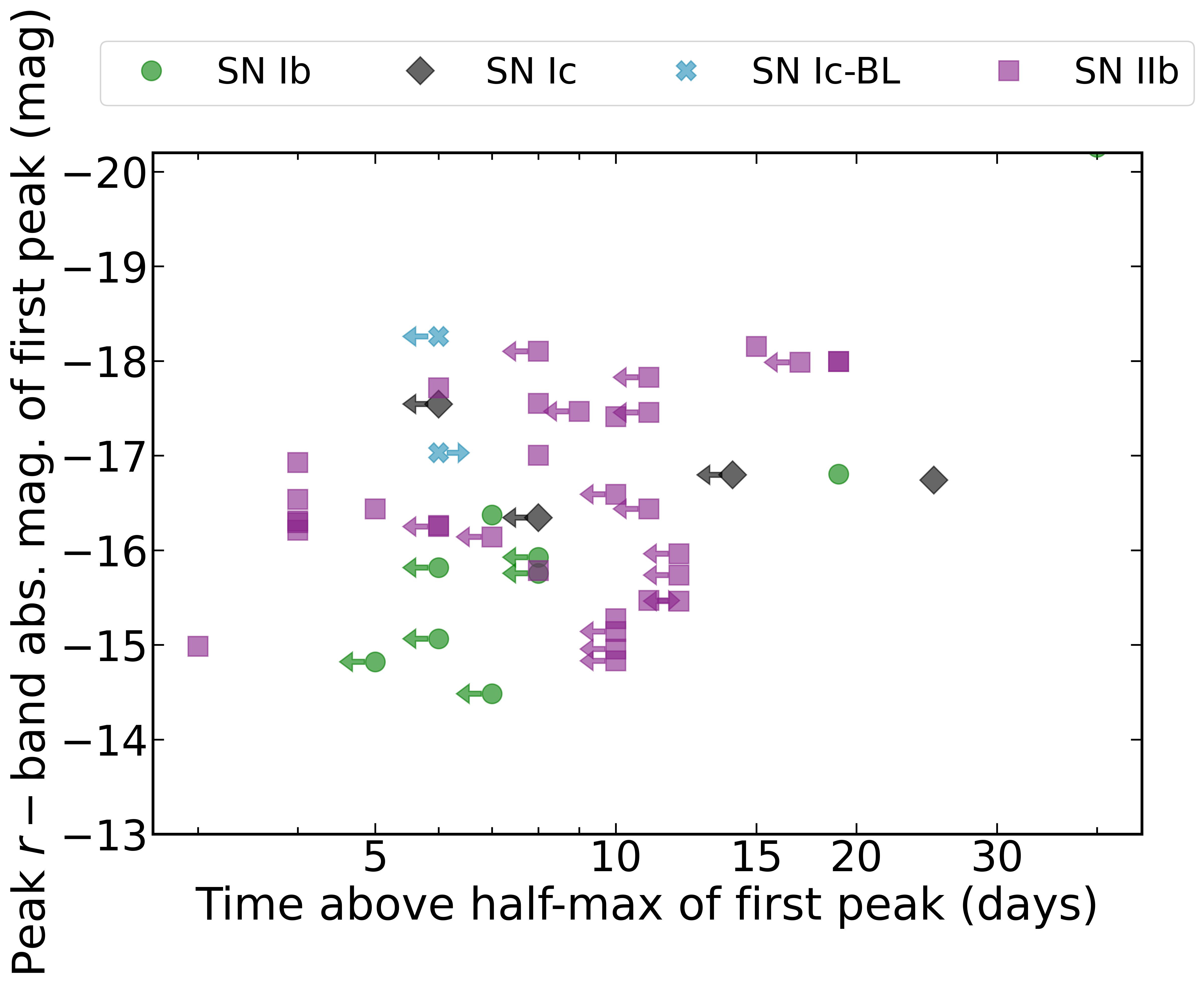}\includegraphics[width=0.49\textwidth]{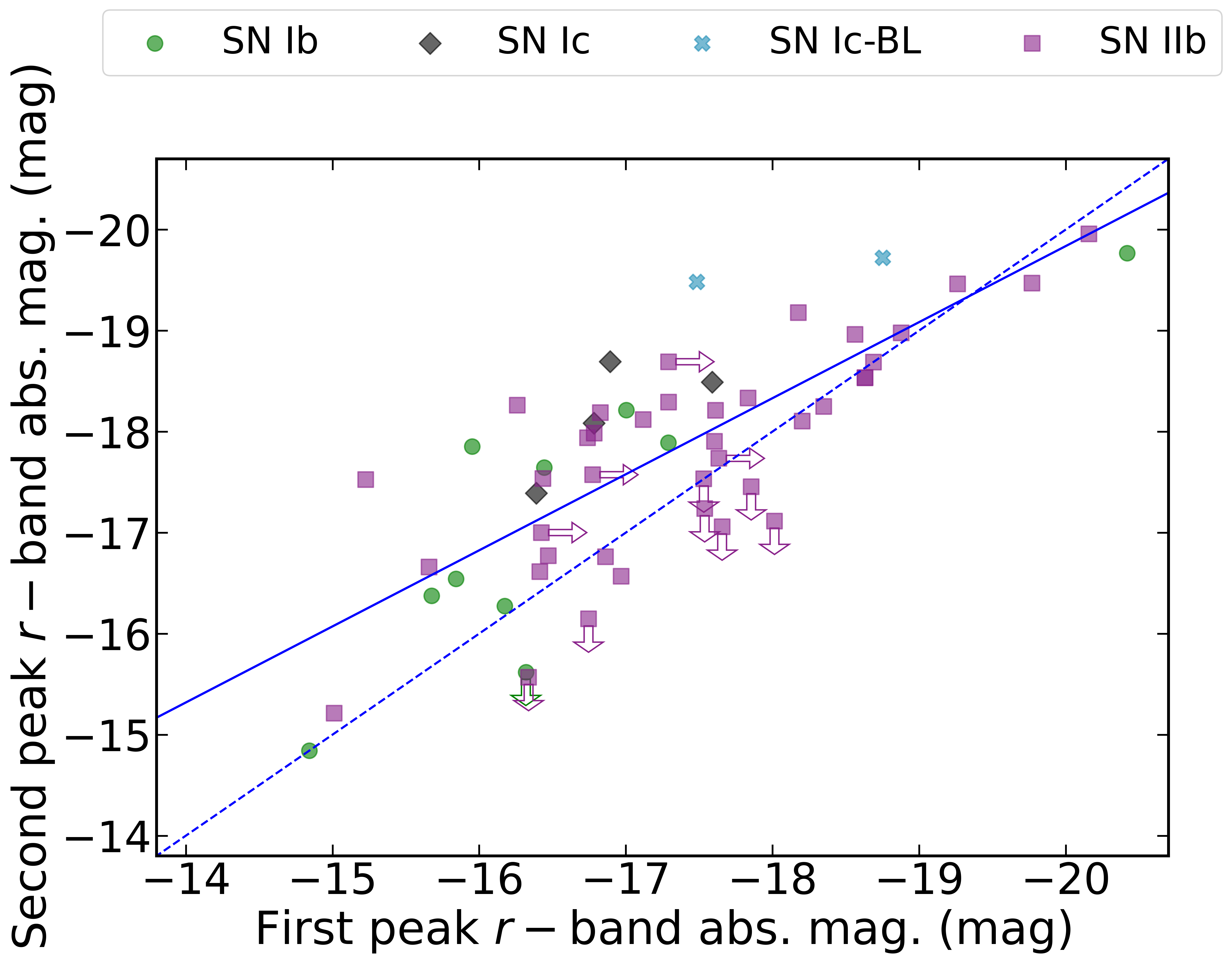}
    \caption{\small {\it Left:} Parameter space of peak magnitude of the first peak versus time above half maximum  of the first peak $t_{1/2}^\mathrm{first}$ for all double-peaked SNe observed by ZTF as part of the BTS and CLU surveys. The Type Ibc(BL) SNe in the figure are part of the sample in this work. {\it Right:}  We see a correlation in the peak magnitude of the first and second peaks of the SNe following $M_2 = 0.8 \times M_1 - 4.7$. $M_1$ and $M_2$ are the peak magnitudes of the first and second peak respectively. The solid line shows the best-fit linear relation. We can infer from the $y=x$ dashed line that the second peak is brighter than the first peak for most sources. The correlation could imply that the SNe that show double-peaked lightcurves have He-star progenitors that shed their envelope in binary interactions.}
    \label{fig:lumtime}
\end{figure*}

We note that the time above half maximum of the first peak ($t_{1/2}^\mathrm{first}$) is $< 8$ days for fourteen SNe, while three SNe have an unusually long first peak with $t_{1/2}^\mathrm{first} > 15$ days (see Figure \ref{fig:lumtime}). The bolometric luminosity for these three sources (SNe 2019cad, 2022hgk, 2021uyv) increases with time for the first peak which is not expected for the shock-cooling phase. Hence, we believe that the powering mechanism for the first peak of these sources is not shock cooling and leave the detailed lightcurve analysis of these SNe for future work.

\begin{figure}
    \centering
    \includegraphics[width=7cm]{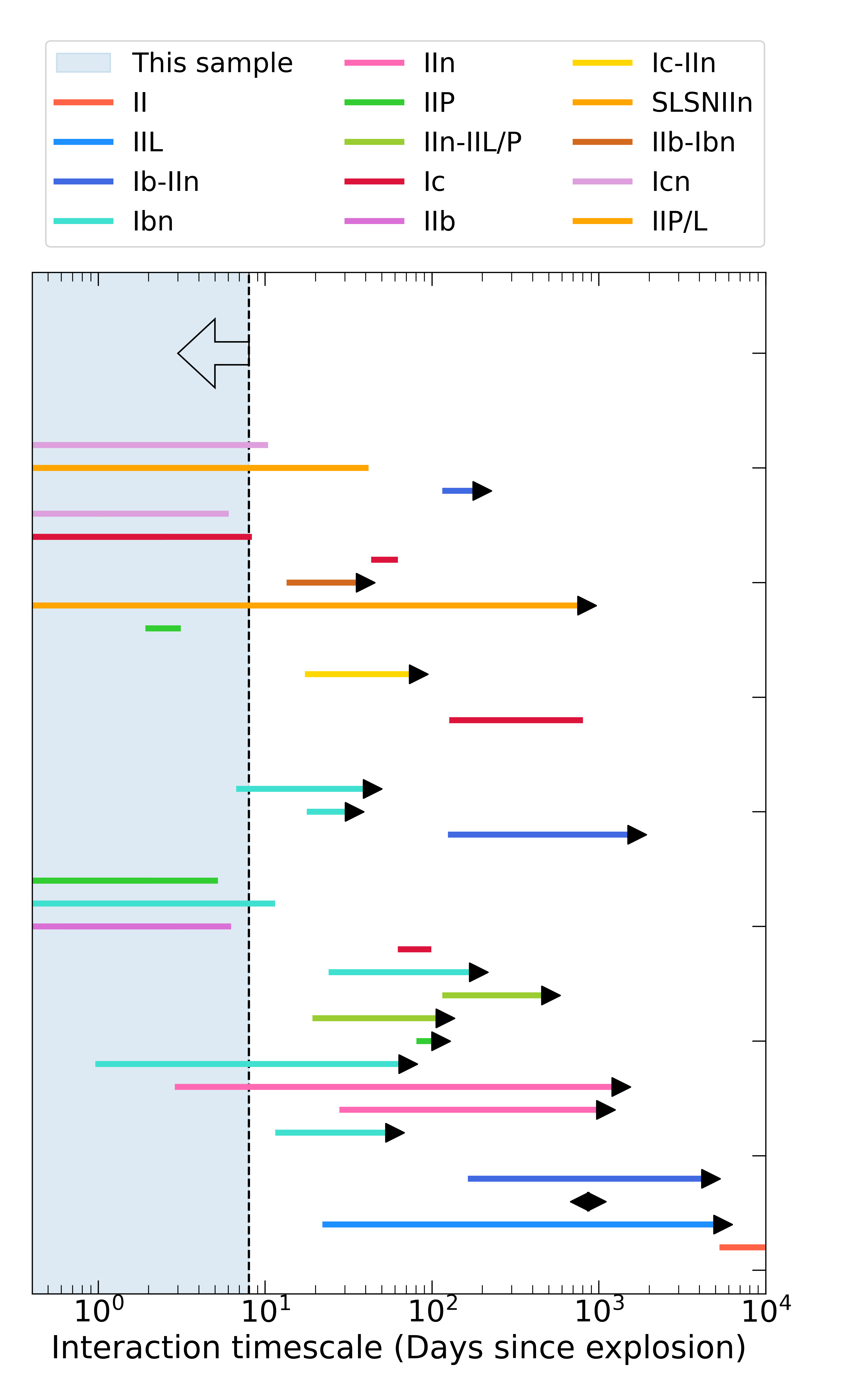}
    \caption{The interaction timescale for various SNe in the literature \citep[from data compiled in][]{Brethaeur2022} showing signatures of CSM interaction including SNe Ibc, Ibn, Icn are shown by the \textcolor{black}{colored lines.} The blue-shaded region shows the range of timescale of the first peak for our sample consistent with shock-cooling ($t_{1/2}^\mathrm{first}$ $<$ 8 days).}
    \label{fig:dptime}
\end{figure}

In Figure \ref{fig:dptime}, we look at the interaction times for some stripped-envelope SNe where interactions were observed \citep{Brethaeur2022}. We note that our sample shows interaction at earlier times compared with CSM interaction signatures for most SESNe in the literature. 

\begin{table*}[htb!]
    \caption{Steps for selecting our sample. See Section \ref{sec:sample} for the details regarding each step.}
    \centering
    \begin{tabular}{ccc}
     \hline
       Step & Criteria  & \# Candidates \\
     \hline
       1 & Classified as SESN (except Type IIb) & 475 \\
       2 & Well-sampled early lightcurve & 192 \\
       3 & Double-peaked & 19 \\
       4 & Candidate has multi-band photometry during the first peak & 17 \\
     \hline
    \end{tabular}
    \label{tab:search}
\end{table*}

\begin{flushleft}
\begin{table*} 
\footnotesize
\begin{center} 
\caption{Summary of the sample of 17 SNe used in this paper. The sources which have been mentioned previously in the literature are labeled with a superscript with a footnote of the list of papers where they were discussed. The absolute magnitudes have been measured by assuming Milky Way extinction ($\mathrm{A_{V,MW}}$) and host galaxy extinction ($\mathrm{A_{V,host}}$) as described in Section \ref{sec:extinction}. Subscripts 1 and 2 refer to the peak parameters of the first peak and second peak respectively. } 
\begin{tabular}{cccccccccccc} 
\hline  \\  

ZTF Name & IAU Name & R.A. & Dec. & Redshift  & Type  & $t_\mathrm{max1}$ & $M_{r1}$ & $t_\mathrm{max2}$ & $M_{r2}$ &  $\mathrm{A_{V,MW}}$  & \textcolor{black}{$\mathrm{A_{V,host}}$} \\ &  & (hh:mm:ss) & (dd:mm:ss) & & & (MJD) &  (mag) &  (MJD) & (mag) & (mag) & \textcolor{black}{(mag)}  \\\\ \hline  \\ 

ZTF21aaqhhfu$^{\textcolor{blue}{a}}$   &  SN2021gno  &  12:12:10.29  &  +13:14:57.0  &  $0.006$  &  Ib & $59294$  &  $-14.5$  &  $59306$  &  $-15.2$  &  $0.10$ & \textcolor{black}{0}\\ \hline \\ 

ZTF21abcgnql  &  SN2021niq  &  15:36:06.70  &  +43:24:21.4  &  $0.018$  &  Ib & $59362$  &  $-15.1$  &  $59371$  &  $-15.8$  &  $0.07$ & \textcolor{black}{0}\\ \hline \\ 

ZTF20abbpkng  &  SN2020kzs  &  17:14:55.02  &  +35:31:13.6  &  $0.037$  &  Ib & $58983$  &  \textcolor{black}{$-17.5$}  &  $59009$  &  \textcolor{black}{$-18.7$}  &  $0.08$ & \textcolor{black}{1.1}\\ \hline \\ 

ZTF21abccaue  &  SN2021nng  &  14:17:22.86  &  +58:44:58.9  &  $0.040$  &  Ib & $59336$  &  \textcolor{black}{$-16.5$}  &  $59381$  &  \textcolor{black}{$-18.4$}  &  $0.03$ & \textcolor{black}{0.6}\\ \hline \\ 

ZTF18achcpwu  &  SN2018ise  &  07:07:16.74  &  +64:03:41.8  &  $0.055$  &  Ic & $58423$  &  $-16.7$  &  $58455$  &  $-18.6$  &  $0.11$ & \textcolor{black}{0}\\ \hline \\ 

ZTF18abmxelh  &  SN2018lqo  &  16:28:43.25  &  +41:07:58.6  &  $0.033$  &  Ib & $58340$  &  $-15.8$  &  $58354$  &  $-16.4$  &  $0.02$ & \textcolor{black}{0}\\ \hline \\ 

ZTF21acekmmm  &  SN2021aabp  &  23:09:55.08  &  +09:41:08.9  &  $0.064$  &  Ic-BL & $59486$  &  $-18.3$  &  $59505$  &  $-19.1$  &  $0.15$ & \textcolor{black}{0}\\ \hline \\ 

ZTF21aasuego$^{\textcolor{blue}{a}}$   &  SN2021inl  &  13:01:33.24  &  +27:49:55.0  &  $0.018$  &  Ib & $59311$  &  $-14.8$  &  $59321$  &  $-14.8$  &  $0.02$ & \textcolor{black}{0}\\ \hline \\ 

ZTF21abdxhgv  &  SN2021qwm  &  15:18:25.73  &  +28:26:04.1  &  $0.070$  &  Ib/c & $59369$  &  $-17.1$  &  $59395$  &  $-18.8$  &  $0.07$ & \textcolor{black}{0}\\ \hline \\ 

ZTF22aapisdk  &  SN2022nwx  &  22:15:43.95  &  +37:16:47.0  &  $0.020$  &  Ib & $59755$  &  $-15.8$  &  $59764$  &  $-15.9$  &  $0.41$ & \textcolor{black}{0}\\ \hline \\ 

ZTF22aasxgjp$^{\textcolor{blue}{b}}$   &  SN2022oqm  &  15:09:08.21  &  +52:32:05.1  &  $0.011$  &  Ic & $59772$  &  $-16.3$  &  $59785$  &  $-17.3$  &  $0.05$ & \textcolor{black}{0}\\ \hline \\ 

ZTF21aacufip  &  SN2021vz  &  15:21:26.85  &  +36:46:04.0  &  $0.045$  &  Ic & $59223$  &  $-17.5$  &  $59232$  &  $-18.4$  &  $0.05$ & \textcolor{black}{0}\\ \hline \\ 


ZTF22aaezyos  &  SN2022hgk  &  14:10:23.70  &  +44:14:01.2  &  $0.033$  &  Ib & $59688$  &  $-16.8$  &  $59713$  &  $-18.0$  &  $0.02$ & \textcolor{black}{0} \\ \hline \\ 

ZTF21abmlldj  &  SN2021uvy  &  00:29:30.87  &  +12:06:21.0  &  $0.094$  &  Ib & $59449$  &  $-20.3$  &  $59536$  &  $-19.6$  &  $0.18$ & \textcolor{black}{0}\\ \hline \\ 

ZTF18abfcmjw$^{\textcolor{blue}{c}}$   &  SN2019dge  &  17:36:46.74  &  +50:32:52.1  &  $0.021$  &  Ib & $58584$  &  $-16.3$  &  $58591$  &  $-15.6$  &  $0.07$ & \textcolor{black}{0}\\ \hline \\ 

ZTF20aalxlis$^{\textcolor{blue}{d}}$   &  SN2020bvc  &  14:33:57.01  &  +40:14:37.6  &  $0.025$  &  Ic-BL & $58883$  &  $-17.0$  &  $58900$  &  $-19.0$  &  $0.03$ & \textcolor{black}{0}\\ \hline \\ 

ZTF19aamsetj$^{\textcolor{blue}{e}}$   &  SN2019cad  &  09:08:42.97  &  +44:48:46.0  &  $0.028$  &  Ic & $58567$  &  \textcolor{black}{$-17.9$}  &  $58594$  &  \textcolor{black}{$-19.2$}  &  $0.05$ & \textcolor{black}{1.1}\\ \hline \\ 

\end{tabular}  \label{tablesample} 
\smallskip\footnotesize
\begin{tablenotes}
\item $^a$\citet{Jacobson2022b}. $^a$\citet{Irani2022}. $^c$\citet{Yao2020}. $^d$\citet{Ho2020}. $^e$\citet{Gutierrez2021}.
\end{tablenotes}


\end{center} 

\end{table*}
\end{flushleft}

\section{Data}
\label{sec:data}

In this section, we describe the photometric and spectroscopic data used.

\subsection{Optical photometry}
\label{section:photdata}

We utilize forced photometry data from the ZTF in the $g$, $r$ and $i$ bands and from the Asteroid Terrestrial-impact Last Alert System \citep[ATLAS; ][]{Tonry2018, Smith2020} in the $c$ and $o$ bands. In addition, photometry data is obtained from the Palomar 60-inch telescope 
\citep[P60;][]{Cenko2006}, Sinistro imager on the 1-meter class and the Spectral imager
on the 2-meter class telescopes operated by Las Cumbres Observatory \citep[LCOGT;][]{Brown2013}, the Liverpool Telescope \citep[LT; ][]{Steele2004} in $g$, $r$ and $i$ bands. We also obtain $u$, $i$, $z$ band photometry for a few sources from the LT. P60 and LT data were processed using the FPipe \citep{Fremling2016} image subtraction pipeline with Sloan Digital Sky Survey \citep[SDSS;][]{Ahn2012a} and PanSTARRS 
\citep[PS1;][]{Chambers2016} reference images. Additionally, we have early-time UV data for some sources acquired from the Ultra-violet Optical Telescope (UVOT; \citealt{Roming2005}), which is deployed on the  \emph{Neil Gehrels Swift Observatory}  \citep{Gehrels2004}. UVOT data are reduced using \HEAsoft\footnote{\href{https://heasarc.gsfc.nasa.gov/docs/software/heasoft}{https://heasarc.gsfc.nasa.gov/docs/software/heasoft}}. The photometry data can be found in Appendix \ref{appendix: photometry}. Figure~\ref{fig:goodfig1} shows the lightcurve of \textcolor{black}{SN~2021gno} as an example. We compare the $r$-band absolute magnitude of the SN with our sample in Figure~\ref{fig:allrband}. Similar plots for the other SNe can be found in Appendix \ref{appendix: lightcurves}. Figure~\ref{fig:alllc} 
shows the lightcurves of all the SNe.

\subsection{Optical spectroscopy}
\label{section:spectra}

We acquired spectroscopy at multiple epochs for the SNe in our sample, covering a range from one day to over 300 days after explosion. Each transient typically has at least one spectrum near peak luminosity for initial classification and additional spectral follow-up was conducted as part of the ZTF surveys. Our primary classification instruments are the Spectral Energy Distribution Machine \citep[SEDM;][]{Blagorodnova2018} on the Palomar 60-inch telescope and the Double Beam Spectrograph \citep[DBSP;][]{Oke1982} on the Palomar 200-inch telescope. The DBSP spectra is reduced using the reduction pipelines described in \citet{Bellm2016} and \citet{Roberson2022}. The SEDM data is reduced using  the pipeline detailed in \citet{Rigault2019}. Additionally, we obtained spectra from the the Alhambra Faint Object Spectrograph and Camera on the Nordic Optical Telescope \citep[NOT;][]{Djupvik2010} and the Spectrograph for the Rapid Acquisition of Transients \citep[SPRAT;][]{Piascik2014}. The NOT data were reduced using the PyNOT\footnote{\href{https://github.com/jkrogager/PyNOT}{https://github.com/jkrogager/PyNOT}} and PypeIt \citep{Prochaska2020a} reduction pipelines, while we use the FrodoSpec pipeline \citep{Barnsley2012} for reduction of SPRAT data. We obtain late-time nebular-phase spectra with the Low-Resolution Imaging Spectrometer \citep[LRIS;][]{Oke1995} on the Keck I telescope, with data reduced using the automated \lpipe{} \citep{Perley2019} pipeline. The log of the observed spectra can be found in Table~\ref{table_vel}. Figure~\ref{spec1} shows the spectral sequence for SN 2021gno as an example. The spectral sequence of all the sources can be found in Appendix \ref{appendix: spectra}.

\section{Methods and Analysis} \label{sec:analysis}

\subsection{Extinction correction}\label{sec:extinction}



For precise estimation of the luminosity and explosion properties of a SN, it is essential to determine the impact of dust extinction along the observer's line of sight. Extinction is commonly divided into two components: the first component represents dust extinction from the Milky Way, while the second component accounts for extinction originating from the SN's host galaxy. To correct for Galactic extinction, we  employ the reddening maps provided by \citet{Schlafly11}. For reddening corrections, we use the extinction law described by \citet{Cardelli1989} with a value of $R_V = 3.1$.

\textcolor{black}{To estimate
the host-galaxy extinction, we measure the equivalent width $(\mathrm{EW})$ of the $\mathrm{Na\ {\sc I}\ D}$ absorption feature \citep{Poznanski:2013}. We measure a $\mathrm{EW}_\mathrm{Na\ {\sc I}\ D}$  of 1.5 $\mathrm{\AA}$, 5.5 $\mathrm{\AA}$ and  0.8 $\mathrm{\AA}$ for SN 2019cad, SN 2020nng and SN 2021nng, respectively. We do not see $\mathrm{Na\ {\sc I}\ D}$ absorption for the other sources in the high signal-to-noise spectra. Thus, we assume zero host extinction for the rest of the sources in our analysis. To compute $A_\mathrm{V}$ from the EW measurements, we use $A_\mathrm{V}^{\mathrm{host}} [\rm mag] = 0.78 \times EW_{Na\ {\sc I}\ D} [\mathrm{\AA}]$ \citep{Stritzingetr2018}. We measure $A_\mathrm{V}^{\mathrm{host}}$ = 1.2 mag for SN 2019cad and  $A_\mathrm{V}^{\mathrm{host}}$ = 0.6 mag for SN 2021nng. However, the empirical relation in \citet{Stritzingetr2018} is not valid for the high $\mathrm{EW}_\mathrm{Na\ {\sc I}\ D}$ measured for SN 2020kzs. Instead, we use the difference in the average color ($g-r$) of SN 2020kzs with the color expected for typical SNe Ib with no host extinction assuming the intrinsic template for Type Ib SNe provided in \citet{Stritzingetr2018}. Based on this, we measure $A_\mathrm{V}^{\mathrm{host}}$ = 1.1 mag for SN 2020kzs.}





\subsection{Measuring velocity in Photospheric Spectra}
\label{sec:photspectra}

As described in Section \ref{section:spectra}, we obtain spectra soon after explosion 
for all sources. We use the \texttt{SuperNova Identification} \citep[\texttt{SNID};][]{Blondin2007} code for the classifications. For spectra contaminated by host galaxy, we utilized the \texttt{superfit} \citep{Howell2005} code for classification. The final classification as Type Ibc SNe was determined through manual inspection of the emission and absorption lines and the best-fit templates matched from \texttt{SNID} or \texttt{superfit}.

We measure the expansion velocities of the He~I $\lambda 5876$ and O~I $\lambda 7774$ lines from the absorption part of the P-Cygni profiles of the spectral lines. To do this, we fit a polynomial function, whose degree is manually tuned for each spectrum (typically 3), to the minima of the P-Cygni profiles. These minima serve as estimates for the expansion velocity. In cases where the spectrum is galaxy lines dominated or has low resolution, we manually inspect the spectrum to determine the minima of the absorption feature.
The measured velocities are documented in Table~\ref{table_vel}. We adopt a Monte Carlo approach to estimate the uncertainties in our velocity measurements. We generate a noise spectrum by subtracting a heavily smoothed version of the spectrum from the original spectrum. The standard deviation of this noise spectrum provides an estimate of the noise of the spectrum. Next, we create simulated noisy spectra by adding noise from a standard Gaussian distribution with the calculated standard deviation. We then add these simulated spectra with the heavily smoothed spectra and recalculate the velocities. The 1$\sigma$ uncertainty in the velocity measurements across all the simulated spectra is considered as the standard deviation. \citet{Fremling2018} analyzed the spectra of a sample of 45 Type Ib SNe, 56 Type Ic SNe, 17 Type Ib/c and 55 Type IIb SNe discovered by the Palomar Transient Factory (PTF) and intermediate PTF (iPTF) surveys. We compare our measured velocities with those from \citet{Fremling2018}.
From Figure~\ref{Hecomp}, we find that the expansion velocities of the He~I $\lambda 5876$ and O~I $\lambda 7774$ lines are consistent with those of canonical Type Ibc SNe.

\subsection{Measuring Oxygen line flux in Nebular Spectra}
\label{section:nebspectra}

We obtained nebular phase spectra for ten sources with Keck and P200. A few of the nebular spectra for SN 2021gno and SN 2021inl were taken from \citet{Jacobson2022b} as noted in Table~\ref{table_neb}. We use interpolated late-time photometry to flux calibrate our nebular spectra. When late-time photometry is not available, we extrapolate the lightcurve by assuming a late-time ($>$ 30 days) $i$-band decline rate of $0.019\ \pm\ 0.004$ mag/day, based on the average late-time decay of the SESNe tabulated in \citet{Wheeler2015}. 
For each spectrum, we manually set the wavelength regions and measure the line fluxes using trapezoidal integration. Uncertainties in this method are estimated by Monte Carlo sampling of the estimated fluxes by adding noise (scaled to nearby regions of the continuum) to the line profile. Table~\ref{table_neb} presents the measured fluxes of [Ca~II] $\lambda \lambda$7291, 7324 and [O~I] $\lambda \lambda$6300, 6364, along with their flux ratio.

\subsection{Modeling light curves}\label{sec:bbfit}

\subsubsection{Blackbody fit}
We estimate the bolometric light curve for epochs where we have detections in at least two filters by fitting a blackbody function. For each epoch, we use a Python \emcee\ package \citep{Foreman-Mackey13} to perform a Markov Chain Monte-Carlo (MCMC) analysis in order to estimate the blackbody temperature ($T_\mathrm{BB}$), radius ($R_\mathrm{BB}$) and luminosity ($L_\mathrm{BB}$). The uncertainties of the model parameters are determined by extracting the  $16^\mathrm{th}$ and $84^\mathrm{th}$ percentiles of the posterior probability distribution. \textcolor{black}{We note that UV coverage is only available for SNe 2020bvc, 2022oqm, 2021gno. For these three SNe, the blackbody fit is done on the available UV$-$optical photometry. For the rest, the blackbody fit is done on the available optical photometry only.} The best-fit parameters can be found in Appendix \ref{appendix: blackbody}.

\subsubsection{Fitting Shock Cooling in first-peak}
\label{sec:scb}

In our sample, all sources exhibit a lightcurve characterized by two distinct peaks. The rapid rise of the first peak, accompanied by an initial blue color and high temperature, indicates that the first peak is likely dominated by cooling emissions from the shock-heated extended envelope \citep{nakarpiro2014, piro2015}. 
We plot the peak luminosity versus time above half maxima in Figure~\ref{fig:lumtime}. The same is shown for the first peak of all double-peaked Type IIb SNe (not part of the sample in this work) from BTS+CLU. We note that 43 Type IIb out of 193 Type IIb SNe had detections of two peaks (Das et al. in prep). In the right panel of Figure~\ref{fig:lumtime}, we plot the peak $r-$band magnitude of the first peak vs the peak $r-$band magnitude of the second peak. For the first time, we find that a correlation exists between the peak magnitudes of the first and the second peak. \textcolor{black}{The Pearson correlation coefficient is 0.79 (p $<10^{-5}$). A similar correlation is also seen for the $g-$band, with a Pearson correlation coefficient of 0.81 (p $<10^{-5}$).} The physical reason for this correlation is not clear. The first peak brightness is primarily dependent on the radius of the progenitor while the peak of the second peak is primarily dependent on the Ni-mass. The correlation could imply that the SNe that show double-peaked lightcurves come from He-star progenitors that shed their envelope in binary interactions. Then, this correlation could be related to the He main sequence \citep[see Figure 5 in][]{Sravan2020}, with the progenitor radius being related to the effective temperature and the Ni-mass being related to the luminosity. This would require that the Ni mass is correlated with the ejecta mass \citep{Lyman2016}. \textcolor{black}{This assumes that stripped-envelope SNe come from He-stars.} Such a correlation could also exist if the first peak is also powered by nickel. This is possible if nickel is not entirely in the core but is also present in the outer envelope. However, it is unlikely that this trace amount of nickel can make a significant contribution to the early luminosity. We note that our survey is biased against sources that have a very faint first peak luminosity.

It is important to note that some of the SNe in our sample do not have well-sampled first peaks in both the rising and fading phases.  To fit the multi-band photometry in the shock-cooling phase, we use the model proposed by \citet{Piro2021}. This model allows us to determine key parameters, such as the explosion time ($\mathrm{t_{exp}}$), extended material mass ($\mathrm{M_{ext}}$),  radius ($\mathrm{R_{ext}}$) and energy ($\mathrm{E_{ext}}$). We use the Python \emcee\ package \citep{Foreman-Mackey13} to perform a multiband photometry data fitting analysis. We add a systematic error of 50\% to account for uncertainties in density and opacity assumptions used in the model. Table~\ref{table_piro} and Appendix \ref{appendix: scb} provide the best-fit values and corresponding fits for each SN. 

In Figure \ref{fig:dplit}, we compare the best-fit parameters with those for some H-poor SNe for which CSM interactions were detected. This CSM radii cover a wide range of distances from the explosion site, from $ \sim 3 \times 10^{13} - 10^{18}$ cm. The range of inferred CSM masses is also broad, spanning  from $\sim 10^{-4}$ \Msun\ up to tens of \Msun\ of material (Figure \ref{fig:dplit}). We note that the physical parameters have been estimated with a variety of observational ``tracers".


\begin{figure*}
    \centering
    \includegraphics[width=13.5cm]{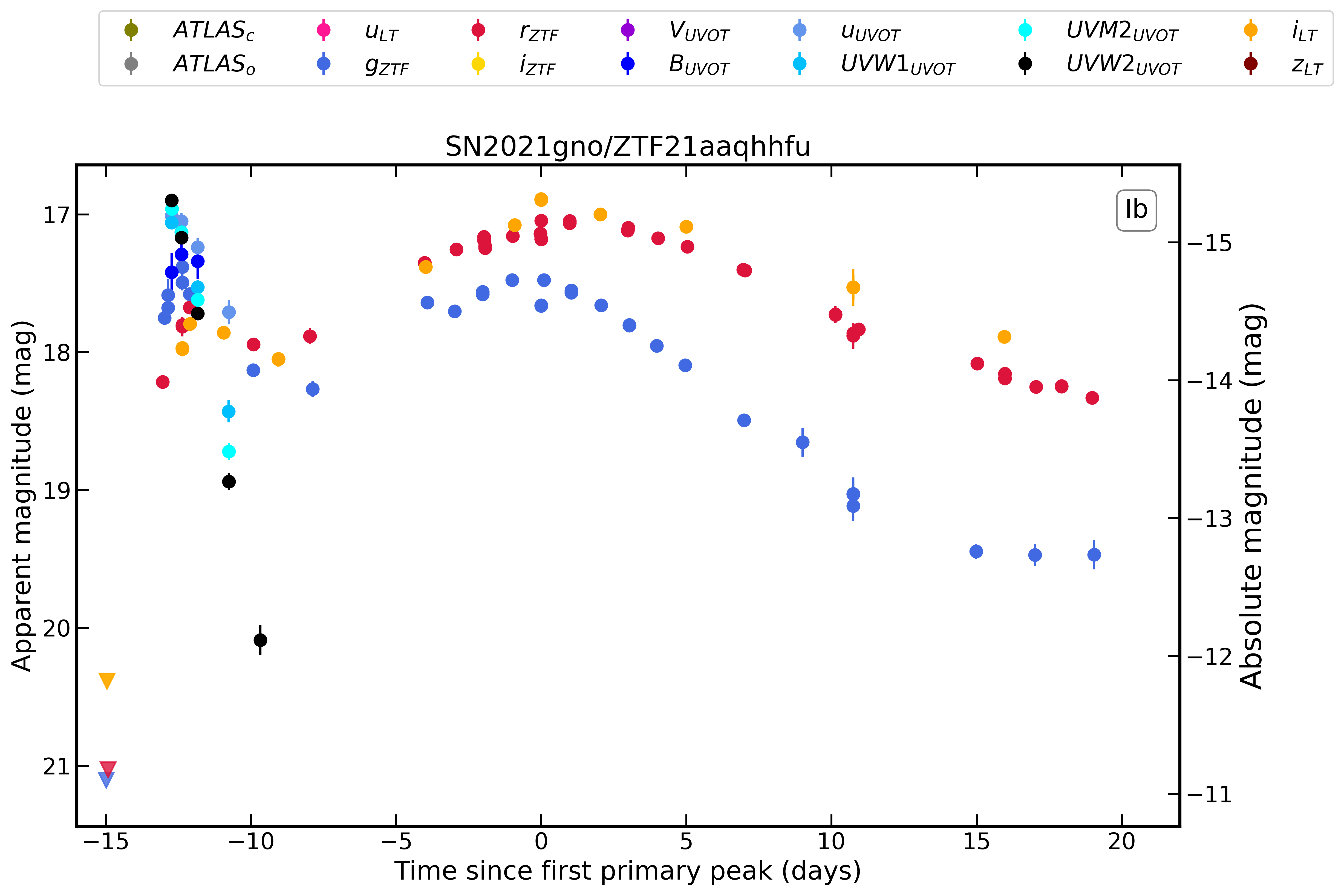}
    \caption{Lightcurve of SN 2021gno ($E(B-V)_\mathrm{MW} = 0.01$). The lightcurves of the other SNe can be found in Appendix \ref{appendix: lightcurves}.}
    \label{fig:goodfig1}
\end{figure*}

\begin{figure*}
    \centering
    \includegraphics[width=13.5cm]{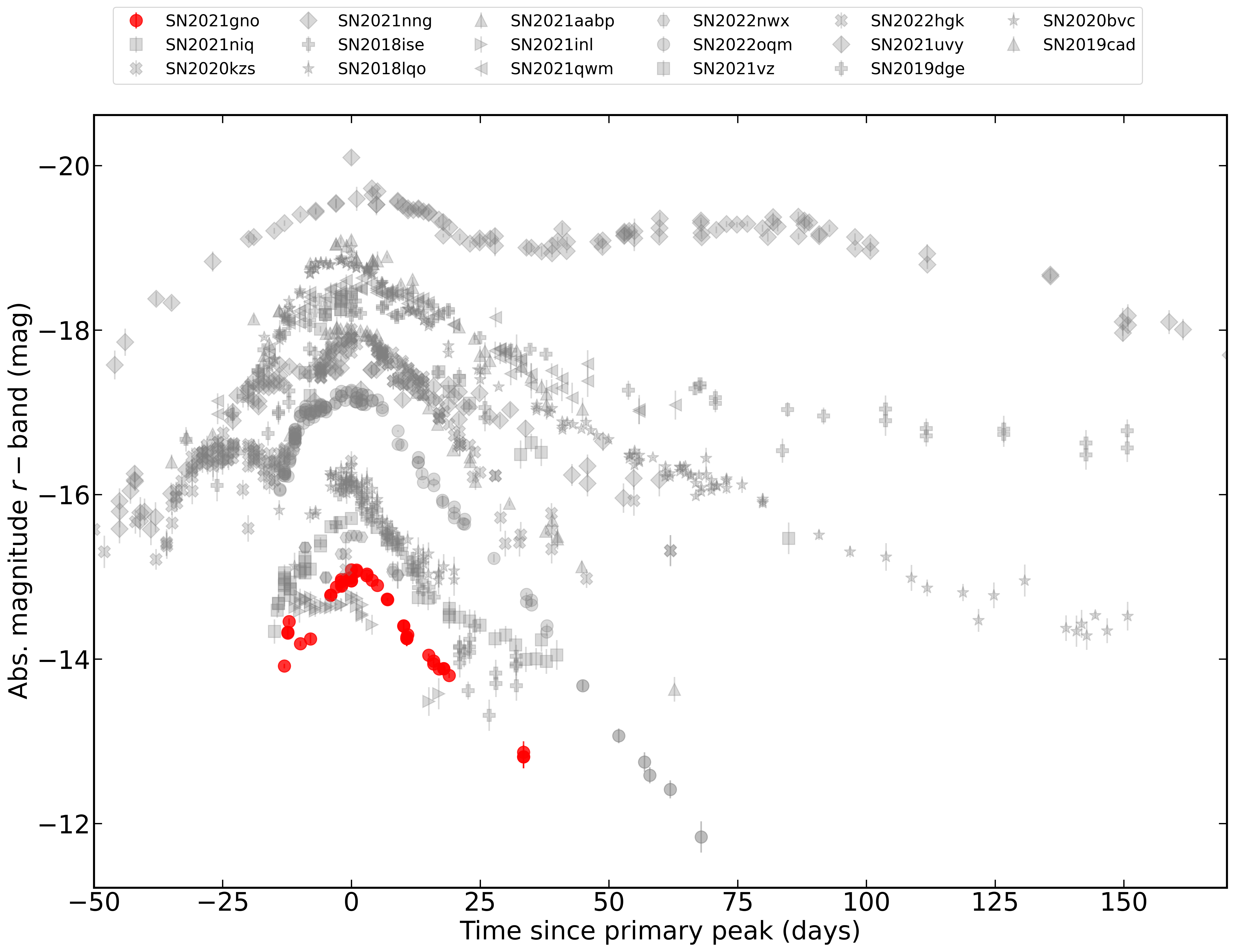}
    \caption{Comparison of the $r$-band absolute magnitude light curve of SN 2021gno to the other sources in the sample. The comparison for the other sources can be found in Appendix \ref{appendix: lightcurves}.}
    \label{fig:allrband}
\end{figure*}

\subsubsection{Shock Cooling order-of-magnitude limits for the first-peak}
\label{sec:shock cooling OOM limits}

We make the assumption that the layer going through shock cooling has a radius $R_\mathrm{ext}$ and mass $M_\mathrm{ext}$. The expansion timescale is $t_\mathrm{exp}\sim R_\mathrm{ext}/v_\mathrm{ext}$, where $v_\mathrm{ext}$ is the velocity of this layer. Photons undergo diffusion from this layer within a timescale approximately given by $t_\mathrm{diff}\sim\tau R_\mathrm{ext}/c$. The bulk of photons emerge from the layer where $t_\mathrm{exp} = t_\mathrm{diff}$ or $\tau \sim c/v_\mathrm{ext}$.

We assume $\rho \sim M_\mathrm{ext}/(4\pi R^3 / 3)$. At a specific radius, the optical depth $\tau$ decreases as a result of expansion: $\tau \sim \kappa \rho R$. The radius increases as $R\sim v_\mathrm{ext} t$, so $\tau \sim 3 \kappa M_\mathrm{ext} / (4 \pi (v_\mathrm{ext} t)^2)$. Setting this equal to $c/v_\mathrm{ext}$,

\begin{equation}
    t \sim \left( \frac{3}{4 \pi } \frac{\kappa M_\mathrm{ext}}{v_\mathrm{ext} c} \right)^{1/2}.
\end{equation}

We have an upper limit on the time to peak of $t_p$ as the epoch of the first peak calculated from the analytical model described in the previous section. We take $\kappa = 0.2\ \mathrm{cm^2 g^{-1}}$ for a hydrogen-poor gas and $v_\mathrm{ext} \sim 0.1c$. Altogether, we find $M_\mathrm{ext} \sim 0.01-1$ \Msun. Note that the predicted values are upper limits because the rise time was likely faster than our measurements. The limits are listed in Table~\ref{table_oom}. The values obtained from the analytical model described in the previous section are consistent with the limits obtained.

Next, we estimate the radius $R_\mathrm{ext}$. If the shock deposits energy $E_\mathrm{dep}$ into the layer, which then cools from expansion, we can estimate the energy $E_\mathrm{cool}\sim E_\mathrm{dep} (R_\mathrm{ext} / v_\mathrm{ext} t)$. Thus, the luminosity from cooling is $L_\mathrm{cool} \sim E_\mathrm{dep} R_\mathrm{ext}/v_\mathrm{ext} t^2$. We assume that the deposited energy is half the kinetic energy $E_\mathrm{KE}$ of the shock, $E_\mathrm{dep} = \pi R_\mathrm{ext}^2 dR \rho v_s^2$, where $\rho$ and $dR$ are the density and width of the layer. Taking $\rho \sim M_\mathrm{ext} / (4 \pi R_\mathrm{ext}^2 dR)$  and $dR \approx R_\mathrm{ext}$, we find 

\begin{equation}
    L_\mathrm{cool} \sim \frac{v_\mathrm{ext} R_\mathrm{ext} M_\mathrm{ext}}{4 t^2}.
\end{equation}

Taking the above $M_\mathrm{ext}$, $t_p$ values, $v_\mathrm{ext}=0.1c$, and lower limits on the peak luminosity from the bolometric blackbody fits,
we find $R_\mathrm{ext}$ in the range $\approx 10-200$ \Rsun.
We can only measure a lower limit on the radius because the true peak luminosity is likely higher than what we can measure.

The limits are listed in Table \ref{table_oom}. The values obtained from the analytical model described in the previous section are consistent with the limits obtained.


\subsubsection{Ruling out Shock breakout from CSM for the first-peak}

In this section, we conduct a rough estimation to determine if the rise time and peak luminosity can be accounted for a model in which shock interaction powers the light curve (``wind shock breakout'').

The shock crossing timescale is $t_\mathrm{cross} \sim R_\mathrm{CSM}/v_s$, which is $\sim$ 0.01 day, assuming shock velocity ($v_s \approx 0.1c$) for the observed radius range, which is around two orders of magnitude less than the observed timescale. The estimated limits are listed in Table \ref{table_oom_csmbo}.


The shock heats the CSM with an energy density that is roughly half of the kinetic energy of the shock, so the energy density of the CSM $\sim (1/2)(\rho v_s^2/2)$.
The luminosity is the total energy deposited divided by $t_\mathrm{cross}$,
\begin{equation}
    L_\mathrm{BO} \sim \frac{v_s^3}{4} \frac{dM}{dR}, 
\end{equation}
which is $> 10^{44} \mathrm{erg}$, again a few orders higher than the observations, assuming a constant density.

    
Therefore, considering shock velocities ($0.1c$) comparable to the observed expansion of the photospheric radius, Table~\ref{table_oom_csmbo} indicates that we would require higher values for $\mathrm{M_{CSM}}$ than what is expected for unbound CSM. Thus we rule out this as a possible explanation for the early bump.

\subsubsection{Modeling the radioactively powered second-peak}
\label{sec:nipeak}

In this section, we describe the modeling of the second peak of the SNe. First, we estimate the contribution of the cooling emission to the bolometric luminosity using the best-fit parameters obtained in Section \ref{sec:scb}. This cooling component is then subtracted from the bolometric lightcurves obtained through blackbody fitting. We employ two methods to fit the peak, assuming it is powered by radioactive decay.
Firstly, we apply the analytical model outlined in \citet{Arnett89}, \citet{Valenti2008}, and \citet{Wheeler2015}. Using this model, we constrain the characteristic photon diffusion timescale ($\tau_m$), characteristic 
 $\gamma$-ray diffusion timescale
($t_o$) and nickel mass (\Mni). Additionally, we use relations from \citet{Wheeler2015} that provide the kinetic energy in the ejecta ($E_\mathrm{kin}$) and the ejecta mass (\Mej) as functions of photospheric velocity ($v_\mathrm{ph}$) and ($\tau_m$). We use the $v_\mathrm{ph}$ measured using the average He I line and O I velocity  from the photospheric spectra within 5 days of the second peak epoch for Type Ib and Type Ic(BL) SNe respectively listed in Section \ref{sec:photspectra}. If there are no velocity measurements available from spectra within 5 days of the second peak, we assume an average velocity of 8000 $\mathrm{km\ s^{-1}}$. For SN~2020bvc, we use $v_\mathrm{ph} = 18000\  \mathrm{km\ s^{-1}}$ derived in \citet{Ho2020}. Secondly, we use the lightcurve analytical models given in \citet{KhatamiKasen2019} to estimate the various explosion parameters. Further details on the model fitting can be found in \citet{Yao2020} (their Appendix B). Figure \ref{fig:mnimej} shows the parameter space occupied by these transients with ejecta mass varying from $\approx$ 0.2 -- 7 \Msun\ and nickel mass varying from 0.01 -- 0.5 \Msun. \textcolor{black}{For SN 2021inl, we note that the estimated ejecta mass and kinetic energy values are higher than those estimated in \citet{Jacobson2022b} as they used a lower photospheric velocity of 7500  $\mathrm{km\ s^{-1}}$, instead of 14350  $\mathrm{km\ s^{-1}}$ used in this work.}  We compare the ejecta mass and nickel mass with those from \citet{Taddia2018} in Figure \ref{fig:mnimej}. The best fit parameters and fits are provided in Table \ref{table_arnett} and Appendix \ref{appendix: arnettfits}.

\begin{figure*}
  \begin{minipage}[b]{\textwidth} 
    \centering
    \includegraphics[width=\textwidth]{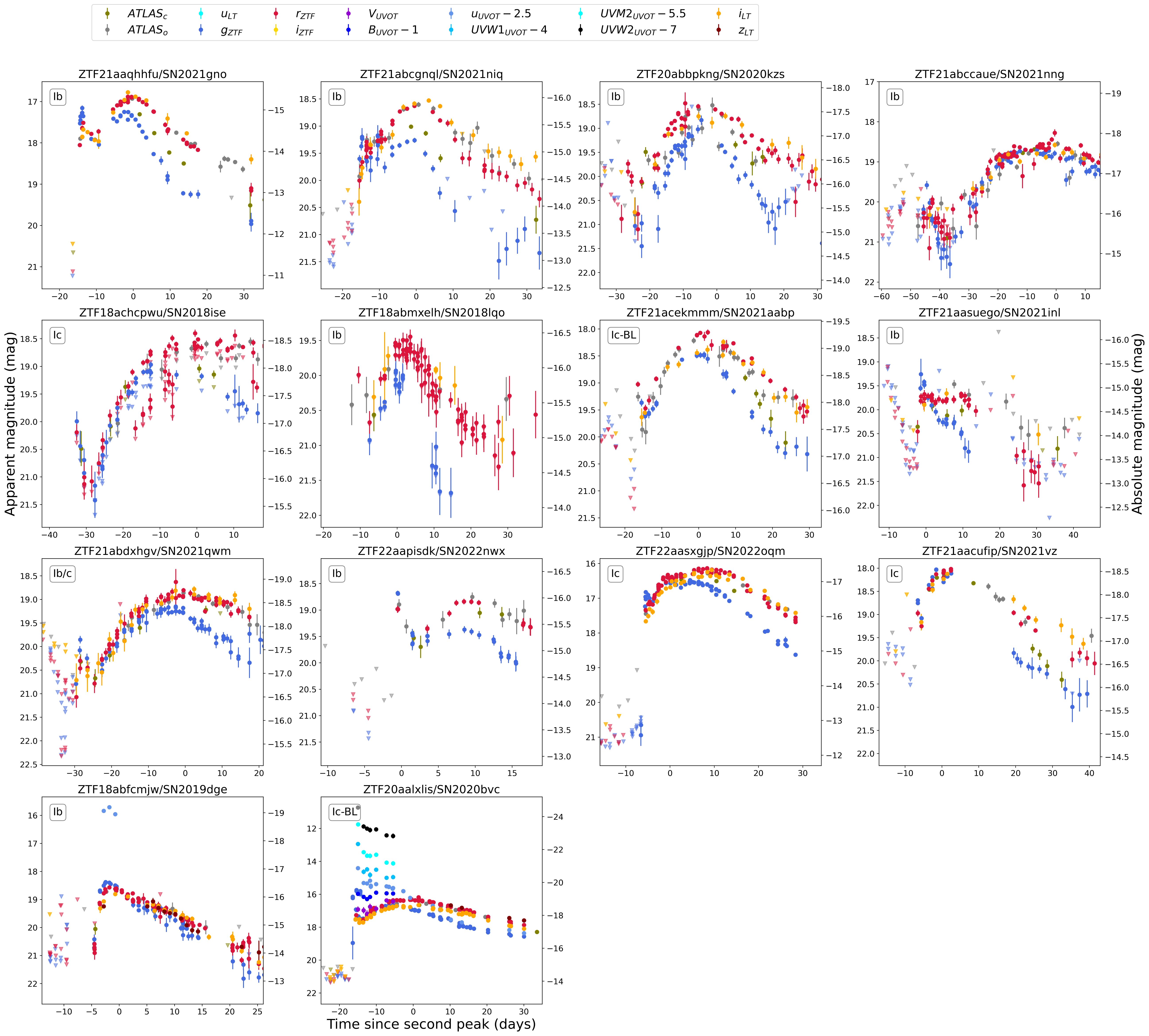} 
  \end{minipage}
  \hfill
  \begin{minipage}[b]{\textwidth} 
    \centering
    \includegraphics[width=\textwidth]{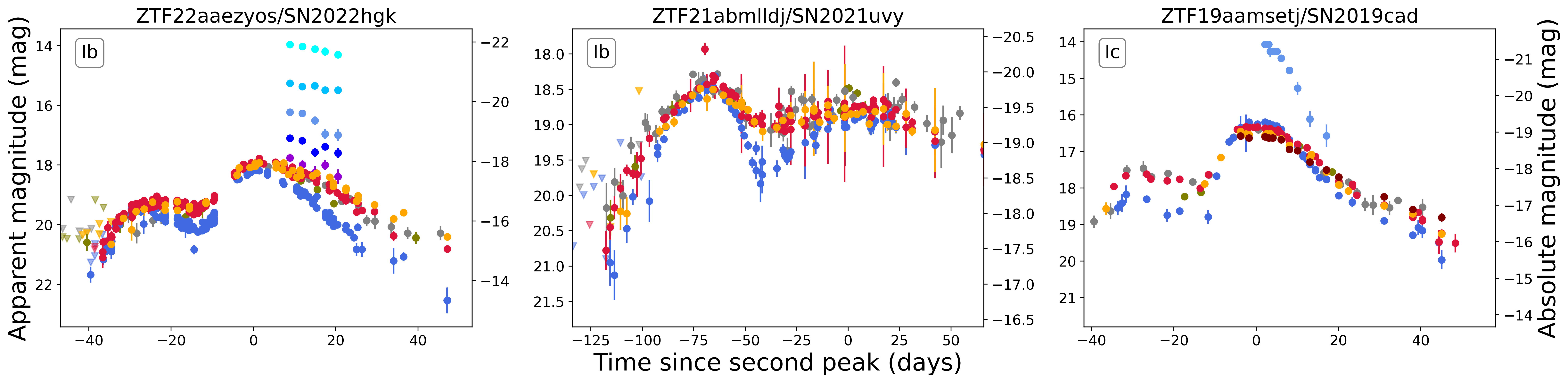} 
  \end{minipage}
  \caption{\textit{Top:} Lightcurves of double-peaked Type Ibc SNe in our sample, obtained through forced-photometry from ZTF, ATLAS, and follow-up observations from various instruments. Further details on the photometry can be found in Section~\ref{section:photdata}. The left y-axis represents the apparent magnitude (mag), while the right y-axis shows the absolute magnitude (mag). The absolute magnitude measurements assume Milky Way and host extinction values from Table~\ref{tablesample}. The x-axis shows the number of rest-frame days since the epoch of the second peak. \textit{Bottom:}    The three SNe in the bottom panel have an unusually long first peak with $t_{1/2}^\mathrm{first} > 15$ days. We leave the detailed lightcurve analysis of these SNe for future work.}
  \label{fig:alllc}
\end{figure*}








\begin{figure*}
    \centering
    \includegraphics[width=13.5cm]{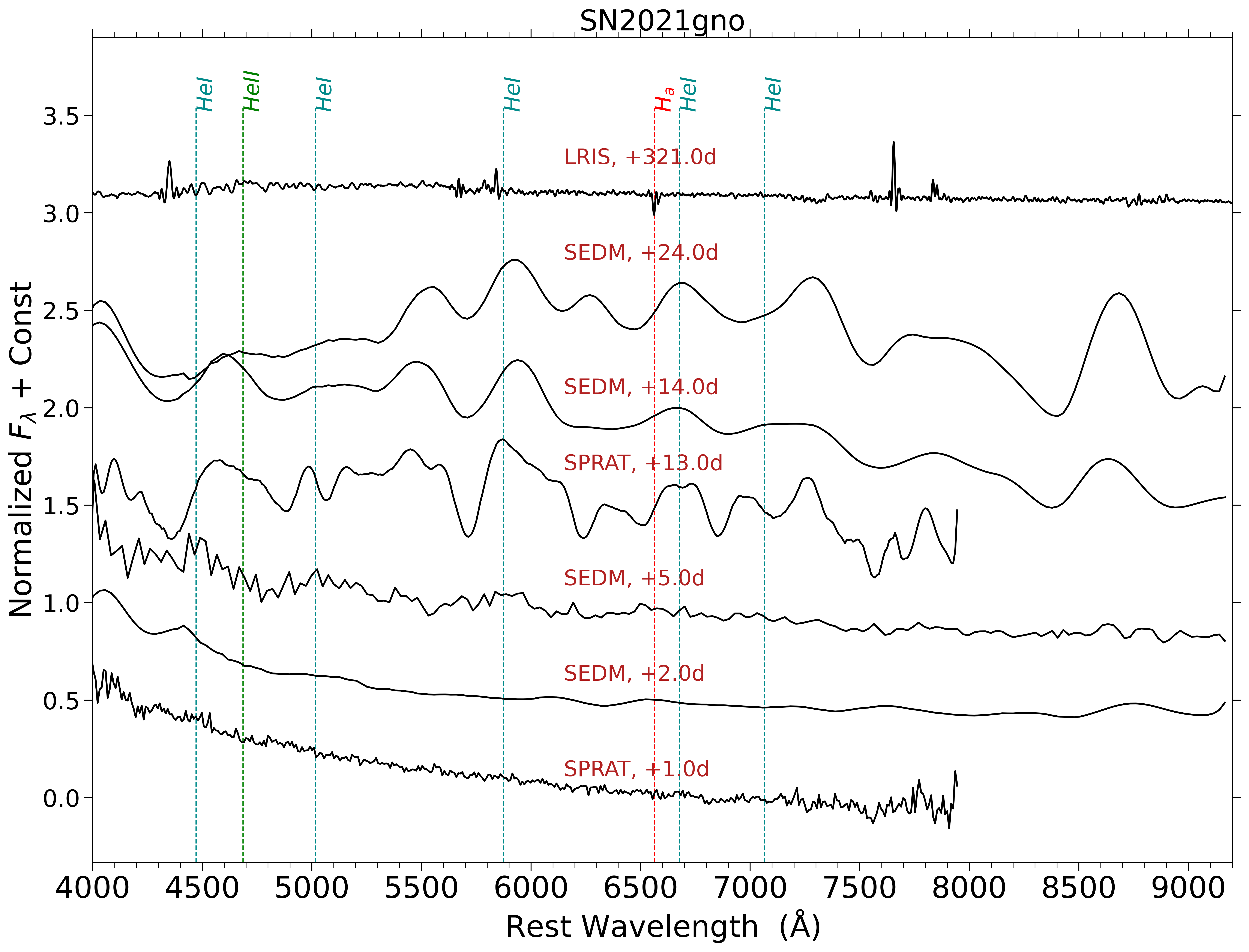}
    \caption{Spectral sequence for SN 2021gno (Type Ib) taken as part of the ZTF and CLU surveys. See Section~\ref{sec:photspectra} for details on the spectra obtained. Spectral sequence for the other sources can be found in Appendix \ref{appendix: spectra}.}
    \label{spec1}
\end{figure*}

\begin{figure*}
    \centering
    \includegraphics[width=0.49\textwidth]{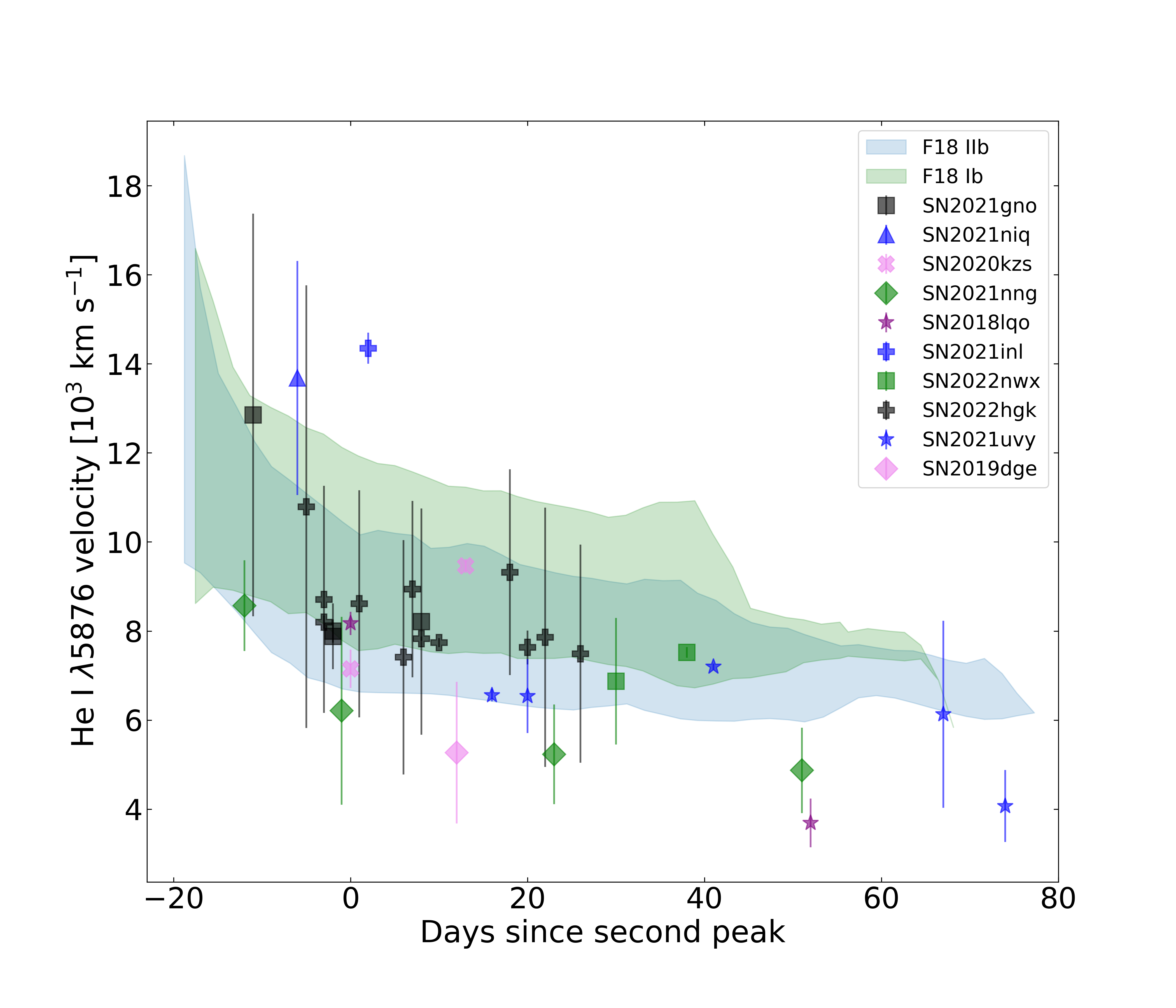}\includegraphics[width=0.49\textwidth]{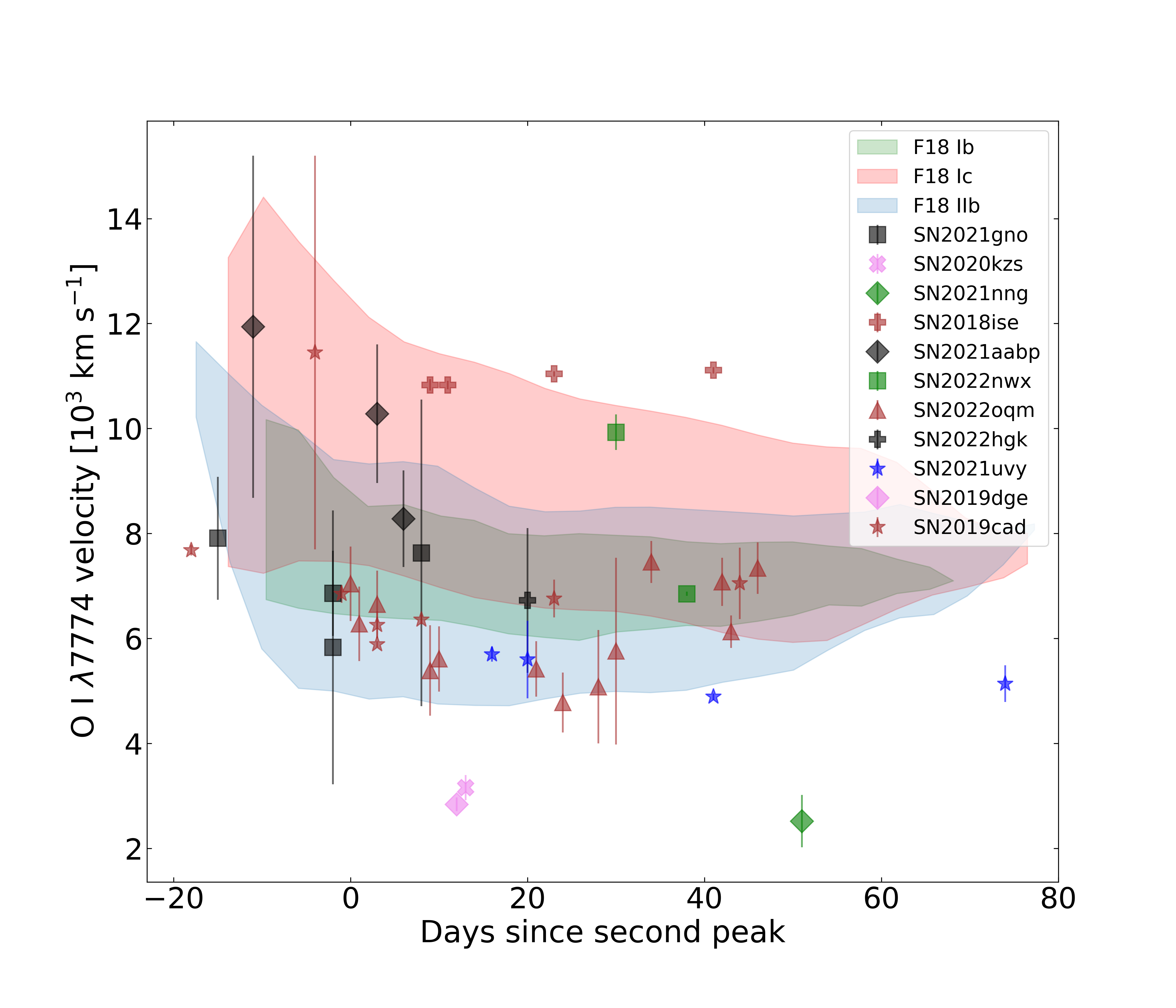}
    \caption{\small {\it Left:} The filled shapes represent the expansion velocity of the He I $\lambda$5876 line for each SN in our sample. The blue-shaded region indicates the  $1\sigma$ range of the ejecta velocities calculated for a sample of canonical Type IIb and Type Ibc SNe from \citet{Fremling2018}. {\it Right:} The filled shapes represent the expansion velocity of the O I $\lambda$7774 line for each SN in our sample. Again, the blue-shaded region represents the  $1\sigma$ range of the ejecta velocities calculated for a sample of normal Type IIb and Type Ibc SNe from \citet{Fremling2018}.}
    \label{Hecomp}
\end{figure*}

\begin{figure*}
    \centering
    \includegraphics[width=13.5cm]{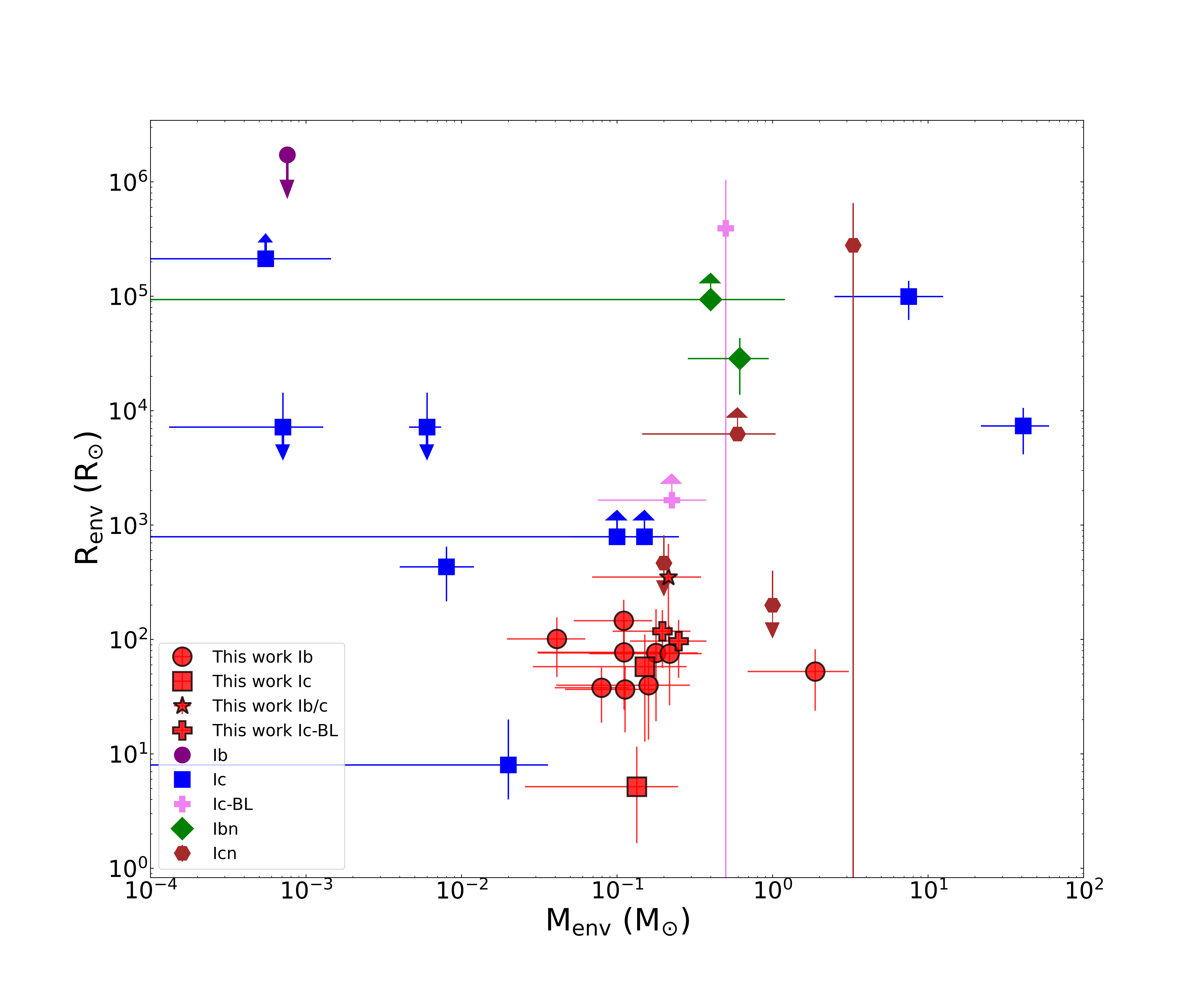}
    \caption{Comparison of the CSM parameters of the sample in this paper with CSM properties of other SNe in the literature \citep[from][]{Brethaeur2022}. \textcolor{black}{We note that the physical parameters have been estimated with a variety of observational ``tracers" and hence  probe different regions of the CSM.}}
    \label{fig:dplit}
\end{figure*}
\begin{figure*}
    \centering
    \includegraphics[width=13.5cm]{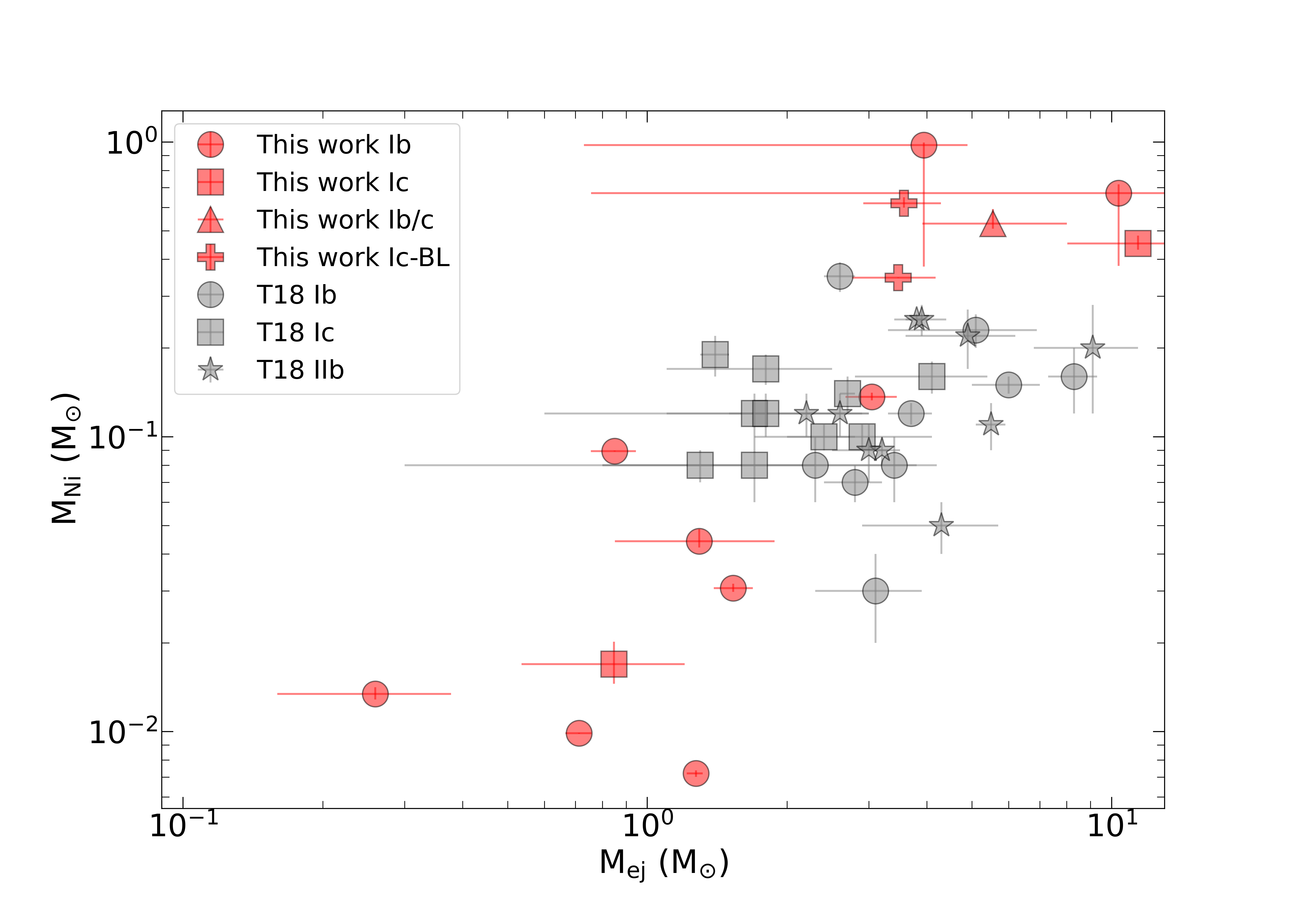}
    \caption{The distribution of measured $^{56}$Ni mass vs ejecta mass for SNe in this sample compared to normal Type Ibc, IIb SNe from \citet{Taddia2018}.}
    \label{fig:mnimej}
\end{figure*}

\begin{figure}
    \centering
    \includegraphics[width=8.9 cm]{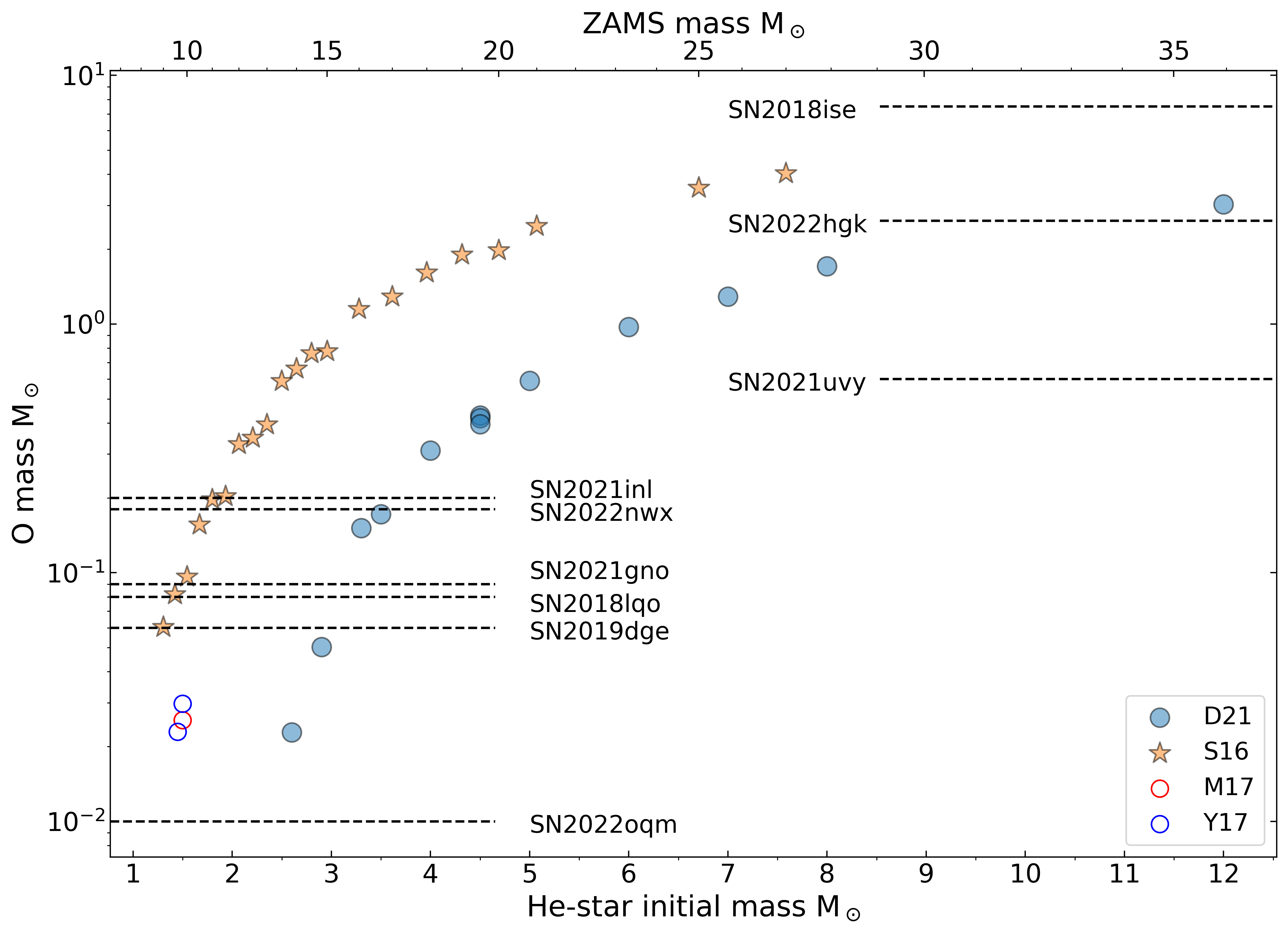}
    \caption{The O mass measurements are depicted by horizontal dashed lines. The O yields from the He-star progenitor models, assuming binary evolution  \citep{Dessart2021}, are represented by solid blue dots. The O yields from the single-star models  \citep{Sukhbold2016} are indicated by filled orange stars. Additionally, we show the O mass predictions from nucleosynthetic models of lower ZAMS stars from USSNe models by \citet{Yoshida2017} and \citet{Moriya2017}.}
    \label{fig:Omass_dessart}
\end{figure}

\begin{figure}
    \centering
    \includegraphics[width=8.9 cm]{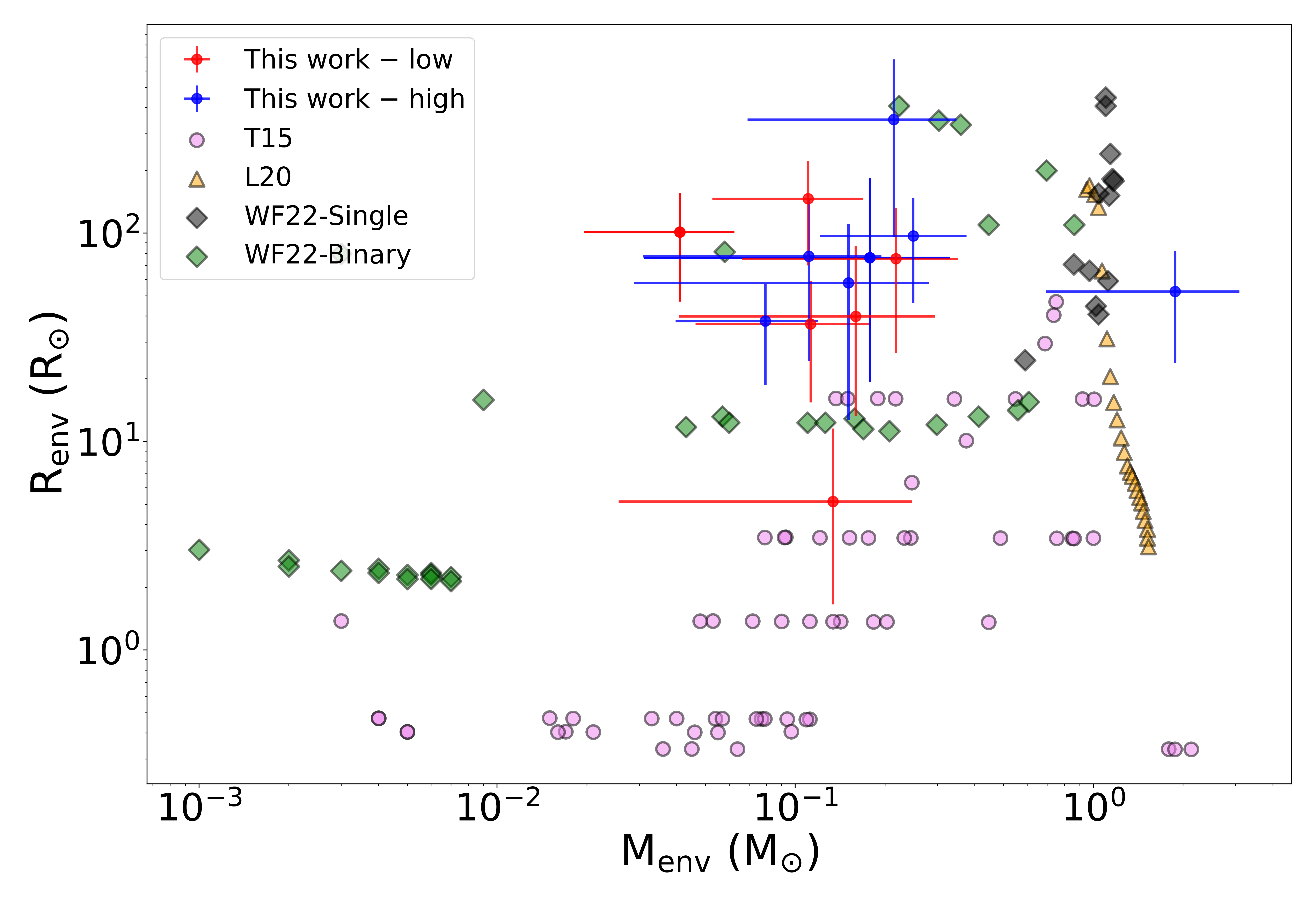}
    \caption{Comparison of the envelope parameters derived using analytical shock-cooling model as described in Section \ref{sec:scb} (in red cross) with \textit{bound} envelope properties from various binary and single star models \citep[WF22;][]{Wu2022b}, \citep[L20;][]{Laplace2020} and \citep[T15;][]{Tauris2015}. The SNe with low progenitor masses  ($\lesssim 12$ \Msun) and higher progenitor masses are shown in red and blue respectively.}
    \label{fig:bound}
\end{figure}



\begin{figure*}
    \centering
    \includegraphics[width=15.5cm]{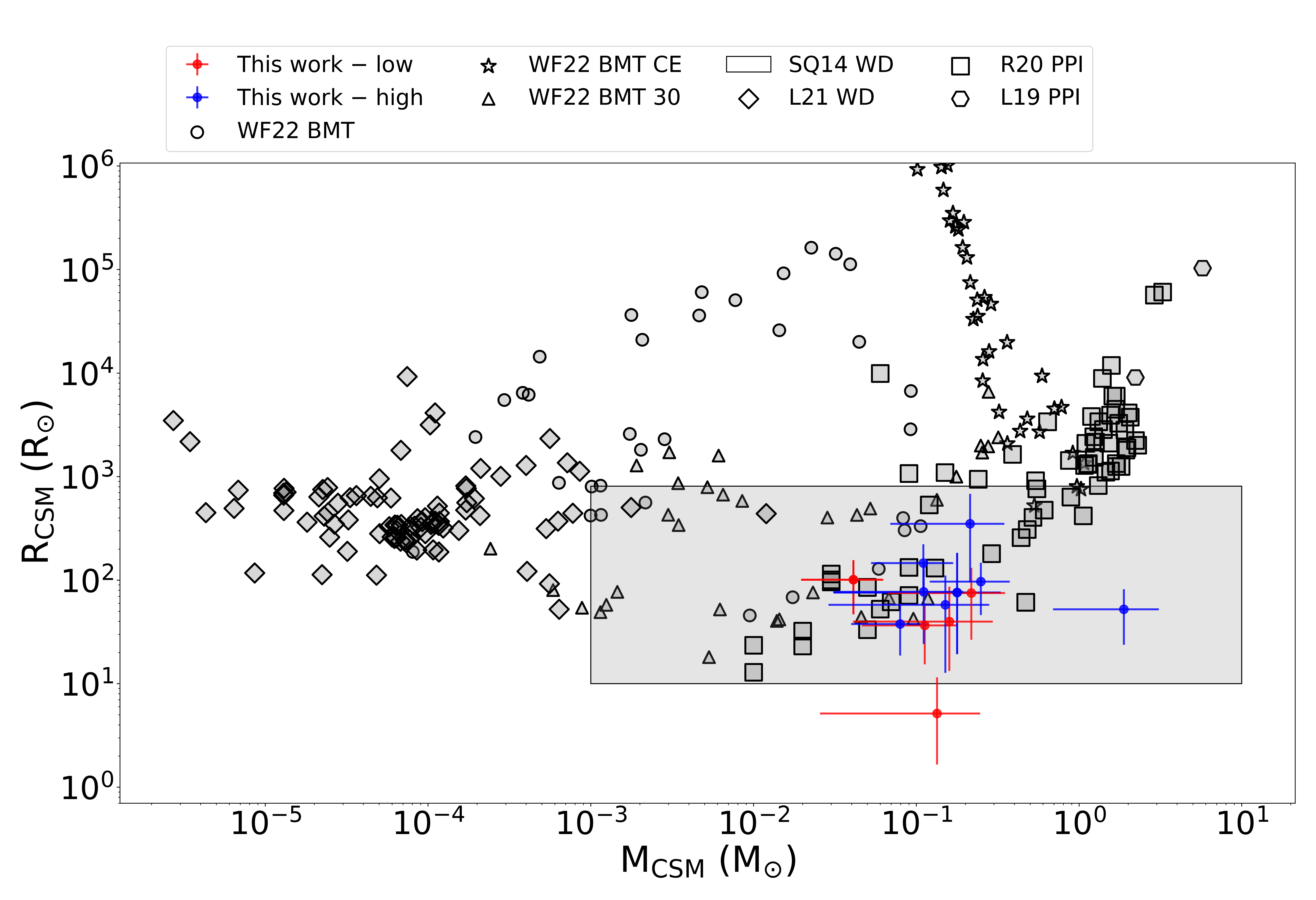}
    \caption{Comparison of the envelope parameters derived using analytical shock-cooling model as described in Section \ref{sec:scb} (in red cross) with \textit{unbound} CSM predictions from various pre-SN mass-loss models: late-time  stable binary mass transfer \citep[BMT;][]{Wu2022b}, late-time  unstable binary mass transfer via common envelope \citep[BMT CE;][]{Wu2022b}, late-time binary mass transfer with shock breakout at optical depth $\sim$ 30 \citep[BMT 30;][]{Wu2022b}, wave-driven mass loss \citep[WD;][]{Leung2021b, Shiode2014}, pulsation-pair instability driven  mass loss \citep[PPI;][]{Renzo2020, Leung2019}. \textcolor{black}{The SNe with low progenitor masses  ($\lesssim 12$ \Msun) and higher progenitor masses are shown in red and blue respectively.}}
    \label{fig:ppi}
\end{figure*}

\section{Constraining progenitor mass}
\label{sec:zams}

The late-time evolution of a star, including pre-SN mass loss is strongly dependent on the progenitor mass. In this section, we try to provide rough estimates of the progenitor mass based on the nebular spectra and the lightcurves.

We have at least one nebular spectrum for ten SNe obtained using LRIS on the Keck I telescope. Using the procedure described in Section \ref{section:nebspectra}, we calculate the [Ca~II] $\lambda \lambda$7291, 7324 to [O~I] $\lambda \lambda$6300, 6364 flux ratio and determine the [O I] $\lambda \lambda$6300, 6364 fluxes  (Table~\ref{table_neb}). Next, we use these [O I] luminosity measurements to compute the oxygen abundance and, subsequently, the progenitor mass. To determine the minimum required oxygen mass for a given [O~I] luminosity, we use the analytical relation in \citet{Uomoto1986}.
This analytical formula is applicable where the electron density is higher than $\sim 7 \times 10^5\ \mathrm{cm^{-3}}$. This is estimated to be valid for our case, with ejecta mass in the range of $0.3-6$ \msun. We use temperature values of $\approx$ $3500-4000$ K estimated in other core-collapse SNe from the [O~I] emission \citep{Sollerman1998, Elmhamdi2011}. Using this, we get an estimate of O mass in our sample in the range of $\approx$ $0.001-1$ \Msun (see Table~\ref{table_neb}). 

We use these O-mass estimates to constrain the progenitor mass.
To achieve this, we refer to the work of \citet{Dessart2021}, who conducted 1D non-local thermodynamic equilibrium radiative transfer calculations specifically for nebular-phase stripped SNe. Strong [Ca~II] and weak [O~I] emission is predicted for lower mass He stars. The high [Ca~II]/[O~I] flux ratio we observe for SNe 2021gno, 2021inl, 2022nwx, 2022oqm, 2018lqo in our sample is indicative of a  low initial He-star mass progenitor. In Figure~\ref{fig:Omass_dessart}, we present a comparison of the measured O mass in our sample and the synthesized O mass obtained from He-star progenitor models from both binary evolution \citep{Dessart2021} and single-star models \citep{Sukhbold2016}. Of the 14 SNe consistent with shock-cooling, we find that the SNe with progenitor mass less than 12 \Msun~are SNe 2019dge, 2021gno, 2021inl, 2022nwx, 2022oqm, 2018lqo. To determine the progenitor mass from the He-star mass, we use the relation provided in \citet{Woosley2015}.  

In order to make a comparison with such low progenitor masses, we also consider estimates of the O synthesized in the case of ultra-stripped SNe (USSNe). USSNe arise from low-mass He stars ($< 3.5\ M_\odot$) that have been highly stripped by a binary companion in a close orbit, leaving behind CO cores with approximate masses ranging from 1.45 to 1.6 $M_\odot$ at the time of the explosion \citep{Tauris2015}. It is worth noting that the CO core mass serves as a reliable indicator of the ZAMS mass, as it remains unaffected by binary stripping \citep{Fransson1989, Jerkstrand2014, Jerkstrand2015}. We find that the O-yields for the CO-cores of USSNe are higher than five SNe in our sample (see Figure \ref{fig:Omass_dessart}).

We caution that these measurements assume that the radioactive energy deposited in the O-rich shells is primarily released through cooling in the [O~I] lines. However, presence of impurity species can affect the [O~I] luminosities. E.g., \citet{Dessart2020} showed that if Ca is mixed into the O-rich regions, the [O~I] line emissions are weakened.
Nevertheless, extensive studies on CCSNe have indicated that mixing is not significant in these events. Detailed modeling of CCSNe has revealed that the [Ca~II] lines are the primary coolant in the Si-rich layers while the emission from [O~I] originates from the outer layers rich in oxygen, formed during the hydrostatic burning phase \citep{Jerkstrand2015, Dessart2020}. Additionally, \cite{Polin2021} have demonstrated that even a contamination of 1\% level of $^{40}$Ca can cool a nebular region entirely through [Ca~II] emission. Thus, if these ejecta regions were mixed, it would be challenging to observe the emission of [O~I] line.

We note that the low ejecta mass ($\lesssim 1.5$ \Msun) for SNe 2021gno, 2021inl, 2022nwx, 2018lqo, 2021niq are consistent with those predicted for the lower end of the He-star mass stars based on predicted ejecta properties of H-poor stars \citep[e.g.,][]{Dessart2021}. Nebular spectra estimates for all the above SNe are also consistent with low ZAMS mass, except for SN~2021niq for which we don't have any nebular spectrum. Also, for SN~2022oqm, the ejecta mass is not consistent with the progenitor mass estimate from the nebular spectra. In this paper, we consider SN~2021gno, SN~2021niq, SN~2021inl, SN~2022nwx, SN~2018lqo, SN~2019dge as potential SNe with progenitor masses less than $\sim$12 \Msun.


\section{Mass-loss scenarios}
\label{sec:massloss}


In the previous sections, we presented the results from the analysis of our double-peaked Type Ibc(BL) sample that included lightcurve and spectral properties. In this section, we try to understand the physical process that gave rise to the first peak. The early bump is most likely due to interaction with the external stellar material that is part of the extended bound envelope of a massive star or unbound material ejected in a pre-SN mass loss event. \textcolor{black}{There are other evidences for CSM interaction for some of the SNe in our sample. For example, \citet{Jacobson2022b} measured a CSM mass of $0.3-1.6 \times 10^{-3}$ \Msun\ that extends upto $5 \times 10^{14}$ cm. Luminous x-Ray and radio counterparts were observed for SN 2020bvc \citep{Ho2020}. \citet{Irani2022} predicts the presence of C/O-rich CSM at $2-5 \times 10^{14}$ cm based on early-time spectra. Similarly, early-time spectra for SN 2019dge \citep{Yao2020} was used to constrain the distance of He-rich CSM at $\gtrsim 3 \times 10^{13}$ cm.} The sample provides a unique opportunity to understand the origin of the CSM from late-time stellar evolution. There are different theoretical models for possible pre-SN mass loss. In this section, we explore these scenarios and compare them to the observations.

\subsection{Pre-SN mass loss for progenitor masses $\lesssim$ 12 \Msun}

\subsubsection{Low mass binary He-stars}

We know that the majority ($\sim$ 70\%) of young massive stars live in interacting binary systems \citep{Mason1998, Sana2012}. Recent evidence suggests that Type Ibc SNe form when less massive stars are stripped due to a binary companion \citep[e.g.,][]{Podsiadlowski1992}. These stripped stars are formed when they lose their hydrogen envelopes through case B mass transfer (MT) after hydrogen burning.
The stripped stars with $\mathrm{M_{He}} \lesssim 4$ \Msun\ expand again and lose a significant amount of their He envelope through Case BC MT. This results in 
stars with low pre-collapse masses which can explain the inferred $M_\mathrm{ej}$ of the sources with low progenitor mass and low ejecta mass constraints.

There have been attempts to model the case BB MT to make predictions for mass loss and the final fate of the progenitor \citep{Tauris2013, Tauris2015, Yoon2010, Laplace2020}. However, these do not predict the significant CSM that we infer in our observations. However, \citet{Wu2022b} find that when the O/Ne-core burning is taken into account, He-stars of masses $\approx 2.5 - 3$ \Msun\  rapidly re-expand. As a result, they undergo high rates of binary MT weeks to decades before core-collapse. In $Part\ A$ and $Part\ B$, we look at the possible cases where the shock passes through this re-expanded bound material before and after the late-time MT. In $Part\ C$, we look at the possible case where the shock passes through the unbound material ejected as part of the late-time MT.



\subsection*{\textit{Part A: Bound stellar material before late-time binary mass transfer}} \label{sec:bound1}

Stripped stars with initial masses 2.5 \Msun\ $\lesssim \mathrm{M_{He}} \lesssim$ 3 \msun\ expand by two orders of magnitude during C-burning
beginning $\sim 10^5$ years before core-collapse. \citet{Wu2022} found that the radius can expand to $\sim 200\ \Rsun$ for low mass He-stars during O/Ne-burning. 

We investigate if the low-luminosity first bumps we see for those with low progenitor mass be produced as the shock from the core-collapse passes through this bound puffed-up stellar envelope. We can see from Table \ref{table_ss} that the models for single star evolution from \citet{Wu2022b} can puff up to a radius that is consistent with what is calculated from the shock-cooling modeling.

Based on the density profiles of these stars (see Figure \ref{fig:ssdensity}), we assume the material at r $ > \mathrm{R_{max}}/3$ \citep{nakarpiro2014} is the bounded envelope responsible for the early bump, where $\mathrm{R_{max}}$ is the radial distance of the star where the density drops below $10^{-7} \mathrm{g\ cm^{-3}}$. We also compare the bound envelope properties of binary and single stars from \citet{Laplace2020, Tauris2015} with those calculated for our sample in Figure \ref{fig:bound}. We find that the expected envelope mass for these models is $\sim$ 1.2 \msun, an order of magnitude greater than the observed values.

We also note that the above scenario would require that the star will not interact afterward with the binary companion after undergoing Case B mass transfer. This is possible when the binary stars have very large periods so that the Roche-lobe is not filled during the expansion. \cite{Wu2022b} find that the highest-mass models $\mathrm{M_{He}} \geq$ 2.8 \Msun\ with orbital period $(\mathrm{P_{orb}}) = 100$ day do not expand enough to fill their Roche lobes. 

However, in these cases, it is more likely that the substantial radius expansion of the stripped stars suggests the possibility of them reoccupying their Roche lobes and experiencing subsequent phases of mass transfer \citep{Dewi2002, Dewi2003, Ivanova2003}. Additional phases of mass transfer can produce stars with low envelope masses, possibly explaining the low ejecta mass we observe for those with low progenitor mass. We discuss this in the next part.

\subsection*{\textit{Part B: Bound stellar material after late-time binary mass transfer}} \label{sec:bound2}


\citet{Wu2022b} calculated the mass-loss rates from the late-time binary transfer described earlier and the accumulated mass loss at $P_\mathrm{orb} =$ 1, 10, and 100 days. After the late-time mass transfer, the final masses range between $\sim 1.4–2.9$ \Msun. As these models reach Si-burning with final masses $\geq$ 1.4 \Msun\, they are expected to undergo core collapse. Assuming $\mathrm{M_{NS}} = 1.4$ \Msun, the implied SN ejecta masses are $\lesssim$ 1.5 \Msun\. The density profiles of these stars after the late-time mass transfer are shown in Appendix \ref{appendix:density}. 
The envelope radius of most of these binary stars (especially those with $\mathrm{P_{orb}}=$10 days) is consistent with the observed values (see Table \ref{table_mt} and Figure \ref{fig:bound}). Using these density profiles and the same procedure used in the previous section, we get an envelope mass range of $0.01-0.1$ \Msun, which is consistent with the measured mass. 

But for those stars which have lost mass through the late-time mass transfer, there should be another sign of interaction when the shock passes through the unbound CSM. It is possible that we did not have high enough cadence spectra to look for these interactions or that any interaction contribution to the lightcurve was too small compared to the Ni-powered lightcurve.



\subsection*{\textit{Part C: Unbound stellar material after late time transfer}} \label{sec:unbound}

\citet{Wu2022b} assume that shells of expelled material form at a distribution of radii around the binary system as a result of the late-time MT. To estimate the properties of this CSM they perform a mass-weighted average of these radii to calculate the characteristic CSM radius. They calculate the total CSM mass in each system as the integrated mass loss rate at core collapse.

We note that the shock-cooling breakout radius is expected to be smaller than the mass-weighted radius reported in \citet{Wu2022b}. Here, we calculate the CSM radius assuming the shock breakouts at an optical depth ($\tau\sim 3c/v_s \sim30$). We assume a CSM wind-density profile of the form $\rho = K r^{-2}$ \citep[used in e.g.,][]{Ofek2010, Chevalier2011}, where $r$ is the distance from the progenitor. $K\equiv \dot M / (4\pi v_{\rm w})$ is the wind density parameter, $v_{\rm w}$ is the wind velocity, and $\dot M$ is the mass-loss rate. $\mathrm{R_{out}}$ is the maximum distance of the unbound CSM ejected during the late-time MT. If we assume that the shock breakout occurs at an optical depth ($\tau\ \sim\ 30$):
\begin{equation}
    \tau(r) = \int_r^{\mathrm{R_{out}}} \kappa \rho dr = \kappa K \bigg(\frac{1}{r} - \frac{1}{\mathrm{R_{out}}}\bigg) \approx \frac{c}{v_s}
\end{equation}
we get the shock-breakout radius ($r_\mathrm{SBO}$) as:
\begin{equation}
    r_\mathrm{SBO} = \frac{R_d\mathrm{R_{out}}}{R_d+\mathrm{R_{out}}}
\end{equation},
where $R_d\equiv \frac{\kappa K v_S}{c}$ \citep[see][]{Chevalier2011}. From Figure \ref{fig:ppi}, we see that the shock-cooling breakout radius (SCB-30) from most models is consistent with our observations. 

These models have small SN ejecta masses $\lesssim 1.5$ \msun, assuming an NS mass of $M_\mathrm{NS} = 1.4$ \msun\ consistent with the measured ejecta masses for SN~2021gno, SN~2021niq, SN~2021inl, SN~2022nwx, SN~2018lqo, SN~2019dge in our sample. We compare the CSM properties for these with those predicted from the late-time MT simulations  (see Figure \ref{fig:ppi}).  We also show the CSM properties expected in case of unstable MT, which leads to a common envelope event (see \citealt{Wu2022} for details). 
We find that the CSM properties across both scenarios are consistent with the CSM masses $(\sim 0.01-1$ \msun) and radii ($\sim 10^{11}-10^{13}$ cm) inferred for those SNe with low ejecta mass. 




However, as mentioned earlier, the late-time mass transfer only occurs for the low progenitor mass SNe. To explain the CSM properties for sources with high ejecta mass, we turn to other pre-SN mass-loss models.

\subsection{Pre-SN mass loss for higher progenitor masses}

\subsubsection{Wave heating process in hydrogen-poor stars}

Wave-driven mass loss \citep{Quataert2012, Shiode2014, Fuller2017, Fuller2018, Wu2021, Leung2021b} occur when convective motions in the massive
star's core excite internal gravity waves during its late-phase nuclear burning. These gravity waves propagate through the radiative core and transmit some percentage of its energy into the envelope via acoustic waves, which can be sufficient to eject a substantial amount of mass.

We compare the CSM properties with mass-loss models in \citet{Leung2021} in Figure \ref{fig:ppi}. The wave heating process in massive hydrogen-poor stars was investigated by \citet{Leung2021}, who surveyed a range of  stellar models with main sequence progenitor masses from $20-70$ \msun\ and metallicity from 0.002 to 0.02.  Most of these models predict CSM masses less than $\sim 10^{-2}$ \msun. The low mass makes just wave driven-driven mass loss an unlikely explanation for all the observed CSM masses. However, a few models predict somewhat higher wave energy fluxes, have larger ejected mass $(\sim 10^{-2}$ \msun) or have a very large $R_\mathrm{CSM} \sim 10^{14}$ cm. These are models with large wave energies or long wave heating time scales, respectively. It requires the merger of nearby burning shells, in their models the carbon and helium shells. The merging of the two shells allows gravity waves to propagate across the star with a lower evanescence. However, numerical models show that such a phenomenon only occurs at individual masses of massive stars rather than a robust mass range. This may be consistent with the rarity of SNe observed in this work.
 
We find that the CSM properties predicted in \citet{Shiode2014} are consistent with our observations (see Figure \ref{fig:ppi}).
\citet{Shiode2014} predicts that wave excitation and damping during Si-burning can inflate nominally compact Wolf-Rayet progenitors to $10^{-3}-1$ \Msun\ of the envelope of Wolf-Rayet stars to tens to hundreds of \Rsun.  These findings indicate that certain supernova (SN) progenitors, often characterized by their compact nature, including those associated with Type Ibc SNe, exhibit a shock cooling signature that differs considerably from conventional assumptions. The authors predict that the outcome of wave energy deposition during silicon fusion in Wolf-Rayet (WR) progenitors would probably manifest as a core-collapse SN classified spectroscopically as Type Ibc (i.e., a compact star), but displaying early thermal emission reminiscent of extended stellar envelopes, which is observed as an early bump in our sample. However, we note that their estimates do not involve hydrodynamical simulations.


\subsubsection{Pulsation Pair Instablility}

For very massive stars ($M_\mathrm{ZAMS} = 80 - 140$ \msun), the electron-positron pair instability (PPI) drives explosive O-burning and mass ejection, which accounts for an outburst of $\sim 0.1$ to tens of \msun \citep{Woosley2019, Leung2019, Renzo2020}. Pulsation-induced mass loss relies on the electron–positron pair-creation catastrophe which happens in very massive stars \citep{Heger2002}. 
The star can experience several mass-loss events, depending on the available carbon and oxygen in the core \citep{Woosley2017, Leung2020}.

In Figure \ref{fig:ppi}, we plot the predicted CSM properties for the PPI-driven mass loss from \citet{Renzo2020, Leung2019}. The models center around $M_\mathrm{CSM} \sim 0.01-10$ \msun\ and $R_\mathrm{CSM} \sim  10^{11}-10^{15}$ cm. These are consistent with the observed CSM properties for our sample (see Figure \ref{fig:ppi}). The objects in this work are consistent with the lower mass PPISNe reported in the literature near a He core mass $\sim 40~M_{\odot}$ (or ZAMS mass $\sim 80~M_{\odot}$). We notice that the PPISN model will be in tension for the objects with a low ejecta mass reported in this work. Given the high progenitor mass for PPISN ($80 - 140~M_{\odot}$), and the production $^{56}$Ni, which indicates a robust explosion, if a spherical explosion is considered the ejecta mass would be much larger. Most PPISN models predict that the star will collapse into a black hole. The low ejecta mass could be explained if most of the star's mass falls into the black hole and only a small fraction of the mass is ejected during the SN explosion. It is also possible that the aspherical explosion plays a role here. Through a jet-like energy deposition, only matter along the jet opening angle acquires the energy deposition, thus the necessary energy deposition and the corresponding ejecta mass can be substantially lower even when the progenitor mass is high. Then, a relatively lower amount of energy is needed for the same ejecta velocity. 
The aspherical shape may lead to strong polarity in the optical signals, which can be checked for such sources in the future. Further samples along the trend may provide further evidence for PPISN being a robust production mechanism for low-mass CSM. However, given its high progenitor mass which is less common in the stellar population according to the Salpeter relation, further comparison with the canonical supernova rate will be important to check the compatibility of this picture with the stellar statistics. If we assume a Salpeter IMF, roughly $\sim$ 2.3\% of CCSNe should undergo PPI, which is roughly consistent with the rate predicted for the double-peaked Type Ibc SNe in Section \ref{sec:sample}.

\subsubsection{Wolf-Rayet+Red Super Giant wind mass loss}

One possibility is that the progenitors of our observed sources underwent a typical phase of Red Super Giant (RSG) with line-driven wind mass loss. 
Assuming a wind velocity ($v_w$) of $\approx \mathrm{10\ km\ s^{-1}}$,  the expected mass-loss rates range from  $\dot{M} \approx 10^{-6}$ \msun\ $\mathrm{yr^{-1}}$ to $\dot{M} \approx 10^{-3}$ \msun\ $\mathrm{yr^{-1}}$ \citep[e.g.,][]{deJager1988, Marshall2004, vanloon2005}. 

Subsequently, there is a relatively brief phase of Wolf-Rayet (WR), characterized by higher wind velocities of a few 1000 $\mathrm{km\ s^{-1}}$ and mass-loss rates around $\sim 10^{-5}$ \msun\ $\mathrm{yr^{-1}}$ \citep[e.g.,][]{Crowther2007}. It is a possibility that the stellar progenitor explodes as a Type Ibc supernova within the bubble formed by its own WR winds interacting with the prior RSG wind phase.
However, the documented cases of WR-RSG wind-wind interaction are associated with ``bubbles'' at typical distances of $\sim 10^{19}$ cm \citep{Marston1997}, significantly farther than the few $<10^{15}$ cm distances inferred for our sources. For our sample, the proximity of the CSM shell implies an extremely short WR phase with a duration of $\sim 10^3\ (v \sim 1000\ \mathrm{km\ s^{-1}})$ yrs, conflicting with the $\sim 10^5$ yr duration of the WR phase expected in the case of isolated massive stars. For such a short lifespan, assuming a mass loss rate of  $\sim 10^{-5}$ \Msun $\mathrm{yr}^{-1}$, the mass loss will be around $0.01$ \Msun, which is an order or so less than what we observe. Thus, wind loss from WR+RSG is not consistent with our observations. 

\subsection{\textcolor{black}{White dwarf progenitor}}

\textcolor{black}{In the earlier sections, we discussed the presence of the early shock-cooling signatures as originating from the extended envelope or CSM of a massive progenitor star. However, we note that an early excess in the lightcurve is also possible in the context of a white dwarf. Such scenarios include white dwarf systems that involve companion interaction through Roche-lobe overflow \citep{Magee2020c, Kasen2010}, clumpy nickel distribution in the ejecta \citep{Magee2020b}, CSM interaction \citep{Piro2016}. The ejecta mass and oxygen mass measured for SN~2021gno, SN~2021niq, SN~2021inl, SN~2019dge, SN~2018lqo are within the mass limit for a white dwarf. Based on the strong [Ca II] emission lines in the nebular spectra, SN~2021gno, SN~2021inl, SN~2018lqo and SN~2022oqm could belong to the thermonuclear group of Ca-rich gap transients, which result from explosive burning of He shells on the surface of low mass white dwarfs \citep{De2020}. \citet{Jacobson2022b} favored a low-mass hybrid He/C/O $+$ C/O WD binary progenitor system for SN~2021gno and SN~2021inl based on the environment of their explosion sites. The disruption of a C/O WD by a heavier WD companion is favored for SN~2022oqm in \cite{Irani2022}. Detailed spectral and lightcurve analysis in the context of the various white dwarf progenitor channels is left for future work.}

\section{Summary and Future Goals}
\label{sec:summary}



\begin{enumerate}   

    \item We present a sample of 17 double-peaked Type Ibc(BL) SNe from ZTF. This was selected from a sample of 475 SNe classified as Ibc(BL) as part of the ZTF and CLU surveys. Out of these 475 SNe, there were 144 SNe with well-sampled early light curves. The rate of this sample is $\sim 3 - 9 \%$ of Type Ibc(BL) SNe. 
    
    The first peak is likely produced after the shockwave runs through an extended envelope, and the layer cools (the “shock cooling” phase). Type Ibc SNe are thought to arise from compact stars, so the envelope is more likely to be stellar material that was ejected in some mass-loss episode.


    \item The peak magnitude of the first peak range from -14.2 to -20.1. We find that the peak magnitude of the first peak and second peak are correlated as $M_2 = 0.8 \times M_1 - 4.7$ where $M_1, M_2$ are the peak magnitudes of the first and second peak, respectively. The correlation could imply that the SNe that show double-peaked lightcurves have He-star progenitors that shed their envelope in binary interactions. The photospheric velocities of the SNe in our sample are consistent with those of canonical Type Ibc SNe. 

    
     \item Based on nebular spectra and lightcurve properties, we divide our sample into two groups: six SNe $-$ SN~2021gno, SN~2018lqo, SN~2021inl, SN~2022nwx, SN~2019dge, SN~2021niq $-$ with progenitor mass less than $\sim$ 12 \Msun\ and ejecta mass less than 1.5 \Msun\ and the rest with higher progenitor mass.

    
    \item The observed CSM properties for SNe with low progenitor and ejecta mass might be explained as due to the binary evolution of low-mass He stars due to late-time mass transfer. The observed CSM properties of SNe with higher ejecta mass are consistent with certain models of wave-driven mass loss due to Si-burning or pulsation-pair instability-driven mass loss.



\end{enumerate}

The sample presented in this paper will enable detailed modeling of the progenitor and supernova, offering insights into their mass-loss histories and envelope structures and thus inform stellar evolution models. The investigation of double-peaked Type Ibc supernovae and the mechanisms behind pre-supernova mass loss have implications across multiple areas of astronomy. These findings have the potential to alter predictions related to ionizing radiation and wind feedback from stellar populations, thereby influencing conclusions about star formation rates and initial mass functions in galaxies beyond our own. Moreover, these discoveries impact our understanding of the origins of diverse compact stellar remnants and shape the way we utilize supernovae as tools for studying stellar evolution throughout cosmic history

While analytical modeling of shock-cooling provides a good estimate of the CSM properties, it might not be able to take into account detailed nuances such as variable opacities, densities, etc. The exact structure of the CSM and its impact on the explosion light curve require detailed hydrodynamics and radiative transfer calculations which we leave for future work.

It is also important to understand the implication of the missing early bump in the majority of Type Ibc SN in understanding the multiplicity of stars, binary evolution, and the extent of stripping in compact binaries including common envelope evolution. Further theoretical work to study this is left for future work.

This sample shows that shock-cooling emission may be very common in H-poor SNe. We might be missing many of them because of poor early-time cadence. Early observations with future wide-field UV surveys such as ULTRASAT \citep{Sagiv2014, Shvartzvald2023} and UVEX \citep{Kulkarni2021} will be critical for the discovery and study of these SNe.  Also, X-ray and radio follow-up observations \citep{Matsuoka2020, Kashiyama2022} of H-poor SNe with well-sampled early optical light curves will help better constrain the mass-loss mechanisms.

\section{Data availability}

All the photometric and spectroscopic data used in this work will be available  
 \href{https://caltech.box.com/s/xu7zet7zb622l3rl4vtth5djkbca6522}{here} after publication.

The optical photometry and spectroscopy will also be made public through WISeREP, the Weizmann Interactive Supernova Data Repository \citep{Yaron2012}.

\section{Acknowldegement}

We thank Anthony L. Piro for insightful discussions and comments. We would also like to thank Daniel Brethauer for providing the data used in \citet{Brethaeur2022}. 
Based on observations obtained with the Samuel Oschin Telescope
48-inch and the 60-inch Telescope at the Palomar Observatory as part of the Zwicky Transient Facility project.
ZTF is supported by the National Science Foundation under Grant No. AST-2034437 and a collaboration including
Caltech, IPAC, the Weizmann Institute of Science, the Oskar Klein Center at Stockholm University, the University
of Maryland, Deutsches Elektronen-Synchrotron and Humboldt University, the TANGO Consortium of Taiwan, the
University of Wisconsin at Milwaukee, Trinity College Dublin, Lawrence Livermore National Laboratories, IN2P3,
France, the University of Warwick, the University of Bochum, and Northwestern University. Operations are conducted
by COO, IPAC, and UW. 

SED Machine is based upon work supported by the National Science Foundation under
Grant No. 1106171. 

The ZTF forced-photometry service was funded under the Heising-Simons Foundation grant \#12540303 (PI: Graham).

The GROWTH Marshal was supported by the GROWTH project funded by the National Science Foundation under Grant No 1545949.

The data presented here were obtained in part with ALFOSC, which is provided by the Instituto de Astrofisica de Andalucia (IAA) under a joint agreement with the University of Copenhagen and NOT.

The Liverpool Telescope is operated on the island of La Palma by Liverpool John Moores University in the Spanish Observatorio del Roque de los Muchachos of the Instituto de Astrofisica de Canarias with financial support from the UK Science and Technology Facilities Council.
Based on observations made with the Italian Telescopio Nazionale Galileo (TNG) operated on the island of La Palma by the Fundación Galileo Galilei of the INAF (Istituto Nazionale di Astrofisica) at the Spanish Observatorio del Roque de los Muchachos of the Instituto de Astrofisica de Canarias.

The W. M. Keck Observatory is operated as a scientific partnership among the California Institute of Technology, the University of California and the National Aeronautics and Space Administration. The Observatory was made possible by the generous financial support of the W. M. Keck Foundation. The authors wish to recognize and acknowledge the very significant cultural role and reverence that the summit of Maunakea has always had within the indigenous Hawaiian community.  We are most fortunate to have the opportunity to conduct observations from this mountain.
The \texttt{ztfquery} code was funded by the European Research Council (ERC) under the European Union's Horizon 2020 research and innovation programme (grant agreement $n^{\circ}759194 -$ USNAC, PI: Rigault).

Shing-Chi Leung acknowledges support by the National Science Foundation under Grant AST-2316807.

\begin{table*} 
\begin{center} 
\caption{Summary of the nebular properties} 
\begin{tabular}{ccccccc} 
\hline  \\  
Source &  Spectra Tel.+Inst. & Phase  & [Ca II]/[O I] flux ratio & [O I] lum. & O mass   \\    &    & (days since primary peak)   & & ($10^{38} \mathrm{erg\ s^{-1}}$) & \textcolor{black}{(\Msun)} \\ \hline  \\ 











ZTF21aaqhhfu/SN~2021gno\footnote{\label{note5} from \citet{Jacobson2022b}.}   &  Keck + LRIS  &  $84$  &  \textcolor{black}{$8.97 \pm 1.79$}  &  \textcolor{black}{$3.9 \pm 0.4$}  &  \textcolor{black}{$0.1 - 0.3$} \\ \hline \\ 

ZTF18achcpwu/SN~2018ise  &  Keck + LRIS  &  $120$  &  \textcolor{black}{$0.93 \pm 0.08$}  &  \textcolor{black}{$303.45 \pm 12.11$}  &  \textcolor{black}{$7.5 - 20.0$} \\ \hline \\ 

ZTF18abmxelh/SN~2018lqo  &  Keck + LRIS  &  $52$  &  \textcolor{black}{$> 50$}  &  \textcolor{black}{$< 1.6$}  &  \textcolor{black}{$< 0.1$} \\ \hline \\ 

ZTF21aasuego/SN~2021inl\textsuperscript{\ref{note5}}  &  Keck + LRIS  &  $111$  &  \textcolor{black}{$4.39 \pm 0.88$}  &  \textcolor{black}{$8.2 \pm 0.4$}  &  \textcolor{black}{$0.2 - 0.6$} \\ \hline \\ 

ZTF22aapisdk/SN~2022nwx  &  Keck + LRIS  &  $86$  &  \textcolor{black}{$10.95 \pm 1.15$}  &  \textcolor{black}{$7.62 \pm 0.90$}  &  \textcolor{black}{$0.2 - 0.5$} \\ \hline \\ 

ZTF22aasxgjp/SN~2022oqm   &  P200 + DBSP  &  $74$  &  \textcolor{black}{$> 22$}  &  \textcolor{black}{$< 0.17$}  &  \textcolor{black}{$< 0.01$} \\ \hline \\ 


ZTF22aaezyos/SN~2022hgk  &  Keck + LRIS  &  $75$  &  \textcolor{black}{$1.13 \pm 0.04$}  &  \textcolor{black}{$104 \pm 1.4$}  &  \textcolor{black}{$2.6 - 7.4$} \\ \hline \\ 

ZTF21abmlldj/SN~2021uvy  &  Keck + LRIS   &  $426$  &  \textcolor{black}{$3.38 \pm 0.15$}  &  \textcolor{black}{$81 \pm 8$}  &  \textcolor{black}{$0.6 - 3.1$}\\ \hline \\ 

ZTF18abfcmjw/SN~2019dge & Keck + LRIS & $83$ & \textcolor{black}{$1.22 \pm 0.19$} & \textcolor{black}{$2.44 \pm 0.33$}  & \textcolor{black}{$0.06 - 0.1$} &   \\ \hline \\ 

\end{tabular}  \label{table_neb} 
\end{center} 
\end{table*}

\begin{table*} 
\begin{center} 
\caption{Shock cooling modeling} 
\begin{tabular}{ccccc} 
\hline  \\  

Source & $E_{ext} $ & $R_{ext}$ & $M_{ext}$ & $t_{exp}$  \\ &  ($\times 10^{49}$ erg) & $R_{\odot}$ & $(\times 10^{-2} M_{\odot})$ & JD \\\\ \hline  \\ 


ZTF21aaqhhfu/SN2021gno &  $0.70^{+0.06}_{-0.06}$  &  $101.19^{+8.32}_{-7.43}$    &  $4.10^{+0.9}_{-0.9}$    &  $59292.29^{+0.05}_{-0.05}$   \\ \hline \\ 

ZTF21abcgnql/SN2021niq  &  $4.71^{+1.00}_{-0.86} $  &  $36.64^{+7.67}_{-5.82}$    &  $11.26^{+1.02}_{-1.01}$    &  $59355.80^{+0.10}_{-0.12}$   \\ \hline \\ 


ZTF20abbpkng/SN2020kzs  &  $34.24^{+33.39}_{-20.59}$  &  $76.11^{+139.24}_{-37.44}$    &  $17.79^{+6.28}_{-5.80}$    &  $58980.47^{+0.31}_{-0.50}$   \\ \hline \\


ZTF21abccaue/SN2021nng  &  $93.44^{+4.84}_{-9.73}$  &  $52.40^{+6.67}_{-4.83}$    &  $188.21^{+26.63}_{-24.96}$    &  $59333.35^{+0.26}_{-0.35}$   \\ \hline \\

ZTF18achcpwu/SN2018ise  &  $23.42^{+25.90}_{-10.53} $  &  $57.76^{+77.12}_{-61.14}$    &  $15.08^{+12.93}_{-12.20}$    &  $58420.81^{+0.81}_{-0.96}$   \\ \hline \\ 

ZTF18abmxelh/SN2018lqo  &  $7.57^{+3.38}_{-2.80} $  &  $75.29^{+75.72}_{-59.76}$    &  $21.77^{+13.34}_{-15.13}$    &  $58336.50^{+0.50}_{-0.35}$   \\ \hline \\ 

ZTF21acekmmm/SN2021aabp  &  $79.40^{+4.47}_{-4.41} $  &  $96.79^{+53.49}_{-52.93}$    &  $24.90^{+12.61}_{-12.81}$    &  $59484.01^{+0.02}_{-0.01}$   \\ \hline \\ 

ZTF21aasuego/SN2021inl  &  $3.90^{+2.91}_{-2.38} $  &  $39.85^{+73.71}_{-33.17}$    &  $15.94^{+13.53}_{-11.87}$    &  $59308.66^{+0.26}_{-0.77}$   \\ \hline \\ 

ZTF21abdxhgv/SN2021qwm  &  $12.62^{+9.28}_{-5.45} $  &  $350.65^{+488.49}_{-334.14}$    &  $21.41^{+13.34}_{-14.49}$    &  $59366.17^{+0.60}_{-0.68}$   \\ \hline \\ 

ZTF22aapisdk/SN2022nwx  &  $45.39^{+36.30}_{-29.55} $  &  $5.16^{+10.20}_{-4.43}$    &  $13.39^{+11.27}_{-10.84}$    &  $59754.55^{+0.21}_{-0.43}$   \\ \hline \\ 


ZTF22aasxgjp/SN2022oqm  &  $15.58^{+0.15}_{-0.15}$  &  $37.81^{+0.41}_{-0.39}$    &  $7.94^{+1.1}_{-2.3}$    &  $59770.23^{+0.1}_{-0.2}$   \\ \hline \\

ZTF21aacufip/SN2021vz  &  $46.36^{+25.01}_{-18.70} $  &  $77.29^{+99.91}_{-67.37}$    &  $11.10^{+8.35}_{-8.02}$    &  $59221.50^{+0.33}_{-0.33}$   \\ \hline \\ 

ZTF18abfcmjw/SN2019dge  &  $5.73^{+0.25}_{-0.27} $  &  $146.04^{+79.39}_{-78.92}$    &  $11.05^{+6.17}_{-6.16}$    &  $58580.25^{+0.02}_{-0.02}$   \\ \hline \\ 

ZTF20aalxlis/SN2020bvc  &  $67.18^{+2.94}_{-3.31} $  &  $118.21^{+65.45}_{-64.20}$    &  $19.55^{+10.13}_{-10.18}$    &  $58881.00^{+0.3}_{-0.3}$   \\ \hline \\ 

\end{tabular}  \label{table_piro} 
\end{center} 
\end{table*}

\begin{table} 
\begin{center} 
\caption{Order of magnitude shock-cooling CSM estimates} 
\begin{tabular}{ccc} 
\hline  \\  

Source & $\mathrm{M_{CSM}}$ & $\mathrm{R_{CSM}}$   \\ & (\Msun) & (\Rsun) \\\\ \hline  \\ 

SN~2021gno  &  $0.02$  &  $10$   \\ \hline \\ 

SN~2021niq  &  $0.20$  &  $10$   \\ \hline \\ 

SN~2020kzs  &  $0.53$  &  $20$   \\ \hline \\ 

SN~2021nng  &  $0.06$  &  $40$   \\ \hline \\ 

SN~2018ise  &  $0.05$  &  $60$   \\ \hline \\ 

SN~2018lqo  &  $1.23$  &  $10$   \\ \hline \\ 

SN~2021aabp  &  $0.34$  &  $60$   \\ \hline \\ 

SN~2021inl  &  $0.04$  &  $10$   \\ \hline \\ 

SN~2021qwm  &  $0.07$  &  $220$   \\ \hline \\ 

SN~2022nwx  &  $0.03$  &  $10$   \\ \hline \\ 

SN~2022oqm  &  $0.01$  &  $30$   \\ \hline \\ 

SN~2021vz  &  $0.28$  &  $1030$   \\ \hline \\ 


SN~2022hgk  &  $0.18$  &  $20$   \\ \hline \\ 

SN~2021uvy  &  $11.31$  &  $260$   \\ \hline \\ 

SN~2019dge  &  $0.06$  &  $30$   \\ \hline \\ 

SN~2020bvc  &  $0.01$  &  $210$   \\ \hline \\ 

SN~2019cad  &  $0.86$  &  $20$   \\ \hline \\ 

\end{tabular}  \label{table_oom} 
\end{center} 
\end{table}

\begin{table} 
\begin{center} 
\caption{Order of magnitude shock-breakout CSM estimates} 
\begin{tabular}{ccc} 
\hline  \\  

Source & $\mathrm{M_{CSM}}$ & $\mathrm{R_{CSM}}$   \\ & ($10^{-3}$ \Msun) & (\Rsun) \\\\ \hline  \\ 

SN~2021gno  &  $0.19$  &  $5900$ \\ \hline \\ 

SN~2021niq  &  $0.11$  &  $19400$ \\ \hline \\ 

SN~2020kzs  &  $0.13$  &  $3100$ \\ \hline \\ 

SN~2021nng  &  $0.69$  &  $10800$ \\ \hline \\ 

SN~2018ise  &  $1.13$  &  $9400$ \\ \hline \\ 

SN~2018lqo  &  $0.03$  &  $47400$ \\ \hline \\ 

SN~2021aabp  &  $0.40$  &  $25000$ \\ \hline \\ 

SN~2021inl  &  $0.15$  &  $8900$ \\ \hline \\ 

SN~2021qwm  &  $3.45$  &  $11300$ \\ \hline \\ 

SN~2022nwx  &  $0.22$  &  $6800$ \\ \hline \\ 

SN~2022oqm  &  $1.26$  &  $4700$ \\ \hline \\ 

SN~2021vz  &  $8.12$  &  $22500$ \\ \hline \\ 


SN~2022hgk  &  $0.15$  &  $18500$ \\ \hline \\ 

SN~2021uvy  &  $0.31$  &  $143700$ \\ \hline \\ 

SN~2019dge  &  $0.53$  &  $10800$ \\ \hline \\ 

SN~2020bvc  &  $7.26$  &  $5000$ \\ \hline \\ 

SN~2019cad  &  $0.1$  &  $39500$ \\ \hline \\ 

\end{tabular}  \label{table_oom_csmbo} 
\end{center} 
\end{table}

\begin{table*} 
\begin{center} 
\caption{Summary of the best-fit physical parameters obtained by radioactive decay modeling of the second peak of the SNe in our sample. The `Ar' subscript refers to the \citet{Arnett89} model while the `KK' subscript refers to the \citet{KhatamiKasen2019} model. The photospheric velocity is estimated as described in Section~\ref{sec:nipeak}.} 
\begin{tabular}{ccccccccc} 
\hline  \\  

Source &  $M_\mathrm{Ni-Ar}$  & $M_\mathrm{Ni-KK}$ & $\mathrm{\tau_m}$ & Velocity & $M_\mathrm{ej-Ar}$ & $M_\mathrm{ej-KK}$  & $E_\mathrm{kin}$   & $t_0$   \\  & (0.01 \msun)  & (0.01 \msun) & (days) & (km/s) & (\msun)  & (\msun)  & ($10^{51}$ erg) & (days)   \\\\ \hline  \\ 

SN2021gno  &  $1.01^{+0.11}_{-0.11}$    &  $1.09$    &  $8.99^{+0.01}_{-0.01}$  &  $7940.0 \pm 500.0$  &  $0.71^{+0.05}_{-0.05}$    &  $0.53$    &  $0.27^{+0.03}_{-0.04}$    &  $38.57^{+2.72}_{-3.68}$   \\ \hline \\ 

SN2021niq  &  $4.42^{+0.49}_{-0.49}$    &  $5.20$    &  $12.07^{+1.45}_{-0.89}$  &  $8000 \pm 1600.0$  &  $1.29^{+0.59}_{-0.44}$    &  $0.88$    &  $0.49^{+0.13}_{-0.07}$    &  $50.24^{+33.85}_{-15.88}$   \\ \hline \\ 


SN2020kzs  &   $13.66^{+1.50}_{-1.50}$    &  $18.31$    &  $19.59^{+0.69}_{-0.62}$  &  $7150.0 \pm 430.0$  &  $3.04^{+0.40}_{-0.37}$    &  $1.68$    &  $0.93^{+0.07}_{-0.06}$    &  $60.55^{+2.68}_{-2.60}$   \\ \hline \\


SN2021nng  &  $67.09^{+4.70}_{-29.11}$    &  $72.17$    &  $38.76^{+1.00}_{-13.86}$  &  $6210.0 \pm 2110.0$  &  $10.35^{+4.06}_{-9.60}$    &  $2.85$    &  $2.38^{+0.12}_{-1.40}$    &  $44.81^{+28.06}_{-2.37}$   \\ \hline \\

SN2018ise  &  $45.24^{+2.86}_{-2.15}$    &  $100.36$    &  $35.83^{+2.10}_{-1.77}$  &  $8000 \pm 1600.0$  &  $11.40^{+3.65}_{-3.37}$    &  $6.05$    &  $4.35^{+0.53}_{-0.42}$    &  $88.85^{+7.83}_{-10.36}$   \\ \hline \\ 

SN2018lqo  &  $3.06^{+0.34}_{-0.34}$    &  $3.61$    &  $13.00^{+0.45}_{-0.40}$  &  $8170.0 \pm 260.0$  &  $1.53^{+0.16}_{-0.14}$    &  $0.99$    &  $0.61^{+0.04}_{-0.04}$    &  $37.60^{+2.15}_{-2.10}$   \\ \hline \\ 

SN2021aabp  &  $61.94^{+2.92}_{-2.10}$    &  $76.66$    &  $17.69^{+0.63}_{-0.48}$  &  $10280.0 \pm 1320.0$  &  $3.57^{+0.72}_{-0.65}$    &  $1.95$    &  $2.25^{+0.16}_{-0.12}$    &  $41.16^{+2.28}_{-2.48}$   \\ \hline \\ 

SN2021inl  &  $0.72^{+0.08}_{-0.08}$    &  $0.77$    &  $8.94^{+0.04}_{-0.09}$  &  $14350.0 \pm 350.0$  &  $1.27^{+0.04}_{-0.06}$    &  $0.91$    &  $1.56^{+0.02}_{-0.03}$    &  $23.89^{+2.73}_{-2.20}$   \\ \hline \\ 

SN2021qwm  &  $52.86^{+6.37}_{-2.11}$    &  $82.22$    &  $25.00^{+2.87}_{-1.22}$  &  $8000 \pm 1600.0$  &  $5.55^{+2.46}_{-1.64}$    &  $2.92$    &  $2.12^{+0.51}_{-0.20}$    &  $68.20^{+13.49}_{-12.05}$   \\ \hline \\ 

SN2022nwx  &  $1.69^{+0.19}_{-0.19}$    &  $1.16$    &  $9.77^{+1.02}_{-0.86}$  &  $8000 \pm 1600.0$  &  $0.85^{+0.36}_{-0.31}$    &  $0.36$    &  $0.32^{+0.07}_{-0.05}$    &  $13.06^{+1.48}_{-1.22}$   \\ \hline \\ 


SN2022oqm  &  $8.94^{+0.98}_{-0.98}$    &  $10.18$    &  $10.73^{+0.04}_{-0.04}$  &  $6660.0 \pm 690.0$  &  $0.85^{+0.09}_{-0.09}$    &  $0.60$    &  $0.23^{+0.00}_{-0.00}$    &  $36.51^{+0.18}_{-0.18}$   \\ \hline \\ 

SN2021vz  &  $97.55^{+10.73}_{-59.80}$    &  $34.23$    &  $21.07^{+0.43}_{-7.99}$  &  $8000 \pm 1600.0$  &  $3.94^{+0.95}_{-3.21}$    &  $0.68$    &  $1.50^{+0.06}_{-0.92}$    &  $14.60^{+16.20}_{-0.54}$   \\ \hline \\ 

SN2019dge  &  $1.34^{+0.15}_{-0.15}$    &  $1.40$    &  $5.41^{+0.65}_{-0.53}$  &  $8000 \pm 1600.0$  &  $0.26^{+0.12}_{-0.10}$    &  $0.23$    &  $0.10^{+0.03}_{-0.02}$    &  $21.77^{+0.65}_{-0.65}$   \\ \hline \\ 

SN2020bvc  &  $34.65^{+3.81}_{-3.81}$    &  $40.44$    &  $13.18^{+0.03}_{-0.03}$  &  $18000 \pm 3600.0$  &  $3.47^{+0.71}_{-0.71}$    &  $2.18$    &  $6.70^{+0.03}_{-0.03}$    &  $41.84^{+0.16}_{-0.16}$   \\ \hline \\

\end{tabular}  \label{table_arnett} 
\end{center} 
\end{table*}

\begin{table} 
\begin{center} 
\caption{Bound stellar material properties of single star.} 
\begin{tabular}{cccc} 
\hline  \\  

Initial mass & $\mathrm{R_{max}}$ & Mass $(>\mathrm{R_{max}}/3)$  & Mass $(>\mathrm{5\ R_\odot})$   \\  (\Msun) & (\Rsun) & (\Msun) & (\Msun)\\\\ \hline  \\ 

$2.51$  &  $446.68$  &  $1.09$  &  $1.10$ \\ \hline \\ 

$2.55$  &  $407.38$  &  $1.09$  &  $1.10$ \\ \hline \\ 

$2.58$  &  $154.88$  &  $1.04$  &  $1.04$ \\ \hline \\ 

$2.62$  &  $239.88$  &  $1.14$  &  $1.14$ \\ \hline \\ 

$2.65$  &  $151.36$  &  $1.12$  &  $1.13$ \\ \hline \\ 

$2.68$  &  $181.97$  &  $1.15$  &  $1.16$ \\ \hline \\ 

$2.72$  &  $177.83$  &  $1.17$  &  $1.17$ \\ \hline \\ 

$2.75$  &  $70.79$  &  $0.82$  &  $0.73$ \\ \hline \\ 

$2.79$  &  $66.07$  &  $0.93$  &  $0.87$ \\ \hline \\ 

$2.82$  &  $44.67$  &  $1.01$  &  $0.92$ \\ \hline \\ 

$2.86$  &  $40.74$  &  $1.04$  &  $0.94$ \\ \hline \\ 

$2.90$  &  $58.88$  &  $1.09$  &  $1.02$ \\ \hline \\ 

$2.92$  &  $24.55$  &  $0.55$  &  $0.20$ \\ \hline \\ 

\end{tabular}  \label{table_ss} 
\end{center} 
\end{table}

\begin{table}

\begin{center} 
\caption{Bound stellar material properties after binary mass transfer} 
\begin{tabular}{ccccc} 
\hline  \\  

Initial mass & Period & $\mathrm{R_{max}}$ & Mass $(>\mathrm{R_{max}}/3)$  & Mass $(>\mathrm{5\ R_\odot})$   \\  (\Msun) & (days) & (\Rsun) & (\Msun) & (\Msun)\\\\ \hline  \\ 

$2.51$  &  $100$  &  $79.43$  &  $0.00$  &  $0.00$ \\ \hline \\ 

$2.55$  &  $100$  &  $81.28$  &  $0.06$  &  $0.05$ \\ \hline \\ 

$2.58$  &  $100$  &  $407.38$  &  $0.22$  &  $0.22$ \\ \hline \\ 

$2.62$  &  $100$  &  $346.74$  &  $0.30$  &  $0.30$ \\ \hline \\ 

$2.65$  &  $100$  &  $109.65$  &  $0.44$  &  $0.44$ \\ \hline \\ 

$2.68$  &  $100$  &  $331.13$  &  $0.35$  &  $0.36$ \\ \hline \\ 

$2.72$  &  $100$  &  $199.53$  &  $0.69$  &  $0.70$ \\ \hline \\ 

$2.75$  &  $100$  &  $109.65$  &  $0.85$  &  $0.85$ \\ \hline \\ 

$2.51$  &  $10$  &  $15.85$  &  $0.01$  &  $0.00$ \\ \hline \\ 

$2.55$  &  $10$  &  $13.18$  &  $0.06$  &  $0.02$ \\ \hline \\ 

$2.58$  &  $10$  &  $12.30$  &  $0.06$  &  $0.01$ \\ \hline \\ 

$2.62$  &  $10$  &  $11.75$  &  $0.04$  &  $0.00$ \\ \hline \\ 

$2.65$  &  $10$  &  $12.30$  &  $0.11$  &  $0.04$ \\ \hline \\ 

$2.68$  &  $10$  &  $12.30$  &  $0.12$  &  $0.05$ \\ \hline \\ 

$2.72$  &  $10$  &  $12.88$  &  $0.16$  &  $0.06$ \\ \hline \\ 

$2.75$  &  $10$  &  $11.48$  &  $0.17$  &  $0.05$ \\ \hline \\ 

$2.79$  &  $10$  &  $11.22$  &  $0.21$  &  $0.07$ \\ \hline \\ 

$2.82$  &  $10$  &  $12.02$  &  $0.30$  &  $0.10$ \\ \hline \\ 

$2.86$  &  $10$  &  $13.18$  &  $0.40$  &  $0.14$ \\ \hline \\ 

$2.90$  &  $10$  &  $14.13$  &  $0.56$  &  $0.25$ \\ \hline \\ 

$2.92$  &  $10$  &  $15.49$  &  $0.60$  &  $0.26$ \\ \hline \\ 

\end{tabular}  \label{table_mt} 
\end{center} 
\end{table}

\begin{table} 
\ContinuedFloat

\begin{center} 
\caption{Continued...} 
\begin{tabular}{ccccc} 
\hline  \\  

Initial mass & Period & $\mathrm{R_{max}}$ & Mass $(>\mathrm{R_{max}}/3)$  & Mass $(>\mathrm{5\ R_\odot})$   \\  (\Msun) & (days) & (\Rsun) & (\Msun) & (\Msun)\\\\ \hline  \\ 

$2.51$  &  $1$  &  $2.69$  &  $0.00$  &  $0.00$ \\ \hline \\ 

$2.55$  &  $1$  &  $3.02$  &  $0.00$  &  $0.00$ \\ \hline \\ 

$2.58$  &  $1$  &  $2.51$  &  $0.00$  &  $0.00$ \\ \hline \\ 

$2.62$  &  $1$  &  $2.40$  &  $0.00$  &  $0.00$ \\ \hline \\ 

$2.65$  &  $1$  &  $2.45$  &  $0.00$  &  $0.00$ \\ \hline \\ 

$2.68$  &  $1$  &  $2.34$  &  $0.01$  &  $0.00$ \\ \hline \\ 

$2.72$  &  $1$  &  $2.34$  &  $0.00$  &  $0.00$ \\ \hline \\ 

$2.75$  &  $1$  &  $2.29$  &  $0.01$  &  $0.00$ \\ \hline \\ 

$2.79$  &  $1$  &  $2.29$  &  $0.01$  &  $0.00$ \\ \hline \\ 

$2.82$  &  $1$  &  $2.24$  &  $0.01$  &  $0.00$ \\ \hline \\ 

$2.86$  &  $1$  &  $2.19$  &  $0.01$  &  $0.00$ \\ \hline \\ 

$2.90$  &  $1$  &  $2.19$  &  $0.01$  &  $0.00$ \\ \hline \\ 

$2.92$  &  $1$  &  $2.14$  &  $0.01$  &  $0.00$ \\ \hline \\ 

\end{tabular}  \label{table_mt} 
\end{center} 
\end{table}

\FloatBarrier
\bibliography{main}

\begin{thebibliography}{}
\expandafter\ifx\csname natexlab\endcsname\relax\def\natexlab#1{#1}\fi
\providecommand{\url}[1]{\href{#1}{#1}}
\providecommand{\dodoi}[1]{doi:~\href{http://doi.org/#1}{\nolinkurl{#1}}}
\providecommand{\doeprint}[1]{\href{http://ascl.net/#1}{\nolinkurl{http://ascl.net/#1}}}
\providecommand{\doarXiv}[1]{\href{https://arxiv.org/abs/#1}{\nolinkurl{https://arxiv.org/abs/#1}}}

\bibitem[{{Ahn} {et~al.}(2012){Ahn}, {Alexandroff}, {Allende Prieto},
  {Anderson}, {Anderton}, {Andrews}, {Aubourg}, {Bailey}, {Balbinot}, {Barnes},
  \& et~al.}]{Ahn2012a}
{Ahn}, C.~P., {Alexandroff}, R., {Allende Prieto}, C., {et~al.} 2012, \apjs,
  203, 21, \dodoi{10.1088/0067-0049/203/2/21}

\bibitem[{{Arnett} {et~al.}(1989){Arnett}, {Bahcall}, {Kirshner}, \&
  {Woosley}}]{Arnett89}
{Arnett}, W.~D., {Bahcall}, J.~N., {Kirshner}, R.~P., \& {Woosley}, S.~E. 1989,
  \araa, 27, 629, \dodoi{10.1146/annurev.aa.27.090189.003213}

\bibitem[{{Arnett} \& {Meakin}(2011)}]{Arnett2011}
{Arnett}, W.~D., \& {Meakin}, C. 2011, \apj, 733, 78,
  \dodoi{10.1088/0004-637X/733/2/78}

\bibitem[{{Barbarino} {et~al.}(2017){Barbarino}, {Botticella}, {Dall'Ora},
  {Della Valle}, {Benetti}, {Lyman}, {Smartt}, {Arcavi}, {Baltay}, {Bersier},
  {Dennefeld}, {Ellman}, {Fraser}, {Gal-Yam}, {Hosseinzadeh}, {Howell},
  {Inserra}, {Kankare}, {Leloudas}, {Maguire}, {McCully}, {Mitra}, {McKinnon},
  {Olivares E.}, {Pignata}, {Rabinowitz}, {Rostami}, {Smith}, {Sullivan},
  {Valenti}, {Yaron}, \& {Young}}]{Barbarino2017}
{Barbarino}, C., {Botticella}, M.~T., {Dall'Ora}, M., {et~al.} 2017, \mnras,
  471, 2463, \dodoi{10.1093/mnras/stx1709}

\bibitem[{{Barnsley} {et~al.}(2012){Barnsley}, {Smith}, \&
  {Steele}}]{Barnsley2012}
{Barnsley}, R.~M., {Smith}, R.~J., \& {Steele}, I.~A. 2012, Astronomische
  Nachrichten, 333, 101, \dodoi{10.1002/asna.201111634}

\bibitem[{{Bellm} \& {Sesar}(2016)}]{Bellm2016}
{Bellm}, E.~C., \& {Sesar}, B. 2016, {pyraf-dbsp: Reduction pipeline for the
  Palomar Double Beam Spectrograph}, Astrophysics Source Code Library.
\newblock \doeprint{1602.002}

\bibitem[{{Bellm} {et~al.}(2019{\natexlab{a}}){Bellm}, {Kulkarni}, {Graham},
  {Dekany}, {Smith}, {Riddle}, {Masci}, {Helou}, {Prince}, {Adams},
  {Barbarino}, {Barlow}, {Bauer}, {Beck}, {Belicki}, {Biswas}, {Blagorodnova},
  {Bodewits}, {Bolin}, {Brinnel}, {Brooke}, {Bue}, {Bulla}, {Burruss}, {Cenko},
  {Chang}, {Connolly}, {Coughlin}, {Cromer}, {Cunningham}, {De}, {Delacroix},
  {Desai}, {Duev}, {Eadie}, {Farnham}, {Feeney}, {Feindt}, {Flynn},
  {Franckowiak}, {Frederick}, {Fremling}, {Gal-Yam}, {Gezari}, {Giomi},
  {Goldstein}, {Golkhou}, {Goobar}, {Groom}, {Hacopians}, {Hale}, {Henning},
  {Ho}, {Hover}, {Howell}, {Hung}, {Huppenkothen}, {Imel}, {Ip}, {Ivezi{\'c}},
  {Jackson}, {Jones}, {Juric}, {Kasliwal}, {Kaspi}, {Kaye}, {Kelley},
  {Kowalski}, {Kramer}, {Kupfer}, {Landry}, {Laher}, {Lee}, {Lin}, {Lin},
  {Lunnan}, {Giomi}, {Mahabal}, {Mao}, {Miller}, {Monkewitz}, {Murphy},
  {Ngeow}, {Nordin}, {Nugent}, {Ofek}, {Patterson}, {Penprase}, {Porter},
  {Rauch}, {Rebbapragada}, {Reiley}, {Rigault}, {Rodriguez}, {van Roestel},
  {Rusholme}, {van Santen}, {Schulze}, {Shupe}, {Singer}, {Soumagnac}, {Stein},
  {Surace}, {Sollerman}, {Szkody}, {Taddia}, {Terek}, {Van Sistine}, {van
  Velzen}, {Vestrand}, {Walters}, {Ward}, {Ye}, {Yu}, {Yan}, \&
  {Zolkower}}]{Bellm2019}
{Bellm}, E.~C., {Kulkarni}, S.~R., {Graham}, M.~J., {et~al.}
  2019{\natexlab{a}}, \pasp, 131, 018002, \dodoi{10.1088/1538-3873/aaecbe}

\bibitem[{{Bellm} {et~al.}(2019{\natexlab{b}}){Bellm}, {Kulkarni}, {Barlow},
  {Feindt}, {Graham}, {Goobar}, {Kupfer}, {Ngeow}, {Nugent}, {Ofek}, {Prince},
  {Riddle}, {Walters}, \& {Ye}}]{Bellm2019b}
{Bellm}, E.~C., {Kulkarni}, S.~R., {Barlow}, T., {et~al.} 2019{\natexlab{b}},
  \pasp, 131, 068003, \dodoi{10.1088/1538-3873/ab0c2a}

\bibitem[{{Ben-Ami} {et~al.}(2014){Ben-Ami}, {Gal-Yam}, {Mazzali}, {Gnat},
  {Modjaz}, {Rabinak}, {Sullivan}, {Bildsten}, {Poznanski}, {Yaron}, {Arcavi},
  {Bloom}, {Horesh}, {Kasliwal}, {Kulkarni}, {Nugent}, {Ofek}, {Perley},
  {Quimby}, \& {Xu}}]{Ben-Ami2014}
{Ben-Ami}, S., {Gal-Yam}, A., {Mazzali}, P.~A., {et~al.} 2014, \apj, 785, 37,
  \dodoi{10.1088/0004-637X/785/1/37}

\bibitem[{{Blagorodnova} {et~al.}(2018){Blagorodnova}, {Neill}, {Walters},
  {Kulkarni}, {Fremling}, {Ben-Ami}, {Dekany}, {Fucik}, {Konidaris}, {Nash},
  {Ngeow}, {Ofek}, {O' Sullivan}, {Quimby}, {Ritter}, \&
  {Vyhmeister}}]{Blagorodnova2018}
{Blagorodnova}, N., {Neill}, J.~D., {Walters}, R., {et~al.} 2018, \pasp, 130,
  035003, \dodoi{10.1088/1538-3873/aaa53f}

\bibitem[{{Blondin} \& {Tonry}(2007)}]{Blondin2007}
{Blondin}, S., \& {Tonry}, J.~L. 2007, in American Institute of Physics
  Conference Series, Vol. 924, The Multicolored Landscape of Compact Objects
  and Their Explosive Origins, ed. T.~{di Salvo}, G.~L. {Israel},
  L.~{Piersant}, L.~{Burderi}, G.~{Matt}, A.~{Tornambe}, \& M.~T. {Menna},
  312--321, \dodoi{10.1063/1.2774875}

\bibitem[{{Bostroem} {et~al.}(2019){Bostroem}, {Valenti}, {Horesh}, {Morozova},
  {Kuin}, {Wyatt}, {Jerkstrand}, {Sand}, {Lundquist}, {Smith}, {Sullivan},
  {Hosseinzadeh}, {Arcavi}, {Callis}, {Cartier}, {Gal-Yam}, {Galbany},
  {Guti{\'e}rrez}, {Howell}, {Inserra}, {Kankare}, {L{\'o}pez}, {McCully},
  {Pignata}, {Piro}, {Rodr{\'\i}guez}, {Smartt}, {Smith}, {Yaron}, \&
  {Young}}]{Bostroem2019}
{Bostroem}, K.~A., {Valenti}, S., {Horesh}, A., {et~al.} 2019, \mnras, 485,
  5120, \dodoi{10.1093/mnras/stz570}

\bibitem[{{Brethauer} {et~al.}(2022){Brethauer}, {Margutti}, {Milisavljevic},
  {Bietenholz}, {Chornock}, {Coppejans}, {Colle}, {Hajela}, {Terreran},
  {Vargas}, {DeMarchi}, {Harris}, {Jacobson-Gal{\'a}n}, {Kamble}, {Patnaude},
  \& {Stroh}}]{Brethaeur2022}
{Brethauer}, D., {Margutti}, R., {Milisavljevic}, D., {et~al.} 2022, \apj, 939,
  105, \dodoi{10.3847/1538-4357/ac8b14}

\bibitem[{{Brown} {et~al.}(2013){Brown}, {Baliber}, {Bianco}, {Bowman},
  {Burleson}, {Conway}, {Crellin}, {Depagne}, {De Vera}, {Dilday}, {Dragomir},
  {Dubberley}, {Eastman}, {Elphick}, {Falarski}, {Foale}, {Ford}, {Fulton},
  {Garza}, {Gomez}, {Graham}, {Greene}, {Haldeman}, {Hawkins}, {Haworth},
  {Haynes}, {Hidas}, {Hjelstrom}, {Howell}, {Hygelund}, {Lister}, {Lobdill},
  {Martinez}, {Mullins}, {Norbury}, {Parrent}, {Paulson}, {Petry}, {Pickles},
  {Posner}, {Rosing}, {Ross}, {Sand}, {Saunders}, {Shobbrook}, {Shporer},
  {Street}, {Thomas}, {Tsapras}, {Tufts}, {Valenti}, {Vander Horst}, {Walker},
  {White}, \& {Willis}}]{Brown2013}
{Brown}, T.~M., {Baliber}, N., {Bianco}, F.~B., {et~al.} 2013, \pasp, 125,
  1031, \dodoi{10.1086/673168}

\bibitem[{{Bruch} {et~al.}(2021){Bruch}, {Gal-Yam}, {Schulze}, {Yaron}, {Yang},
  {Soumagnac}, {Rigault}, {Strotjohann}, {Ofek}, {Sollerman}, {Masci},
  {Barbarino}, {Ho}, {Fremling}, {Perley}, {Nordin}, {Cenko}, {Adams},
  {Adreoni}, {Bellm}, {Blagorodnova}, {Bulla}, {Burdge}, {De}, {Dhawan},
  {Drake}, {Duev}, {Dugas}, {Graham}, {Graham}, {Irani}, {Jencson},
  {Karamehmetoglu}, {Kasliwal}, {Kim}, {Kulkarni}, {Kupfer}, {Liang},
  {Mahabal}, {Miller}, {Prince}, {Riddle}, {Sharma}, {Smith}, {Taddia},
  {Taggart}, {Walters}, \& {Yan}}]{Bruch2021}
{Bruch}, R.~J., {Gal-Yam}, A., {Schulze}, S., {et~al.} 2021, \apj, 912, 46,
  \dodoi{10.3847/1538-4357/abef05}

\bibitem[{{Cardelli} {et~al.}(1989){Cardelli}, {Clayton}, \&
  {Mathis}}]{Cardelli1989}
{Cardelli}, J.~A., {Clayton}, G.~C., \& {Mathis}, J.~S. 1989, \apj, 345, 245,
  \dodoi{10.1086/167900}

\bibitem[{{Cenko} {et~al.}(2006){Cenko}, {Fox}, {Moon}, {Harrison}, {Kulkarni},
  {Henning}, {Guzman}, {Bonati}, {Smith}, {Thicksten}, {Doyle}, {Petrie},
  {Gal-Yam}, {Soderberg}, {Anagnostou}, \& {Laity}}]{Cenko2006}
{Cenko}, S.~B., {Fox}, D.~B., {Moon}, D.-S., {et~al.} 2006, \pasp, 118, 1396,
  \dodoi{10.1086/508366}

\bibitem[{{Chambers} {et~al.}(2016){Chambers}, {Magnier}, {Metcalfe},
  {Flewelling}, {Huber}, {Waters}, {Denneau}, {Draper}, {Farrow}, {Finkbeiner},
  {Holmberg}, {Koppenhoefer}, {Price}, {Rest}, {Saglia}, {Schlafly}, {Smartt},
  {Sweeney}, {Wainscoat}, {Burgett}, {Chastel}, {Grav}, {Heasley}, {Hodapp},
  {Jedicke}, {Kaiser}, {Kudritzki}, {Luppino}, {Lupton}, {Monet}, {Morgan},
  {Onaka}, {Shiao}, {Stubbs}, {Tonry}, {White}, {Ba{\~n}ados}, {Bell},
  {Bender}, {Bernard}, {Boegner}, {Boffi}, {Botticella}, {Calamida},
  {Casertano}, {Chen}, {Chen}, {Cole}, {Deacon}, {Frenk}, {Fitzsimmons},
  {Gezari}, {Gibbs}, {Goessl}, {Goggia}, {Gourgue}, {Goldman}, {Grant},
  {Grebel}, {Hambly}, {Hasinger}, {Heavens}, {Heckman}, {Henderson}, {Henning},
  {Holman}, {Hopp}, {Ip}, {Isani}, {Jackson}, {Keyes}, {Koekemoer}, {Kotak},
  {Le}, {Liska}, {Long}, {Lucey}, {Liu}, {Martin}, {Masci}, {McLean}, {Mindel},
  {Misra}, {Morganson}, {Murphy}, {Obaika}, {Narayan}, {Nieto-Santisteban},
  {Norberg}, {Peacock}, {Pier}, {Postman}, {Primak}, {Rae}, {Rai}, {Riess},
  {Riffeser}, {Rix}, {R{\"o}ser}, {Russel}, {Rutz}, {Schilbach}, {Schultz},
  {Scolnic}, {Strolger}, {Szalay}, {Seitz}, {Small}, {Smith}, {Soderblom},
  {Taylor}, {Thomson}, {Taylor}, {Thakar}, {Thiel}, {Thilker}, {Unger},
  {Urata}, {Valenti}, {Wagner}, {Walder}, {Walter}, {Watters}, {Werner},
  {Wood-Vasey}, \& {Wyse}}]{Chambers2016}
{Chambers}, K.~C., {Magnier}, E.~A., {Metcalfe}, N., {et~al.} 2016, arXiv
  e-prints.
\newblock \doarXiv{1612.05560}

\bibitem[{{Chevalier} \& {Irwin}(2011)}]{Chevalier2011}
{Chevalier}, R.~A., \& {Irwin}, C.~M. 2011, \apjl, 729, L6,
  \dodoi{10.1088/2041-8205/729/1/L6}

\bibitem[{{Cook} {et~al.}(2019){Cook}, {Kasliwal}, {Van Sistine}, {Kaplan},
  {Sutter}, {Kupfer}, {Shupe}, {Laher}, {Masci}, {Dale}, {Sesar}, {Brady},
  {Yan}, {Ofek}, {Reitze}, \& {Kulkarni}}]{Cook2019}
{Cook}, D.~O., {Kasliwal}, M.~M., {Van Sistine}, A., {et~al.} 2019, \apj, 880,
  7, \dodoi{10.3847/1538-4357/ab2131}

\bibitem[{{Crowther}(2007)}]{Crowther2007}
{Crowther}, P.~A. 2007, \araa, 45, 177,
  \dodoi{10.1146/annurev.astro.45.051806.110615}

\bibitem[{{De} {et~al.}(2021){De}, {Fremling}, {Gal-Yam}, {Yaron}, {Kasliwal},
  \& {Kulkarni}}]{De2021}
{De}, K., {Fremling}, U.~C., {Gal-Yam}, A., {et~al.} 2021, \apjl, 907, L18,
  \dodoi{10.3847/2041-8213/abd627}

\bibitem[{{De} {et~al.}(2018){De}, {Kasliwal}, {Cantwell}, {Cao}, {Cenko},
  {Gal-Yam}, {Johansson}, {Kong}, {Kulkarni}, {Lunnan}, {Masci}, {Matuszewski},
  {Mooley}, {Neill}, {Nugent}, {Ofek}, {Perrott}, {Rebbapragada}, {Rubin}, {O'
  Sullivan}, \& {Yaron}}]{De2018a}
{De}, K., {Kasliwal}, M.~M., {Cantwell}, T., {et~al.} 2018, \apj, 866, 72,
  \dodoi{10.3847/1538-4357/aadf8e}

\bibitem[{{De} {et~al.}(2020){De}, {Kasliwal}, {Tzanidakis}, {Fremling},
  {Adams}, {Aloisi}, {Andreoni}, {Bagdasaryan}, {Bellm}, {Bildsten},
  {Cannella}, {Cook}, {Delacroix}, {Drake}, {Duev}, {Dugas}, {Frederick},
  {Gal-Yam}, {Goldstein}, {Golkhou}, {Graham}, {Hale}, {Hankins}, {Helou},
  {Ho}, {Irani}, {Jencson}, {Kaplan}, {Kaye}, {Kulkarni}, {Kupfer}, {Laher},
  {Leadbeater}, {Lunnan}, {Masci}, {Miller}, {Neill}, {Ofek}, {Perley},
  {Polin}, {Prince}, {Quataert}, {Reiley}, {Riddle}, {Rusholme}, {Sharma},
  {Shupe}, {Sollerman}, {Tartaglia}, {Walters}, {Yan}, \& {Yao}}]{De2020}
{De}, K., {Kasliwal}, M.~M., {Tzanidakis}, A., {et~al.} 2020, \apj, 905, 58,
  \dodoi{10.3847/1538-4357/abb45c}

\bibitem[{{de Jager} {et~al.}(1988){de Jager}, {Nieuwenhuijzen}, \& {van der
  Hucht}}]{deJager1988}
{de Jager}, C., {Nieuwenhuijzen}, H., \& {van der Hucht}, K.~A. 1988, \aaps,
  72, 259

\bibitem[{{Dekany} {et~al.}(2020){Dekany}, {Smith}, {Riddle}, {Feeney},
  {Porter}, {Hale}, {Zolkower}, {Belicki}, {Kaye}, {Henning}, {Walters},
  {Cromer}, {Delacroix}, {Rodriguez}, {Reiley}, {Mao}, {Hover}, {Murphy},
  {Burruss}, {Baker}, {Kowalski}, {Reif}, {Mueller}, {Bellm}, {Graham}, \&
  {Kulkarni}}]{Dekany20}
{Dekany}, R., {Smith}, R.~M., {Riddle}, R., {et~al.} 2020, \pasp, 132, 038001,
  \dodoi{10.1088/1538-3873/ab4ca2}

\bibitem[{{Dessart} \& {Hillier}(2020)}]{Dessart2020}
{Dessart}, L., \& {Hillier}, D.~J. 2020, \aap, 642, A33,
  \dodoi{10.1051/0004-6361/202038148}

\bibitem[{{Dessart} {et~al.}(2021){Dessart}, {Hillier}, {Sukhbold}, {Woosley},
  \& {Janka}}]{Dessart2021}
{Dessart}, L., {Hillier}, D.~J., {Sukhbold}, T., {Woosley}, S.~E., \& {Janka},
  H.~T. 2021, \aap, 656, A61, \dodoi{10.1051/0004-6361/202141927}

\bibitem[{{Dewi} \& {Pols}(2003)}]{Dewi2003}
{Dewi}, J.~D.~M., \& {Pols}, O.~R. 2003, \mnras, 344, 629,
  \dodoi{10.1046/j.1365-8711.2003.06844.x}

\bibitem[{{Dewi} {et~al.}(2002){Dewi}, {Pols}, {Savonije}, \& {van den
  Heuvel}}]{Dewi2002}
{Dewi}, J.~D.~M., {Pols}, O.~R., {Savonije}, G.~J., \& {van den Heuvel},
  E.~P.~J. 2002, \mnras, 331, 1027, \dodoi{10.1046/j.1365-8711.2002.05257.x}

\bibitem[{{Djupvik} \& {Andersen}(2010)}]{Djupvik2010}
{Djupvik}, A.~A., \& {Andersen}, J. 2010, in Astrophysics and Space Science
  Proceedings, Vol.~14, Highlights of Spanish Astrophysics V, 211,
  \dodoi{10.1007/978-3-642-11250-8\_21}

\bibitem[{{Elmhamdi}(2011)}]{Elmhamdi2011}
{Elmhamdi}, A. 2011, \actaa, 61, 179.
\newblock \doarXiv{1109.2318}

\bibitem[{{Foreman-Mackey} {et~al.}(2013){Foreman-Mackey}, {Hogg}, {Lang}, \&
  {Goodman}}]{Foreman-Mackey13}
{Foreman-Mackey}, D., {Hogg}, D.~W., {Lang}, D., \& {Goodman}, J. 2013, \pasp,
  125, 306, \dodoi{10.1086/670067}

\bibitem[{{Fransson} \& {Chevalier}(1989)}]{Fransson1989}
{Fransson}, C., \& {Chevalier}, R.~A. 1989, \apj, 343, 323,
  \dodoi{10.1086/167707}

\bibitem[{{Fremling} {et~al.}(2016){Fremling}, {Sollerman}, {Taddia}, {Ergon},
  {Fraser}, {Karamehmetoglu}, {Valenti}, {Jerkstrand}, {Arcavi}, {Bufano},
  {Elias Rosa}, {Filippenko}, {Fox}, {Gal-Yam}, {Howell}, {Kotak}, {Mazzali},
  {Milisavljevic}, {Nugent}, {Nyholm}, {Pian}, \& {Smartt}}]{Fremling2016}
{Fremling}, C., {Sollerman}, J., {Taddia}, F., {et~al.} 2016, \aap, 593, A68,
  \dodoi{10.1051/0004-6361/201628275}

\bibitem[{{Fremling} {et~al.}(2018){Fremling}, {Sollerman}, {Kasliwal},
  {Kulkarni}, {Barbarino}, {Ergon}, {Karamehmetoglu}, {Taddia}, {Arcavi},
  {Cenko}, {Clubb}, {De Cia}, {Duggan}, {Filippenko}, {Gal-Yam}, {Graham},
  {Horesh}, {Hosseinzadeh}, {Howell}, {Kuesters}, {Lunnan}, {Matheson},
  {Nugent}, {Perley}, {Quimby}, \& {Saunders}}]{Fremling2018}
{Fremling}, C., {Sollerman}, J., {Kasliwal}, M.~M., {et~al.} 2018, \aap, 618,
  A37, \dodoi{10.1051/0004-6361/201731701}

\bibitem[{{Fremling} {et~al.}(2020){Fremling}, {Miller}, {Sharma}, {Dugas},
  {Perley}, {Taggart}, {Sollerman}, {Goobar}, {Graham}, {Neill}, {Nordin},
  {Rigault}, {Walters}, {Andreoni}, {Bagdasaryan}, {Belicki}, {Cannella},
  {Bellm}, {Cenko}, {De}, {Dekany}, {Frederick}, {Golkhou}, {Graham}, {Helou},
  {Ho}, {Kasliwal}, {Kupfer}, {Laher}, {Mahabal}, {Masci}, {Riddle},
  {Rusholme}, {Schulze}, {Shupe}, {Smith}, {van Velzen}, {Yan}, {Yao},
  {Zhuang}, \& {Kulkarni}}]{Fremling2020}
{Fremling}, C., {Miller}, A.~A., {Sharma}, Y., {et~al.} 2020, \apj, 895, 32,
  \dodoi{10.3847/1538-4357/ab8943}

\bibitem[{{Fuller}(2017)}]{Fuller2017}
{Fuller}, J. 2017, \mnras, 470, 1642, \dodoi{10.1093/mnras/stx1314}

\bibitem[{{Fuller} \& {Ro}(2018)}]{Fuller2018}
{Fuller}, J., \& {Ro}, S. 2018, \mnras, 476, 1853, \dodoi{10.1093/mnras/sty369}

\bibitem[{{Gehrels} {et~al.}(2004){Gehrels}, {Chincarini}, {Giommi}, {Mason},
  {Nousek}, {Wells}, {White}, {Barthelmy}, {Burrows}, {Cominsky}, {Hurley},
  {Marshall}, {M{\'e}sz{\'a}ros}, {Roming}, {Angelini}, {Barbier}, {Belloni},
  {Campana}, {Caraveo}, {Chester}, {Citterio}, {Cline}, {Cropper}, {Cummings},
  {Dean}, {Feigelson}, {Fenimore}, {Frail}, {Fruchter}, {Garmire}, {Gendreau},
  {Ghisellini}, {Greiner}, {Hill}, {Hunsberger}, {Krimm}, {Kulkarni}, {Kumar},
  {Lebrun}, {Lloyd-Ronning}, {Markwardt}, {Mattson}, {Mushotzky}, {Norris},
  {Osborne}, {Paczynski}, {Palmer}, {Park}, {Parsons}, {Paul}, {Rees},
  {Reynolds}, {Rhoads}, {Sasseen}, {Schaefer}, {Short}, {Smale}, {Smith},
  {Stella}, {Tagliaferri}, {Takahashi}, {Tashiro}, {Townsley}, {Tueller},
  {Turner}, {Vietri}, {Voges}, {Ward}, {Willingale}, {Zerbi}, \&
  {Zhang}}]{Gehrels2004}
{Gehrels}, N., {Chincarini}, G., {Giommi}, P., {et~al.} 2004, \apj, 611, 1005,
  \dodoi{10.1086/422091}

\bibitem[{{Graham} {et~al.}(2019){Graham}, {Kulkarni}, {Bellm}, {Adams},
  {Barbarino}, {Blagorodnova}, {Bodewits}, {Bolin}, {Brady}, {Cenko}, {Chang},
  {Coughlin}, {De}, {Eadie}, {Farnham}, {Feindt}, {Franckowiak}, {Fremling},
  {Gezari}, {Ghosh}, {Goldstein}, {Golkhou}, {Goobar}, {Ho}, {Huppenkothen},
  {Ivezi{\'c}}, {Jones}, {Juric}, {Kaplan}, {Kasliwal}, {Kelley}, {Kupfer},
  {Lee}, {Lin}, {Lunnan}, {Mahabal}, {Miller}, {Ngeow}, {Nugent}, {Ofek},
  {Prince}, {Rauch}, {van Roestel}, {Schulze}, {Singer}, {Sollerman}, {Taddia},
  {Yan}, {Ye}, {Yu}, {Barlow}, {Bauer}, {Beck}, {Belicki}, {Biswas}, {Brinnel},
  {Brooke}, {Bue}, {Bulla}, {Burruss}, {Connolly}, {Cromer}, {Cunningham},
  {Dekany}, {Delacroix}, {Desai}, {Duev}, {Feeney}, {Flynn}, {Frederick},
  {Gal-Yam}, {Giomi}, {Groom}, {Hacopians}, {Hale}, {Helou}, {Henning},
  {Hover}, {Hillenbrand}, {Howell}, {Hung}, {Imel}, {Ip}, {Jackson}, {Kaspi},
  {Kaye}, {Kowalski}, {Kramer}, {Kuhn}, {Landry}, {Laher}, {Mao}, {Masci},
  {Monkewitz}, {Murphy}, {Nordin}, {Patterson}, {Penprase}, {Porter},
  {Rebbapragada}, {Reiley}, {Riddle}, {Rigault}, {Rodriguez}, {Rusholme}, {van
  Santen}, {Shupe}, {Smith}, {Soumagnac}, {Stein}, {Surace}, {Szkody}, {Terek},
  {Van Sistine}, {van Velzen}, {Vestrand}, {Walters}, {Ward}, {Zhang}, \&
  {Zolkower}}]{Graham2019}
{Graham}, M.~J., {Kulkarni}, S.~R., {Bellm}, E.~C., {et~al.} 2019, \pasp, 131,
  078001, \dodoi{10.1088/1538-3873/ab006c}

\bibitem[{{Guti{\'e}rrez} {et~al.}(2021){Guti{\'e}rrez}, {Bersten}, {Orellana},
  {Pastorello}, {Ertini}, {Folatelli}, {Pignata}, {Anderson}, {Smartt},
  {Sullivan}, {Pursiainen}, {Inserra}, {Elias-Rosa}, {Fraser}, {Kankare},
  {Moran}, {Reguitti}, {Reynolds}, {Stritzinger}, {Burke}, {Frohmaier},
  {Galbany}, {Hiramatsu}, {Howell}, {Kuncarayakti}, {Mattila},
  {M{\"u}ller-Bravo}, {Pellegrino}, \& {Smith}}]{Gutierrez2021}
{Guti{\'e}rrez}, C.~P., {Bersten}, M.~C., {Orellana}, M., {et~al.} 2021,
  \mnras, 504, 4907, \dodoi{10.1093/mnras/stab1009}

\bibitem[{{Heger} \& {Woosley}(2002)}]{Heger2002}
{Heger}, A., \& {Woosley}, S.~E. 2002, \apj, 567, 532, \dodoi{10.1086/338487}

\bibitem[{{Ho} {et~al.}(2020){Ho}, {Kulkarni}, {Perley}, {Cenko}, {Corsi},
  {Schulze}, {Lunnan}, {Sollerman}, {Gal-Yam}, {Anand}, {Barbarino}, {Bellm},
  {Bruch}, {Burns}, {De}, {Dekany}, {Delacroix}, {Duev}, {Frederiks},
  {Fremling}, {Goldstein}, {Golkhou}, {Graham}, {Hale}, {Kasliwal}, {Kupfer},
  {Laher}, {Martikainen}, {Masci}, {Neill}, {Ridnaia}, {Rusholme}, {Savchenko},
  {Shupe}, {Soumagnac}, {Strotjohann}, {Svinkin}, {Taggart}, {Tartaglia},
  {Yan}, \& {Zolkower}}]{Ho2020}
{Ho}, A. Y.~Q., {Kulkarni}, S.~R., {Perley}, D.~A., {et~al.} 2020, \apj, 902,
  86, \dodoi{10.3847/1538-4357/aba630}

\bibitem[{{Howell} {et~al.}(2005){Howell}, {Sullivan}, {Perrett}, {Bronder},
  {Hook}, {Astier}, {Aubourg}, {Balam}, {Basa}, {Carlberg}, {Fabbro},
  {Fouchez}, {Guy}, {Lafoux}, {Neill}, {Pain}, {Palanque-Delabrouille},
  {Pritchet}, {Regnault}, {Rich}, {Taillet}, {Knop}, {McMahon}, {Perlmutter},
  \& {Walton}}]{Howell2005}
{Howell}, D.~A., {Sullivan}, M., {Perrett}, K., {et~al.} 2005, \apj, 634, 1190,
  \dodoi{10.1086/497119}

\bibitem[{{Irani} {et~al.}(2022){Irani}, {Chen}, {Morag}, {Schulze}, {Gal-Yam},
  {Strotjohann}, {Yaron}, {Zimmerman}, {Sharon}, {Perley}, {Sollerman},
  {Tohuvavohu}, {Das}, {Kasliwal}, {Bruch}, {Brink}, {Zheng}, {Patra},
  {Vasylyev}, {Filippenko}, {Yang}, {Graham}, {Bloom}, {Purdum}, {Laher},
  {Wold}, {Sharma}, {Lacroix}, \& {Medford}}]{Irani2022}
{Irani}, I., {Chen}, P., {Morag}, J., {et~al.} 2022, arXiv e-prints,
  arXiv:2210.02554.
\newblock \doarXiv{2210.02554}

\bibitem[{{Ivanova} {et~al.}(2003){Ivanova}, {Belczynski}, {Kalogera}, {Rasio},
  \& {Taam}}]{Ivanova2003}
{Ivanova}, N., {Belczynski}, K., {Kalogera}, V., {Rasio}, F.~A., \& {Taam},
  R.~E. 2003, \apj, 592, 475, \dodoi{10.1086/375578}

\bibitem[{{Jacobson-Gal{\'a}n} {et~al.}(2020){Jacobson-Gal{\'a}n}, {Margutti},
  {Kilpatrick}, {Hiramatsu}, {Perets}, {Khatami}, {Foley}, {Raymond}, {Yoon},
  {Bobrick}, {Zenati}, {Galbany}, {Andrews}, {Brown}, {Cartier}, {Coppejans},
  {Dimitriadis}, {Dobson}, {Hajela}, {Howell}, {Kuncarayakti}, {Milisavljevic},
  {Rahman}, {Rojas-Bravo}, {Sand}, {Shepherd}, {Smartt}, {Stacey}, {Stroh},
  {Swift}, {Terreran}, {Vinko}, {Wang}, {Anderson}, {Baron}, {Berger},
  {Blanchard}, {Burke}, {Coulter}, {DeMarchi}, {DerKacy}, {Fremling}, {Gomez},
  {Gromadzki}, {Hosseinzadeh}, {Kasen}, {Kriskovics}, {McCully},
  {M{\"u}ller-Bravo}, {Nicholl}, {Ordasi}, {Pellegrino}, {Piro}, {P{\'a}l},
  {Ren}, {Rest}, {Rich}, {Sai}, {S{\'a}rneczky}, {Shen}, {Short}, {Siebert},
  {Stauffer}, {Szak{\'a}ts}, {Zhang}, {Zhang}, \& {Zhang}}]{Jacobson2020a}
{Jacobson-Gal{\'a}n}, W.~V., {Margutti}, R., {Kilpatrick}, C.~D., {et~al.}
  2020, \apj, 898, 166, \dodoi{10.3847/1538-4357/ab9e66}

\bibitem[{{Jacobson-Gal{\'a}n}
  {et~al.}(2022{\natexlab{a}}){Jacobson-Gal{\'a}n}, {Dessart}, {Jones},
  {Margutti}, {Coppejans}, {Dimitriadis}, {Foley}, {Kilpatrick}, {Matthews},
  {Rest}, {Terreran}, {Aleo}, {Auchettl}, {Blanchard}, {Coulter}, {Davis}, {de
  Boer}, {DeMarchi}, {Drout}, {Earl}, {Gagliano}, {Gall}, {Hjorth}, {Huber},
  {Ibik}, {Milisavljevic}, {Pan}, {Rest}, {Ridden-Harper}, {Rojas-Bravo},
  {Siebert}, {Smith}, {Taggart}, {Tinyanont}, {Wang}, \&
  {Zenati}}]{Jacobson2022}
{Jacobson-Gal{\'a}n}, W.~V., {Dessart}, L., {Jones}, D.~O., {et~al.}
  2022{\natexlab{a}}, \apj, 924, 15, \dodoi{10.3847/1538-4357/ac3f3a}

\bibitem[{{Jacobson-Gal{\'a}n}
  {et~al.}(2022{\natexlab{b}}){Jacobson-Gal{\'a}n}, {Venkatraman}, {Margutti},
  {Khatami}, {Terreran}, {Foley}, {Angulo}, {Angus}, {Auchettl}, {Blanchard},
  {Bobrick}, {Bright}, {Brout}, {Chambers}, {Couch}, {Coulter}, {Clever},
  {Davis}, {de Boer}, {DeMarchi}, {Dodd}, {Jones}, {Johnson}, {Kilpatrick},
  {Khetan}, {Lai}, {Langeroodi}, {Lin}, {Magnier}, {Milisavljevic}, {Perets},
  {Pierel}, {Raymond}, {Rest}, {Rest}, {Ridden-Harper}, {Shen}, {Siebert},
  {Smith}, {Taggart}, {Tinyanont}, {Valdes}, {Villar}, {Wang}, {Yadavalli},
  {Zenati}, \& {Zenteno}}]{Jacobson2022b}
{Jacobson-Gal{\'a}n}, W.~V., {Venkatraman}, P., {Margutti}, R., {et~al.}
  2022{\natexlab{b}}, \apj, 932, 58, \dodoi{10.3847/1538-4357/ac67dc}

\bibitem[{{Jerkstrand} {et~al.}(2015){Jerkstrand}, {Ergon}, {Smartt},
  {Fransson}, {Sollerman}, {Taubenberger}, {Bersten}, \&
  {Spyromilio}}]{Jerkstrand2015}
{Jerkstrand}, A., {Ergon}, M., {Smartt}, S.~J., {et~al.} 2015, \aap, 573, A12,
  \dodoi{10.1051/0004-6361/201423983}

\bibitem[{{Jerkstrand} {et~al.}(2014){Jerkstrand}, {Smartt}, {Fraser},
  {Fransson}, {Sollerman}, {Taddia}, \& {Kotak}}]{Jerkstrand2014}
{Jerkstrand}, A., {Smartt}, S.~J., {Fraser}, M., {et~al.} 2014, \mnras, 439,
  3694, \dodoi{10.1093/mnras/stu221}

\bibitem[{{Kasen}(2010)}]{Kasen2010}
{Kasen}, D. 2010, \apj, 708, 1025, \dodoi{10.1088/0004-637X/708/2/1025}

\bibitem[{{Kashiyama} {et~al.}(2022){Kashiyama}, {Sawada}, \&
  {Suwa}}]{Kashiyama2022}
{Kashiyama}, K., {Sawada}, R., \& {Suwa}, Y. 2022, \apj, 935, 86,
  \dodoi{10.3847/1538-4357/ac7ff7}

\bibitem[{{Khatami} \& {Kasen}(2023)}]{Khatami2023}
{Khatami}, D., \& {Kasen}, D. 2023, arXiv e-prints, arXiv:2304.03360,
  \dodoi{10.48550/arXiv.2304.03360}

\bibitem[{{Khatami} \& {Kasen}(2019)}]{KhatamiKasen2019}
{Khatami}, D.~K., \& {Kasen}, D.~N. 2019, \apj, 878, 56,
  \dodoi{10.3847/1538-4357/ab1f09}

\bibitem[{{Kulkarni} {et~al.}(2021){Kulkarni}, {Harrison}, {Grefenstette},
  {Earnshaw}, {Andreoni}, {Berg}, {Bloom}, {Cenko}, {Chornock}, {Christiansen},
  {Coughlin}, {Wuollet Criswell}, {Darvish}, {Das}, {De}, {Dessart}, {Dixon},
  {Dorsman}, {El-Badry}, {Evans}, {Ford}, {Fremling}, {Gansicke}, {Gezari},
  {Gotberg}, {Green}, {Graham}, {Heida}, {Ho}, {Jaodand}, {Johns-Krull},
  {Kasliwal}, {Lazzarini}, {Lu}, {Margutti}, {Martin}, {Masters}, {McKernan},
  {Nissanke}, {Parazin}, {Perley}, {Phinney}, {Piro}, {Raaijmakers},
  {Rodriguez}, {Senchyna}, {Singer}, {Spake}, {Stassun}, {Stern}, {Teplitz},
  {Weisz}, \& {Yao}}]{Kulkarni2021}
{Kulkarni}, S.~R., {Harrison}, F.~A., {Grefenstette}, B.~W., {et~al.} 2021,
  arXiv e-prints, arXiv:2111.15608.
\newblock \doarXiv{2111.15608}

\bibitem[{{Laplace} {et~al.}(2020){Laplace}, {G{\"o}tberg}, {de Mink},
  {Justham}, \& {Farmer}}]{Laplace2020}
{Laplace}, E., {G{\"o}tberg}, Y., {de Mink}, S.~E., {Justham}, S., \& {Farmer},
  R. 2020, \aap, 637, A6, \dodoi{10.1051/0004-6361/201937300}

\bibitem[{{Leung} \& {Fuller}(2020)}]{Leung2020}
{Leung}, S.-C., \& {Fuller}, J. 2020, \apj, 900, 99,
  \dodoi{10.3847/1538-4357/abac5d}

\bibitem[{{Leung} {et~al.}(2021{\natexlab{a}}){Leung}, {Fuller}, \&
  {Nomoto}}]{Leung2021b}
{Leung}, S.-C., {Fuller}, J., \& {Nomoto}, K. 2021{\natexlab{a}}, \apj, 915,
  80, \dodoi{10.3847/1538-4357/abfcbe}

\bibitem[{{Leung} {et~al.}(2019){Leung}, {Nomoto}, \& {Blinnikov}}]{Leung2019}
{Leung}, S.-C., {Nomoto}, K., \& {Blinnikov}, S. 2019, \apj, 887, 72,
  \dodoi{10.3847/1538-4357/ab4fe5}

\bibitem[{{Leung} {et~al.}(2021{\natexlab{b}}){Leung}, {Wu}, \&
  {Fuller}}]{Leung2021}
{Leung}, S.-C., {Wu}, S., \& {Fuller}, J. 2021{\natexlab{b}}, \apj, 923, 41,
  \dodoi{10.3847/1538-4357/ac2c63}

\bibitem[{{Lyman} {et~al.}(2016){Lyman}, {Levan}, {James}, {Angus}, {Church},
  {Davies}, \& {Tanvir}}]{Lyman2016}
{Lyman}, J.~D., {Levan}, A.~J., {James}, P.~A., {et~al.} 2016, \mnras, 458,
  1768, \dodoi{10.1093/mnras/stw477}

\bibitem[{{Magee} \& {Maguire}(2020)}]{Magee2020b}
{Magee}, M.~R., \& {Maguire}, K. 2020, \aap, 642, A189,
  \dodoi{10.1051/0004-6361/202037870}

\bibitem[{{Magee} {et~al.}(2020){Magee}, {Maguire}, {Kotak}, {Sim},
  {Gillanders}, {Prentice}, \& {Skillen}}]{Magee2020c}
{Magee}, M.~R., {Maguire}, K., {Kotak}, R., {et~al.} 2020, \aap, 634, A37,
  \dodoi{10.1051/0004-6361/201936684}

\bibitem[{{Marshall} {et~al.}(2004){Marshall}, {van Loon}, {Matsuura}, {Wood},
  {Zijlstra}, \& {Whitelock}}]{Marshall2004}
{Marshall}, J.~R., {van Loon}, J.~T., {Matsuura}, M., {et~al.} 2004, \mnras,
  355, 1348, \dodoi{10.1111/j.1365-2966.2004.08417.x}

\bibitem[{{Marston}(1997)}]{Marston1997}
{Marston}, A.~P. 1997, \apj, 475, 188, \dodoi{10.1086/303534}

\bibitem[{{Masci} {et~al.}(2019){Masci}, {Laher}, {Rusholme}, {Shupe}, {Groom},
  {Surace}, {Jackson}, {Monkewitz}, {Beck}, {Flynn}, {Terek}, {Landry},
  {Hacopians}, {Desai}, {Howell}, {Brooke}, {Imel}, {Wachter}, {Ye}, {Lin},
  {Cenko}, {Cunningham}, {Rebbapragada}, {Bue}, {Miller}, {Mahabal}, {Bellm},
  {Patterson}, {Juri{\'c}}, {Golkhou}, {Ofek}, {Walters}, {Graham}, {Kasliwal},
  {Dekany}, {Kupfer}, {Burdge}, {Cannella}, {Barlow}, {Van Sistine}, {Giomi},
  {Fremling}, {Blagorodnova}, {Levitan}, {Riddle}, {Smith}, {Helou}, {Prince},
  \& {Kulkarni}}]{Masci2019}
{Masci}, F.~J., {Laher}, R.~R., {Rusholme}, B., {et~al.} 2019, \pasp, 131,
  018003, \dodoi{10.1088/1538-3873/aae8ac}

\bibitem[{{Mason} {et~al.}(1998){Mason}, {Henry}, {Hartkopf}, {ten Brummelaar},
  \& {Soderblom}}]{Mason1998}
{Mason}, B.~D., {Henry}, T.~J., {Hartkopf}, W.~I., {ten Brummelaar}, T., \&
  {Soderblom}, D.~R. 1998, \aj, 116, 2975, \dodoi{10.1086/300654}

\bibitem[{{Matsuoka} \& {Maeda}(2020)}]{Matsuoka2020}
{Matsuoka}, T., \& {Maeda}, K. 2020, \apj, 898, 158,
  \dodoi{10.3847/1538-4357/ab9c1b}

\bibitem[{{Modjaz} {et~al.}(2006){Modjaz}, {Stanek}, {Garnavich}, {Berlind},
  {Blondin}, {Brown}, {Calkins}, {Challis}, {Diamond-Stanic}, {Hao}, {Hicken},
  {Kirshner}, \& {Prieto}}]{Modjaz2006}
{Modjaz}, M., {Stanek}, K.~Z., {Garnavich}, P.~M., {et~al.} 2006, \apjl, 645,
  L21, \dodoi{10.1086/505906}

\bibitem[{{Morag} {et~al.}(2023){Morag}, {Sapir}, \& {Waxman}}]{Morag2023}
{Morag}, J., {Sapir}, N., \& {Waxman}, E. 2023, \mnras,
  \dodoi{10.1093/mnras/stad899}

\bibitem[{{Moriya} {et~al.}(2017){Moriya}, {Mazzali}, {Tominaga}, {Hachinger},
  {Blinnikov}, {Tauris}, {Takahashi}, {Tanaka}, {Langer}, \&
  {Podsiadlowski}}]{Moriya2017}
{Moriya}, T.~J., {Mazzali}, P.~A., {Tominaga}, N., {et~al.} 2017, \mnras, 466,
  2085, \dodoi{10.1093/mnras/stw3225}

\bibitem[{{Morozova} {et~al.}(2018){Morozova}, {Piro}, \&
  {Valenti}}]{Morozova2018}
{Morozova}, V., {Piro}, A.~L., \& {Valenti}, S. 2018, \apj, 858, 15,
  \dodoi{10.3847/1538-4357/aab9a6}

\bibitem[{{Nakaoka} {et~al.}(2021){Nakaoka}, {Maeda}, {Yamanaka}, {Tanaka},
  {Kawabata}, {Moriya}, {Kawabata}, {Tominaga}, {Takagi}, {Imazato},
  {Morokuma}, {Sako}, {Ohsawa}, {Nagao}, {Jiang}, {Burgaz}, {Taguchi},
  {Uemura}, {Akitaya}, {Sasada}, {Isogai}, {Otsuka}, \&
  {Maehara}}]{Nakaoka2021}
{Nakaoka}, T., {Maeda}, K., {Yamanaka}, M., {et~al.} 2021, \apj, 912, 30,
  \dodoi{10.3847/1538-4357/abe765}

\bibitem[{{Nakar} \& {Piro}(2014)}]{nakarpiro2014}
{Nakar}, E., \& {Piro}, A.~L. 2014, \apj, 788, 193,
  \dodoi{10.1088/0004-637X/788/2/193}

\bibitem[{{Nicholl} \& {Smartt}(2016)}]{Nicholl2016}
{Nicholl}, M., \& {Smartt}, S.~J. 2016, \mnras, 457, L79,
  \dodoi{10.1093/mnrasl/slv210}

\bibitem[{{Nicholl} {et~al.}(2015){Nicholl}, {Smartt}, {Jerkstrand}, {Sim},
  {Inserra}, {Anderson}, {Baltay}, {Benetti}, {Chambers}, {Chen}, {Elias-Rosa},
  {Feindt}, {Flewelling}, {Fraser}, {Gal-Yam}, {Galbany}, {Huber}, {Kangas},
  {Kankare}, {Kotak}, {Kr{\"u}hler}, {Maguire}, {McKinnon}, {Rabinowitz},
  {Rostami}, {Schulze}, {Smith}, {Sullivan}, {Tonry}, {Valenti}, \&
  {Young}}]{Nicholl2015}
{Nicholl}, M., {Smartt}, S.~J., {Jerkstrand}, A., {et~al.} 2015, \apjl, 807,
  L18, \dodoi{10.1088/2041-8205/807/1/L18}

\bibitem[{{Ofek} {et~al.}(2010){Ofek}, {Rabinak}, {Neill}, {Arcavi}, {Cenko},
  {Waxman}, {Kulkarni}, {Gal-Yam}, {Nugent}, {Bildsten}, {Bloom}, {Filippenko},
  {Forster}, {Howell}, {Jacobsen}, {Kasliwal}, {Law}, {Martin}, {Poznanski},
  {Quimby}, {Shen}, {Sullivan}, {Dekany}, {Rahmer}, {Hale}, {Smith},
  {Zolkower}, {Velur}, {Walters}, {Henning}, {Bui}, \& {McKenna}}]{Ofek2010}
{Ofek}, E.~O., {Rabinak}, I., {Neill}, J.~D., {et~al.} 2010, \apj, 724, 1396,
  \dodoi{10.1088/0004-637X/724/2/1396}

\bibitem[{{Oke} \& {Gunn}(1982)}]{Oke1982}
{Oke}, J.~B., \& {Gunn}, J.~E. 1982, \pasp, 94, 586, \dodoi{10.1086/131027}

\bibitem[{{Oke} {et~al.}(1995){Oke}, {Cohen}, {Carr}, {Cromer}, {Dingizian},
  {Harris}, {Labrecque}, {Lucinio}, {Schaal}, {Epps}, \& {Miller}}]{Oke1995}
{Oke}, J.~B., {Cohen}, J.~G., {Carr}, M., {et~al.} 1995, \pasp, 107, 375,
  \dodoi{10.1086/133562}

\bibitem[{{Pastorello} {et~al.}(2007){Pastorello}, {Smartt}, {Mattila},
  {Eldridge}, {Young}, {Itagaki}, {Yamaoka}, {Navasardyan}, {Valenti}, {Patat},
  {Agnoletto}, {Augusteijn}, {Benetti}, {Cappellaro}, {Boles}, {Bonnet-Bidaud},
  {Botticella}, {Bufano}, {Cao}, {Deng}, {Dennefeld}, {Elias-Rosa},
  {Harutyunyan}, {Keenan}, {Iijima}, {Lorenzi}, {Mazzali}, {Meng}, {Nakano},
  {Nielsen}, {Smoker}, {Stanishev}, {Turatto}, {Xu}, \&
  {Zampieri}}]{pastorello07-06jc}
{Pastorello}, A., {Smartt}, S.~J., {Mattila}, S., {et~al.} 2007, \nat, 447,
  829, \dodoi{10.1038/nature05825}

\bibitem[{{Pastorello} {et~al.}(2008){Pastorello}, {Mattila}, {Zampieri},
  {Della Valle}, {Smartt}, {Valenti}, {Agnoletto}, {Benetti}, {Benn}, {Branch},
  {Cappellaro}, {Dennefeld}, {Eldridge}, {Gal-Yam}, {Harutyunyan}, {Hunter},
  {Kjeldsen}, {Lipkin}, {Mazzali}, {Milne}, {Navasardyan}, {Ofek}, {Pian},
  {Shemmer}, {Spiro}, {Stathakis}, {Taubenberger}, {Turatto}, \&
  {Yamaoka}}]{pastorello2008}
{Pastorello}, A., {Mattila}, S., {Zampieri}, L., {et~al.} 2008, \mnras, 389,
  113, \dodoi{10.1111/j.1365-2966.2008.13602.x}

\bibitem[{{Perley}(2019)}]{Perley2019}
{Perley}, D.~A. 2019, \pasp, 131, 084503, \dodoi{10.1088/1538-3873/ab215d}

\bibitem[{{Perley} {et~al.}(2020){Perley}, {Fremling}, {Sollerman}, {Miller},
  {Dahiwale}, {Sharma}, {Bellm}, {Biswas}, {Brink}, {Bruch}, {De}, {Dekany},
  {Drake}, {Duev}, {Filippenko}, {Gal-Yam}, {Goobar}, {Graham}, {Graham}, {Ho},
  {Irani}, {Kasliwal}, {Kim}, {Kulkarni}, {Mahabal}, {Masci}, {Modak}, {Neill},
  {Nordin}, {Riddle}, {Soumagnac}, {Strotjohann}, {Schulze}, {Taggart},
  {Tzanidakis}, {Walters}, \& {Yan}}]{Perley2020}
{Perley}, D.~A., {Fremling}, C., {Sollerman}, J., {et~al.} 2020, \apj, 904, 35,
  \dodoi{10.3847/1538-4357/abbd98}

\bibitem[{{Perley} {et~al.}(2022){Perley}, {Sollerman}, {Schulze}, {Yao},
  {Fremling}, {Gal-Yam}, {Ho}, {Yang}, {Kool}, {Irani}, {Yan}, {Andreoni},
  {Baade}, {Bellm}, {Brink}, {Chen}, {Cikota}, {Coughlin}, {Dahiwale},
  {Dekany}, {Duev}, {Filippenko}, {Hoeflich}, {Kasliwal}, {Kulkarni}, {Lunnan},
  {Masci}, {Maund}, {Medford}, {Riddle}, {Rosnet}, {Shupe}, {Strotjohann},
  {Tzanidakis}, \& {Zheng}}]{perley2022}
{Perley}, D.~A., {Sollerman}, J., {Schulze}, S., {et~al.} 2022, \apj, 927, 180,
  \dodoi{10.3847/1538-4357/ac478e}

\bibitem[{{Piascik} {et~al.}(2014){Piascik}, {Steele}, {Bates}, {Mottram},
  {Smith}, {Barnsley}, \& {Bolton}}]{Piascik2014}
{Piascik}, A.~S., {Steele}, I.~A., {Bates}, S.~D., {et~al.} 2014, in Society of
  Photo-Optical Instrumentation Engineers (SPIE) Conference Series, Vol. 9147,
  Ground-based and Airborne Instrumentation for Astronomy V, ed. S.~K.
  {Ramsay}, I.~S. {McLean}, \& H.~{Takami}, 91478H, \dodoi{10.1117/12.2055117}

\bibitem[{{Piro}(2015)}]{piro2015}
{Piro}, A.~L. 2015, \apjl, 808, L51, \dodoi{10.1088/2041-8205/808/2/L51}

\bibitem[{{Piro} {et~al.}(2021){Piro}, {Haynie}, \& {Yao}}]{Piro2021}
{Piro}, A.~L., {Haynie}, A., \& {Yao}, Y. 2021, \apj, 909, 209,
  \dodoi{10.3847/1538-4357/abe2b1}

\bibitem[{{Piro} \& {Morozova}(2016)}]{Piro2016}
{Piro}, A.~L., \& {Morozova}, V.~S. 2016, \apj, 826, 96,
  \dodoi{10.3847/0004-637X/826/1/96}

\bibitem[{{Podsiadlowski} {et~al.}(1992){Podsiadlowski}, {Joss}, \&
  {Hsu}}]{Podsiadlowski1992}
{Podsiadlowski}, P., {Joss}, P.~C., \& {Hsu}, J.~J.~L. 1992, \apj, 391, 246,
  \dodoi{10.1086/171341}

\bibitem[{{Polin} {et~al.}(2021){Polin}, {Nugent}, \& {Kasen}}]{Polin2021}
{Polin}, A., {Nugent}, P., \& {Kasen}, D. 2021, \apj, 906, 65,
  \dodoi{10.3847/1538-4357/abcccc}

\bibitem[{{Poznanski}(2013)}]{Poznanski:2013}
{Poznanski}, D. 2013, \mnras, 436, 3224, \dodoi{10.1093/mnras/stt1800}

\bibitem[{{Prochaska} {et~al.}(2020){Prochaska}, {Hennawi}, {Westfall},
  {Cooke}, {Wang}, {Hsyu}, {Davies}, {Farina}, \& {Pelliccia}}]{Prochaska2020a}
{Prochaska}, J., {Hennawi}, J., {Westfall}, K., {et~al.} 2020, The Journal of
  Open Source Software, 5, 2308, \dodoi{10.21105/joss.02308}

\bibitem[{{Quataert} \& {Shiode}(2012{\natexlab{a}})}]{Quatert2012}
{Quataert}, E., \& {Shiode}, J. 2012{\natexlab{a}}, \mnras, 423, L92,
  \dodoi{10.1111/j.1745-3933.2012.01264.x}

\bibitem[{{Quataert} \& {Shiode}(2012{\natexlab{b}})}]{Quataert2012}
---. 2012{\natexlab{b}}, \mnras, 423, L92,
  \dodoi{10.1111/j.1745-3933.2012.01264.x}

\bibitem[{{Rabinak} \& {Waxman}(2011)}]{Rabinak2011}
{Rabinak}, I., \& {Waxman}, E. 2011, \apj, 728, 63,
  \dodoi{10.1088/0004-637X/728/1/63}

\bibitem[{{Renzo} {et~al.}(2020){Renzo}, {Farmer}, {Justham}, {G{\"o}tberg},
  {de Mink}, {Zapartas}, {Marchant}, \& {Smith}}]{Renzo2020}
{Renzo}, M., {Farmer}, R., {Justham}, S., {et~al.} 2020, \aap, 640, A56,
  \dodoi{10.1051/0004-6361/202037710}

\bibitem[{{Rigault} {et~al.}(2019){Rigault}, {Neill}, {Blagorodnova}, {Dugas},
  {Feeney}, {Walters}, {Brinnel}, {Copin}, {Fremling}, {Nordin}, \&
  {Sollerman}}]{Rigault2019}
{Rigault}, M., {Neill}, J.~D., {Blagorodnova}, N., {et~al.} 2019, \aap, 627,
  A115, \dodoi{10.1051/0004-6361/201935344}

\bibitem[{{Roberson} {et~al.}(2022){Roberson}, {Fremling}, \&
  {Kasliwal}}]{Roberson2022}
{Roberson}, M., {Fremling}, C., \& {Kasliwal}, M. 2022, The Journal of Open
  Source Software, 7, 3612, \dodoi{10.21105/joss.03612}

\bibitem[{{Roming} {et~al.}(2005){Roming}, {Kennedy}, {Mason}, {Nousek}, {Ahr},
  {Bingham}, {Broos}, {Carter}, {Hancock}, {Huckle}, {Hunsberger}, {Kawakami},
  {Killough}, {Koch}, {McLelland}, {Smith}, {Smith}, {Soto}, {Boyd},
  {Breeveld}, {Holland}, {Ivanushkina}, {Pryzby}, {Still}, \&
  {Stock}}]{Roming2005}
{Roming}, P. W.~A., {Kennedy}, T.~E., {Mason}, K.~O., {et~al.} 2005, \ssr, 120,
  95, \dodoi{10.1007/s11214-005-5095-4}

\bibitem[{{Sagiv} {et~al.}(2014){Sagiv}, {Gal-Yam}, {Ofek}, {Waxman},
  {Aharonson}, {Kulkarni}, {Nakar}, {Maoz}, {Trakhtenbrot}, {Phinney}, {Topaz},
  {Beichman}, {Murthy}, \& {Worden}}]{Sagiv2014}
{Sagiv}, I., {Gal-Yam}, A., {Ofek}, E.~O., {et~al.} 2014, \aj, 147, 79,
  \dodoi{10.1088/0004-6256/147/4/79}

\bibitem[{{Sana} {et~al.}(2012){Sana}, {de Mink}, {de Koter}, {Langer},
  {Evans}, {Gieles}, {Gosset}, {Izzard}, {Le Bouquin}, \&
  {Schneider}}]{Sana2012}
{Sana}, H., {de Mink}, S.~E., {de Koter}, A., {et~al.} 2012, Science, 337, 444,
  \dodoi{10.1126/science.1223344}

\bibitem[{{Schlafly} \& {Finkbeiner}(2011)}]{Schlafly11}
{Schlafly}, E.~F., \& {Finkbeiner}, D.~P. 2011, \apj, 737, 103,
  \dodoi{10.1088/0004-637X/737/2/103}

\bibitem[{{Shiode} \& {Quataert}(2014)}]{Shiode2014}
{Shiode}, J.~H., \& {Quataert}, E. 2014, \apj, 780, 96,
  \dodoi{10.1088/0004-637X/780/1/96}

\bibitem[{{Shvartzvald} {et~al.}(2023){Shvartzvald}, {Waxman}, {Gal-Yam},
  {Ofek}, {Ben-Ami}, {Berge}, {Kowalski}, {B{\"u}hler}, {Worm}, {Rhoads},
  {Arcavi}, {Maoz}, {Polishook}, {Stone}, {Trakhtenbrot}, {Ackermann},
  {Aharonson}, {Birnholtz}, {Chelouche}, {Guetta}, {Hallakoun}, {Horesh},
  {Kushnir}, {Mazeh}, {Nordin}, {Ofir}, {Ohm}, {Parsons}, {Pe'er}, {Perets},
  {Perdelwitz}, {Poznanski}, {Sadeh}, {Sagiv}, {Shahaf}, {Soumagnac}, {Tal-Or},
  {Van Santen}, {Zackay}, {Guttman}, {Rekhi}, {Townsend}, {Weinstein}, \&
  {Wold}}]{Shvartzvald2023}
{Shvartzvald}, Y., {Waxman}, E., {Gal-Yam}, A., {et~al.} 2023, arXiv e-prints,
  arXiv:2304.14482, \dodoi{10.48550/arXiv.2304.14482}

\bibitem[{{Smith} {et~al.}(2020){Smith}, {Smartt}, {Young}, {Tonry}, {Denneau},
  {Flewelling}, {Heinze}, {Weiland}, {Stalder}, {Rest}, {Stubbs}, {Anderson},
  {Chen}, {Clark}, {Do}, {F{\"o}rster}, {Fulton}, {Gillanders}, {McBrien},
  {O'Neill}, {Srivastav}, \& {Wright}}]{Smith2020}
{Smith}, K.~W., {Smartt}, S.~J., {Young}, D.~R., {et~al.} 2020, \pasp, 132,
  085002, \dodoi{10.1088/1538-3873/ab936e}

\bibitem[{{Smith} {et~al.}(2016){Smith}, {Sullivan}, {D'Andrea}, {Castander},
  {Casas}, {Prajs}, {Papadopoulos}, {Nichol}, {Karpenka}, {Bernard}, {Brown},
  {Cartier}, {Cooke}, {Curtin}, {Davis}, {Finley}, {Foley}, {Gal-Yam},
  {Goldstein}, {Gonz{\'a}lez-Gait{\'a}n}, {Gupta}, {Howell}, {Inserra},
  {Kessler}, {Lidman}, {Marriner}, {Nugent}, {Pritchard}, {Sako}, {Smartt},
  {Smith}, {Spinka}, {Thomas}, {Wolf}, {Zenteno}, {Abbott}, {Benoit-L{\'e}vy},
  {Bertin}, {Brooks}, {Buckley-Geer}, {Carnero Rosell}, {Carrasco Kind},
  {Carretero}, {Crocce}, {Cunha}, {da Costa}, {Desai}, {Diehl}, {Doel},
  {Estrada}, {Evrard}, {Flaugher}, {Fosalba}, {Frieman}, {Gerdes}, {Gruen},
  {Gruendl}, {James}, {Kuehn}, {Kuropatkin}, {Lahav}, {Li}, {Marshall},
  {Martini}, {Miller}, {Miquel}, {Nord}, {Ogando}, {Plazas}, {Reil}, {Romer},
  {Roodman}, {Rykoff}, {Sanchez}, {Scarpine}, {Schubnell}, {Sevilla-Noarbe},
  {Soares-Santos}, {Sobreira}, {Suchyta}, {Swanson}, {Tarle}, {Walker},
  {Wester}, \& {DES Collaboration}}]{Smith2016}
{Smith}, M., {Sullivan}, M., {D'Andrea}, C.~B., {et~al.} 2016, \apjl, 818, L8,
  \dodoi{10.3847/2041-8205/818/1/L8}

\bibitem[{{Smith}(2014)}]{Smith2014}
{Smith}, N. 2014, \araa, 52, 487, \dodoi{10.1146/annurev-astro-081913-040025}

\bibitem[{{Smith}(2017)}]{smith2017}
---. 2017, in Handbook of Supernovae, ed. A.~W. {Alsabti} \& P.~{Murdin}, 403,
  \dodoi{10.1007/978-3-319-21846-5\_38}

\bibitem[{{Sollerman} {et~al.}(1998){Sollerman}, {Leibundgut}, \&
  {Spyromilio}}]{Sollerman1998}
{Sollerman}, J., {Leibundgut}, B., \& {Spyromilio}, J. 1998, \aap, 337, 207

\bibitem[{{Sravan} {et~al.}(2020){Sravan}, {Marchant}, {Kalogera},
  {Milisavljevic}, \& {Margutti}}]{Sravan2020}
{Sravan}, N., {Marchant}, P., {Kalogera}, V., {Milisavljevic}, D., \&
  {Margutti}, R. 2020, \apj, 903, 70, \dodoi{10.3847/1538-4357/abb8d5}

\bibitem[{{Steele} {et~al.}(2004){Steele}, {Smith}, {Rees}, {Baker}, {Bates},
  {Bode}, {Bowman}, {Carter}, {Etherton}, {Ford}, {Fraser}, {Gomboc}, {Lett},
  {Mansfield}, {Marchant}, {Medrano-Cerda}, {Mottram}, {Raback}, {Scott},
  {Tomlinson}, \& {Zamanov}}]{Steele2004}
{Steele}, I.~A., {Smith}, R.~J., {Rees}, P.~C., {et~al.} 2004, in Society of
  Photo-Optical Instrumentation Engineers (SPIE) Conference Series, Vol. 5489,
  Ground-based Telescopes, ed. J.~{Oschmann}, Jacobus~M., 679--692,
  \dodoi{10.1117/12.551456}

\bibitem[{{Stritzinger} {et~al.}(2018){Stritzinger}, {Taddia}, {Burns},
  {Phillips}, {Bersten}, {Contreras}, {Folatelli}, {Holmbo}, {Hsiao},
  {Hoeflich}, {Leloudas}, {Morrell}, {Sollerman}, \&
  {Suntzeff}}]{Stritzingetr2018}
{Stritzinger}, M.~D., {Taddia}, F., {Burns}, C.~R., {et~al.} 2018, \aap, 609,
  A135, \dodoi{10.1051/0004-6361/201730843}

\bibitem[{{Strotjohann} {et~al.}(2015){Strotjohann}, {Ofek}, {Gal-Yam},
  {Sullivan}, {Kulkarni}, {Shaviv}, {Fremling}, {Kasliwal}, {Nugent}, {Cao},
  {Arcavi}, {Sollerman}, {Filippenko}, {Yaron}, {Laher}, \&
  {Surace}}]{Nora2015}
{Strotjohann}, N.~L., {Ofek}, E.~O., {Gal-Yam}, A., {et~al.} 2015, \apj, 811,
  117, \dodoi{10.1088/0004-637X/811/2/117}

\bibitem[{{Strotjohann} {et~al.}(2021){Strotjohann}, {Ofek}, {Gal-Yam},
  {Bruch}, {Schulze}, {Shaviv}, {Sollerman}, {Filippenko}, {Yaron}, {Fremling},
  {Nordin}, {Kool}, {Perley}, {Ho}, {Yang}, {Yao}, {Soumagnac}, {Graham},
  {Barbarino}, {Tartaglia}, {De}, {Goldstein}, {Cook}, {Brink}, {Taggart},
  {Yan}, {Lunnan}, {Kasliwal}, {Kulkarni}, {Nugent}, {Masci}, {Rosnet},
  {Adams}, {Andreoni}, {Bagdasaryan}, {Bellm}, {Burdge}, {Duev}, {Dugas},
  {Frederick}, {Goldwasser}, {Hankins}, {Irani}, {Karambelkar}, {Kupfer},
  {Liang}, {Neill}, {Porter}, {Riddle}, {Sharma}, {Short}, {Taddia},
  {Tzanidakis}, {van Roestel}, {Walters}, \& {Zhuang}}]{Nora2021}
---. 2021, \apj, 907, 99, \dodoi{10.3847/1538-4357/abd032}

\bibitem[{{Strotjohann} {et~al.}(2023){Strotjohann}, {Ofek}, {Gal-Yam},
  {Sollerman}, {Chen}, {Yaron}, {Zackay}, {Rehemtulla}, {Gris}, {Masci},
  {Rusholme}, \& {Purdum}}]{Strotjohann2023}
---. 2023, arXiv e-prints, arXiv:2303.00010, \dodoi{10.48550/arXiv.2303.00010}

\bibitem[{{Sukhbold} {et~al.}(2016){Sukhbold}, {Ertl}, {Woosley}, {Brown}, \&
  {Janka}}]{Sukhbold2016}
{Sukhbold}, T., {Ertl}, T., {Woosley}, S.~E., {Brown}, J.~M., \& {Janka}, H.~T.
  2016, \apj, 821, 38, \dodoi{10.3847/0004-637X/821/1/38}

\bibitem[{{Taddia} {et~al.}(2016){Taddia}, {Fremling}, {Sollerman}, {Corsi},
  {Gal-Yam}, {Karamehmetoglu}, {Lunnan}, {Bue}, {Ergon}, {Kasliwal},
  {Vreeswijk}, \& {Wozniak}}]{Taddia2016}
{Taddia}, F., {Fremling}, C., {Sollerman}, J., {et~al.} 2016, \aap, 592, A89,
  \dodoi{10.1051/0004-6361/201628703}

\bibitem[{{Taddia} {et~al.}(2018){Taddia}, {Stritzinger}, {Bersten}, {Baron},
  {Burns}, {Contreras}, {Holmbo}, {Hsiao}, {Morrell}, {Phillips}, {Sollerman},
  \& {Suntzeff}}]{Taddia2018}
{Taddia}, F., {Stritzinger}, M.~D., {Bersten}, M., {et~al.} 2018, \aap, 609,
  A136, \dodoi{10.1051/0004-6361/201730844}

\bibitem[{{Tauris} {et~al.}(2013){Tauris}, {Langer}, {Moriya}, {Podsiadlowski},
  {Yoon}, \& {Blinnikov}}]{Tauris2013}
{Tauris}, T.~M., {Langer}, N., {Moriya}, T.~J., {et~al.} 2013, \apjl, 778, L23,
  \dodoi{10.1088/2041-8205/778/2/L23}

\bibitem[{{Tauris} {et~al.}(2015){Tauris}, {Langer}, \&
  {Podsiadlowski}}]{Tauris2015}
{Tauris}, T.~M., {Langer}, N., \& {Podsiadlowski}, P. 2015, \mnras, 451, 2123,
  \dodoi{10.1093/mnras/stv990}

\bibitem[{{Tonry} {et~al.}(2018){Tonry}, {Denneau}, {Heinze}, {Stalder},
  {Smith}, {Smartt}, {Stubbs}, {Weiland}, \& {Rest}}]{Tonry2018}
{Tonry}, J.~L., {Denneau}, L., {Heinze}, A.~N., {et~al.} 2018, \pasp, 130,
  064505, \dodoi{10.1088/1538-3873/aabadf}

\bibitem[{{Uomoto}(1986)}]{Uomoto1986}
{Uomoto}, A. 1986, \apjl, 310, L35, \dodoi{10.1086/184777}

\bibitem[{{Valenti} {et~al.}(2008){Valenti}, {Benetti}, {Cappellaro}, {Patat},
  {Mazzali}, {Turatto}, {Hurley}, {Maeda}, {Gal-Yam}, {Foley}, {Filippenko},
  {Pastorello}, {Challis}, {Frontera}, {Harutyunyan}, {Iye}, {Kawabata},
  {Kirshner}, {Li}, {Lipkin}, {Matheson}, {Nomoto}, {Ofek}, {Ohyama}, {Pian},
  {Poznanski}, {Salvo}, {Sauer}, {Schmidt}, {Soderberg}, \&
  {Zampieri}}]{Valenti2008}
{Valenti}, S., {Benetti}, S., {Cappellaro}, E., {et~al.} 2008, \mnras, 383,
  1485, \dodoi{10.1111/j.1365-2966.2007.12647.x}

\bibitem[{{van Loon} {et~al.}(2005){van Loon}, {Cioni}, {Zijlstra}, \&
  {Loup}}]{vanloon2005}
{van Loon}, J.~T., {Cioni}, M. R.~L., {Zijlstra}, A.~A., \& {Loup}, C. 2005,
  \aap, 438, 273, \dodoi{10.1051/0004-6361:20042555}

\bibitem[{{Vreeswijk} {et~al.}(2017){Vreeswijk}, {Leloudas}, {Gal-Yam}, {De
  Cia}, {Perley}, {Quimby}, {Waldman}, {Sullivan}, {Yan}, {Ofek}, {Fremling},
  {Taddia}, {Sollerman}, {Valenti}, {Arcavi}, {Howell}, {Filippenko}, {Cenko},
  {Yaron}, {Kasliwal}, {Cao}, {Ben-Ami}, {Horesh}, {Rubin}, {Lunnan}, {Nugent},
  {Laher}, {Rebbapragada}, {Wo{\'z}niak}, \& {Kulkarni}}]{Vreeswijk2017}
{Vreeswijk}, P.~M., {Leloudas}, G., {Gal-Yam}, A., {et~al.} 2017, \apj, 835,
  58, \dodoi{10.3847/1538-4357/835/1/58}

\bibitem[{{Waxman} \& {Katz}(2017)}]{Waxman2017}
{Waxman}, E., \& {Katz}, B. 2017, in Handbook of Supernovae, ed. A.~W.
  {Alsabti} \& P.~{Murdin}, 967, \dodoi{10.1007/978-3-319-21846-5_33}

\bibitem[{{Wheeler} {et~al.}(2015){Wheeler}, {Johnson}, \&
  {Clocchiatti}}]{Wheeler2015}
{Wheeler}, J.~C., {Johnson}, V., \& {Clocchiatti}, A. 2015, \mnras, 450, 1295,
  \dodoi{10.1093/mnras/stv650}

\bibitem[{{Woosley}(2017)}]{Woosley2017}
{Woosley}, S.~E. 2017, \apj, 836, 244, \dodoi{10.3847/1538-4357/836/2/244}

\bibitem[{{Woosley}(2019)}]{Woosley2019}
---. 2019, \apj, 878, 49, \dodoi{10.3847/1538-4357/ab1b41}

\bibitem[{{Woosley} \& {Heger}(2015)}]{Woosley2015}
{Woosley}, S.~E., \& {Heger}, A. 2015, \apj, 810, 34,
  \dodoi{10.1088/0004-637X/810/1/34}

\bibitem[{{Wu} \& {Fuller}(2021)}]{Wu2021}
{Wu}, S., \& {Fuller}, J. 2021, \apj, 906, 3, \dodoi{10.3847/1538-4357/abc87c}

\bibitem[{{Wu} \& {Fuller}(2022{\natexlab{a}})}]{Wu2022}
{Wu}, S.~C., \& {Fuller}, J. 2022{\natexlab{a}}, \apj, 930, 119,
  \dodoi{10.3847/1538-4357/ac660c}

\bibitem[{{Wu} \& {Fuller}(2022{\natexlab{b}})}]{Wu2022b}
---. 2022{\natexlab{b}}, \apjl, 940, L27, \dodoi{10.3847/2041-8213/ac9b3d}

\bibitem[{{Xiang} {et~al.}(2019){Xiang}, {Wang}, {Mo}, {Wang}, {Smartt},
  {Fraser}, {Ehgamberdiev}, {Mirzaqulov}, {Zhang}, {Zhang}, {Vinko}, {Wheeler},
  {Hosseinzadeh}, {Howell}, {McCully}, {DerKacy}, {Baron}, {Brown}, {Zhang},
  {Bi}, {Song}, {Zhang}, {Rest}, {Nomoto}, {Tolstov}, \&
  {Blinnikov}}]{Xiang2019}
{Xiang}, D., {Wang}, X., {Mo}, J., {et~al.} 2019, \apj, 871, 176,
  \dodoi{10.3847/1538-4357/aaf8b0}

\bibitem[{{Yao} {et~al.}(2020){Yao}, {De}, {Kasliwal}, {Ho}, {Schulze}, {Li},
  {Kulkarni}, {Fruchter}, {Rubin}, {Perley}, {Fuller}, {Piro}, {Fremling},
  {Bellm}, {Burruss}, {Duev}, {Feeney}, {Gal-Yam}, {Golkhou}, {Graham},
  {Helou}, {Kupfer}, {Laher}, {Masci}, {Miller}, {Rusholme}, {Shupe}, {Smith},
  {Sollerman}, {Soumagnac}, \& {Zolkower}}]{Yao2020}
{Yao}, Y., {De}, K., {Kasliwal}, M.~M., {et~al.} 2020, \apj, 900, 46,
  \dodoi{10.3847/1538-4357/abaa3d}

\bibitem[{{Yaron} \& {Gal-Yam}(2012)}]{Yaron2012}
{Yaron}, O., \& {Gal-Yam}, A. 2012, \pasp, 124, 668, \dodoi{10.1086/666656}

\bibitem[{{Yoon}(2015)}]{Yoon2015}
{Yoon}, S.-C. 2015, \pasa, 32, e015, \dodoi{10.1017/pasa.2015.16}

\bibitem[{{Yoon} {et~al.}(2010){Yoon}, {Woosley}, \& {Langer}}]{Yoon2010}
{Yoon}, S.~C., {Woosley}, S.~E., \& {Langer}, N. 2010, \apj, 725, 940,
  \dodoi{10.1088/0004-637X/725/1/940}

\bibitem[{{Yoshida} {et~al.}(2017){Yoshida}, {Suwa}, {Umeda}, {Shibata}, \&
  {Takahashi}}]{Yoshida2017}
{Yoshida}, T., {Suwa}, Y., {Umeda}, H., {Shibata}, M., \& {Takahashi}, K. 2017,
  \mnras, 471, 4275, \dodoi{10.1093/mnras/stx1738}

\end{thebibliography}
\bibliographystyle{aasjournal}

\appendix


The complete datasets in Appendices \ref{appendix: photometry}, \ref{appendix: lightcurves}, \ref{appendix: spectra}, \ref{appendix: blackbody}, and \ref{appendix:density} can be found in Zenodo (\!\dataset[DOI: 10.5281/zenodo.11505429]{https://doi.org/10.5281/zenodo.11505429}) and GitHub\footnote{\url{https://github.com/kaustavkdas/doublepeaked_Ibc}}.

\section{Photometry Data} \label{appendix: photometry}
Summary of the photometry data used for SN 2018lqo (Truncated) is provided in Table \ref{phot_table}. The photometry data for all sources are provided as machine-readable tables in Zenodo via\dataset[DOI: 10.5281/zenodo.11505429]{https://doi.org/10.5281/zenodo.11505429}.

\begin{table*}[h!]
\begin{center} 
\caption{Summary of the photometry data used for SN 2018lqo (Truncated).} 
\begin{tabular}{ccccc} 
\hline  \\  

 Date  &  filter & instrument & mag & limiting mag   \\   (JD)  & & & (AB mag) & (AB mag) \\\\ \hline  \\ 

$2458340.68$  &  r  &  P48+ZTF  &  $20.11 \pm 0.17$  &  $20.31$ \\ \hline \\ 

$2458343.66$  &  r  &  P48+ZTF  &  $20.68 \pm 0.23$  &  $20.64$ \\ \hline \\ 

$2458343.68$  &  g  &  P48+ZTF  &  $20.98 \pm 0.37$  &  $20.71$ \\ \hline \\ 

$2458346.66$  &  g  &  P48+ZTF  &  $20.49 \pm 0.30$  &  $20.49$ \\ \hline \\ 

$2458346.68$  &  r  &  P48+ZTF  &  $20.16 \pm 0.16$  &  $20.53$ \\ \hline \\ 

$2458346.68$  &  r  &  P48+ZTF  &  $20.16 \pm 0.16$  &  $20.53$ \\ \hline \\ 

$2458347.76$  &  r  &  P60+SEDM  &  $20.02 \pm 0.05$  &  $21.69$ \\ \hline \\ 

$2458347.76$  &  r  &  P60+SEDM  &  $20.06 \pm 0.04$  &  $99.00$ \\ \hline \\ 

$2458347.76$  &  g  &  P60+SEDM  &  $20.40 \pm 0.06$  &  $21.81$ \\ \hline \\ 

$2458347.76$  &  g  &  P60+SEDM  &  $20.47 \pm 0.07$  &  $99.00$ \\ \hline \\ 

$2458347.76$  &  i  &  P60+SEDM  &  $19.92 \pm 0.10$  &  $21.48$ \\ \hline \\ 

$2458347.76$  &  i  &  P60+SEDM  &  $19.92 \pm 0.53$  &  $99.00$ \\ \hline \\ 

$2458348.68$  &  i  &  P48+ZTF  &  $19.82 \pm 0.23$  &  $19.94$ \\ \hline \\ 

$2458350.65$  &  r  &  P48+ZTF  &  $19.68 \pm 0.16$  &  $20.45$ \\ \hline \\

\end{tabular}  \label{phot_table} 
\end{center} 
\end{table*}

\section{Lightcurves} \label{appendix: lightcurves}
The lightcurves of all sources can be found in Zenodo via\dataset[DOI: 10.5281/zenodo.11505429]{https://doi.org/10.5281/zenodo.11505429}.
\section{Spectra} \label{appendix: spectra}
The spectra of all sources are provided as machine-readable tables in Zenodo via\dataset[DOI: 10.5281/zenodo.11505429]{https://doi.org/10.5281/zenodo.11505429}.

\section{Blackbody Fits} \label{appendix: blackbody}

Summary of the blackbody properties for SN 2018lqo (Truncated)  is provided in Table \ref{bb_table}. All the best-fit parameters including bolometric luminosity, radius and temperature for each object are provided as machine-readable tables in Zenodo via\dataset[DOI: 10.5281/zenodo.11505429]{https://doi.org/10.5281/zenodo.11505429}.

\begin{table*}[h!]
\begin{center} 
\caption{Summary of the blackbody properties for SN 2018lqo (Truncated).} 
\begin{tabular}{cccc} 
\hline  \\  

Phase &  Log Luminosity  &  Temperature & Radius   \\    (days since first detection)   & ($\mathrm{erg\ s^{-1}}$)  & (K) & (\Rsun) \\\\

\hline  \\
$3.17$  &  $41.55^{+0.11}_{-0.07}$  &  $7048^{+3206}_{-1691}$  &  $6197^{+4753}_{-2832}$ \\ \hline \\ 

$4.33$  &  $41.67^{+0.50}_{-0.10}$  &  $7197^{+10256}_{-1965}$  &  $6751^{+5648}_{-4616}$ \\ \hline \\ 

$6.17$  &  $41.76^{+0.05}_{-0.04}$  &  $5908^{+1167}_{-915}$  &  $11467^{+5377}_{-3439}$ \\ \hline \\ 

$7.26$  &  $41.76^{+0.01}_{-0.01}$  &  $6165^{+250}_{-244}$  &  $10699^{+942}_{-836}$ \\ \hline \\ 

$8.24$  &  $41.90^{+0.54}_{-0.11}$  &  $5171^{+10212}_{-1856}$  &  $15557^{+29245}_{-11786}$ \\ \hline \\ 

$10.21$  &  $41.93^{+0.02}_{-0.02}$  &  $5880^{+405}_{-382}$  &  $14321^{+2247}_{-1852}$ \\ \hline \\ 

$11.22$  &  $41.92^{+0.03}_{-0.02}$  &  $5384^{+443}_{-391}$  &  $16918^{+3269}_{-2683}$ \\ \hline \\ 

$12.21$  &  $41.93^{+0.03}_{-0.03}$  &  $5955^{+635}_{-544}$  &  $13900^{+3299}_{-2635}$ \\ \hline \\ 

$20.22$  &  $41.90^{+0.28}_{-0.16}$  &  $3633^{+712}_{-683}$  &  $36441^{+39393}_{-15074}$ \\ \hline \\ 

$21.17$  &  $41.72^{+0.01}_{-0.01}$  &  $4195^{+70}_{-70}$  &  $22190^{+1028}_{-917}$ \\ \hline \\ 

$22.22$  &  $41.76^{+0.08}_{-0.07}$  &  $3926^{+391}_{-358}$  &  $26476^{+8634}_{-6183}$ \\ \hline \\ 

$25.18$  &  $42.00^{+0.40}_{-0.22}$  &  $3175^{+609}_{-630}$  &  $53509^{+77779}_{-24058}$ \\ \hline \\

\end{tabular}  \label{bb_table} 
\end{center} 
\end{table*}

\section{First-peak fits} \label{appendix: scb}
\begin{figure*}[h!]
    \centering
    \includegraphics[width=13.5cm]{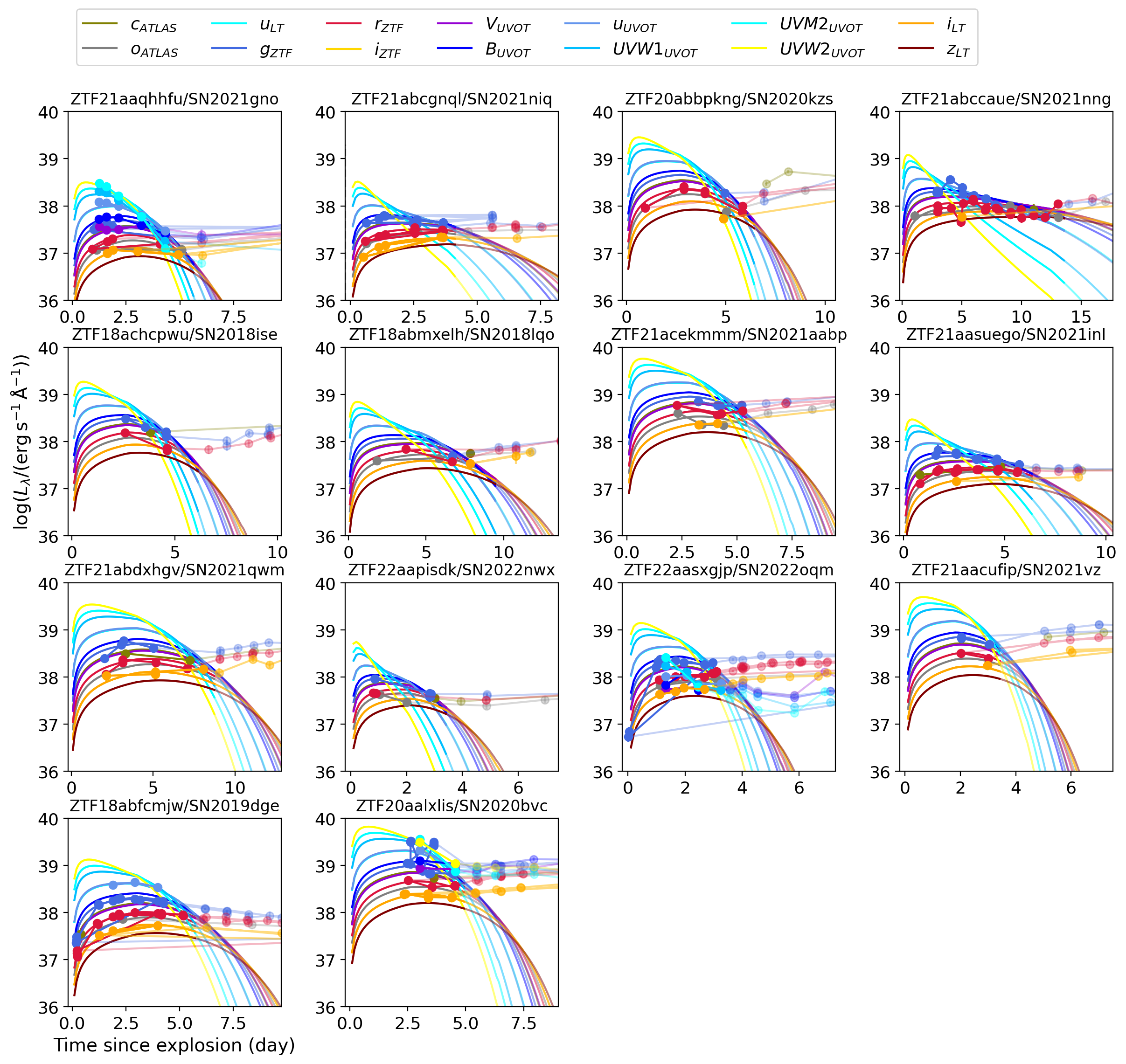}
    \caption{Shock cooling model \citep{Piro2021} fits to multi-band data.}
    \label{pirofit}
\end{figure*}

Figure \ref{pirofit} shows a collage of the best-fit lightcurves for the shock cooling model \citep{Piro2021} fits to the multi-band photometry data.

\section{Second-peak fits} \label{appendix: arnettfits}
\begin{figure*}[h!]
    \centering
    \includegraphics[width=13.5cm]{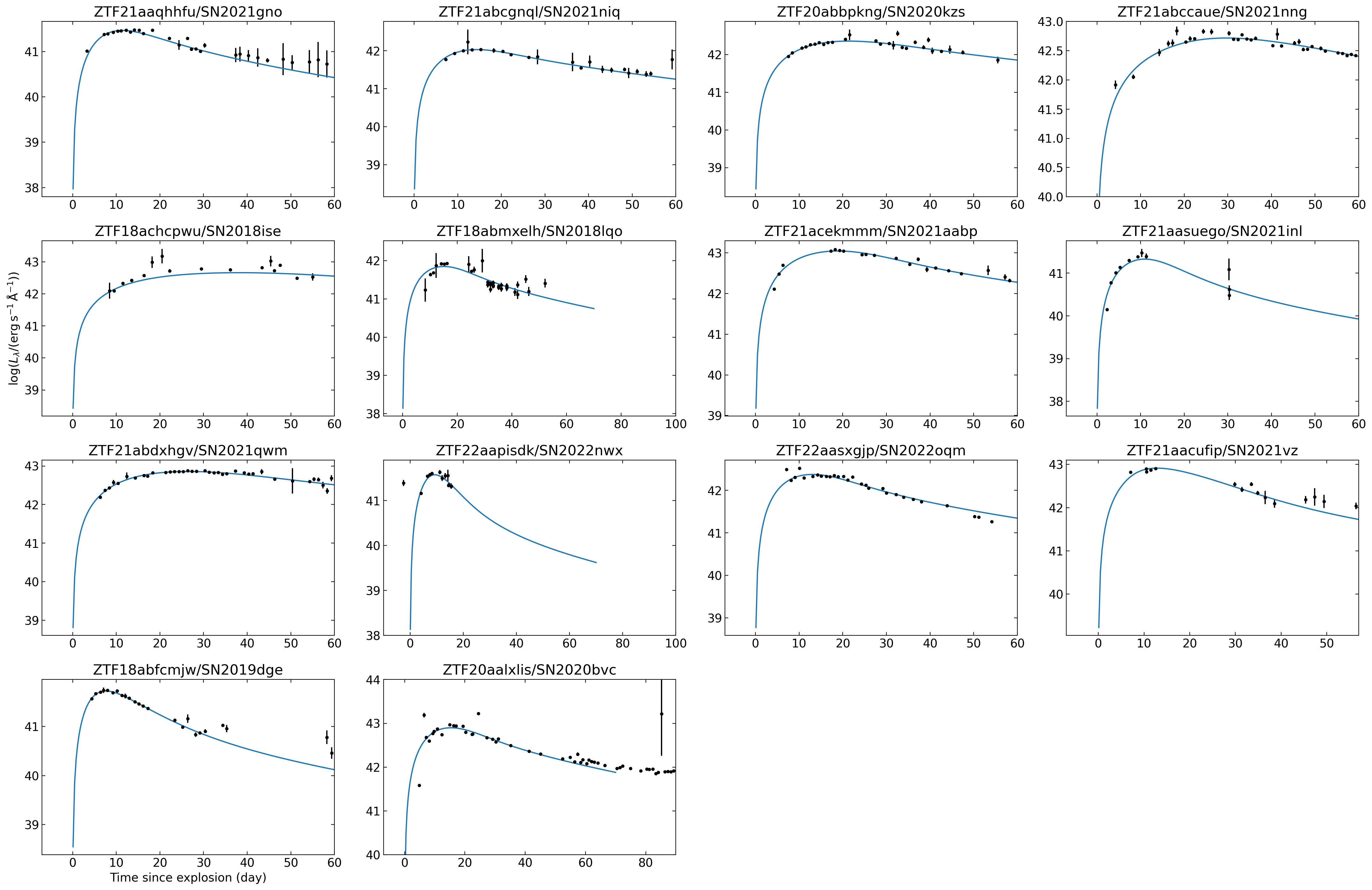}
    \caption{Radioactive \citep{Arnett89} fits to bolometric luminosity data.}
    \label{Nifit}
\end{figure*}

Figure \ref{Nifit} shows a collage of the best-fit lightcurves for the radioactive peak model \citep{Arnett2011} fits to the bolometric luminosity data.

\section{Spectral log}
The spectral log and the velocity measurements are listed in Table \ref{table_vel}.

\begin{table*}[h!] 
\begin{center} 
\caption{Spectral log and ejecta velocity measurements.} 
\begin{tabular}{cccccc} 
\hline  \\  

Source & Date & Phase & Inst. & He I $\lambda$5876 & O I $\lambda$7774  \\ & & (days) &  & (km s$^{-1}$) & (km s$^{-1}$)  \\\\ \hline  \\ 

ZTF21aaqhhfu/SN~2021gno  &  2021-03-20  &  -15.0  &  SPRAT  &  $-$    &  $7910 \pm 1170$   \\ \hline \\ 

ZTF21aaqhhfu/SN~2021gno  &  2021-03-21  &  -14.0  &  SEDM  &  $-$    &  $-$   \\ \hline \\ 

ZTF21aaqhhfu/SN~2021gno  &  2021-03-24  &  -11.0  &  SEDM  &  $12850 \pm 4520$    &  $-$   \\ \hline \\ 

ZTF21aaqhhfu/SN~2021gno  &  2021-04-02  &  -2.0  &  SPRAT  &  $8000 \pm 260$    &  $6860 \pm 810$   \\ \hline \\ 

ZTF21aaqhhfu/SN~2021gno  &  2021-04-02  &  -2.0  &  SEDM  &  $7880 \pm 740$    &  $5830 \pm 2610$   \\ \hline \\ 

ZTF21aaqhhfu/SN~2021gno  &  2021-04-12  &  8.0  &  SEDM  &  $8210 \pm 2540$    &  $7630 \pm 2920$   \\ \hline \\ 

ZTF21aaqhhfu/SN~2021gno  &  2022-02-04  &  306.0  &  LRIS  &  $-$    &  $-$   \\ \hline \\ 

ZTF21abcgnql/SN~2021niq  &  2021-05-31  &  -6.0  &  DBSP  &  $13680 \pm 2630$    &  $-$   \\ \hline \\ 

ZTF21abcgnql/SN~2021niq  &  2022-04-13  &  310.0  &  LRIS  &  $-$    &  $-$   \\ \hline \\ 

ZTF20abbpkng/SN~2020kzs  &  2020-06-01  &  -9.0  &  SEDM  &  $-$    &  $-$   \\ \hline \\ 

\end{tabular}  \label{table_vel} 
\end{center} 
\tablenotetext{*}{The full version of this table is available in machine-readable format in the online journal. A portion is shown here for guidance regarding its form and function.}
\end{table*}

\section{Density profile of He-stars used in the late-time mass transfer models} \label{appendix:density}

\begin{figure*}[h!]
    \centering
    \includegraphics[width=9.5cm]{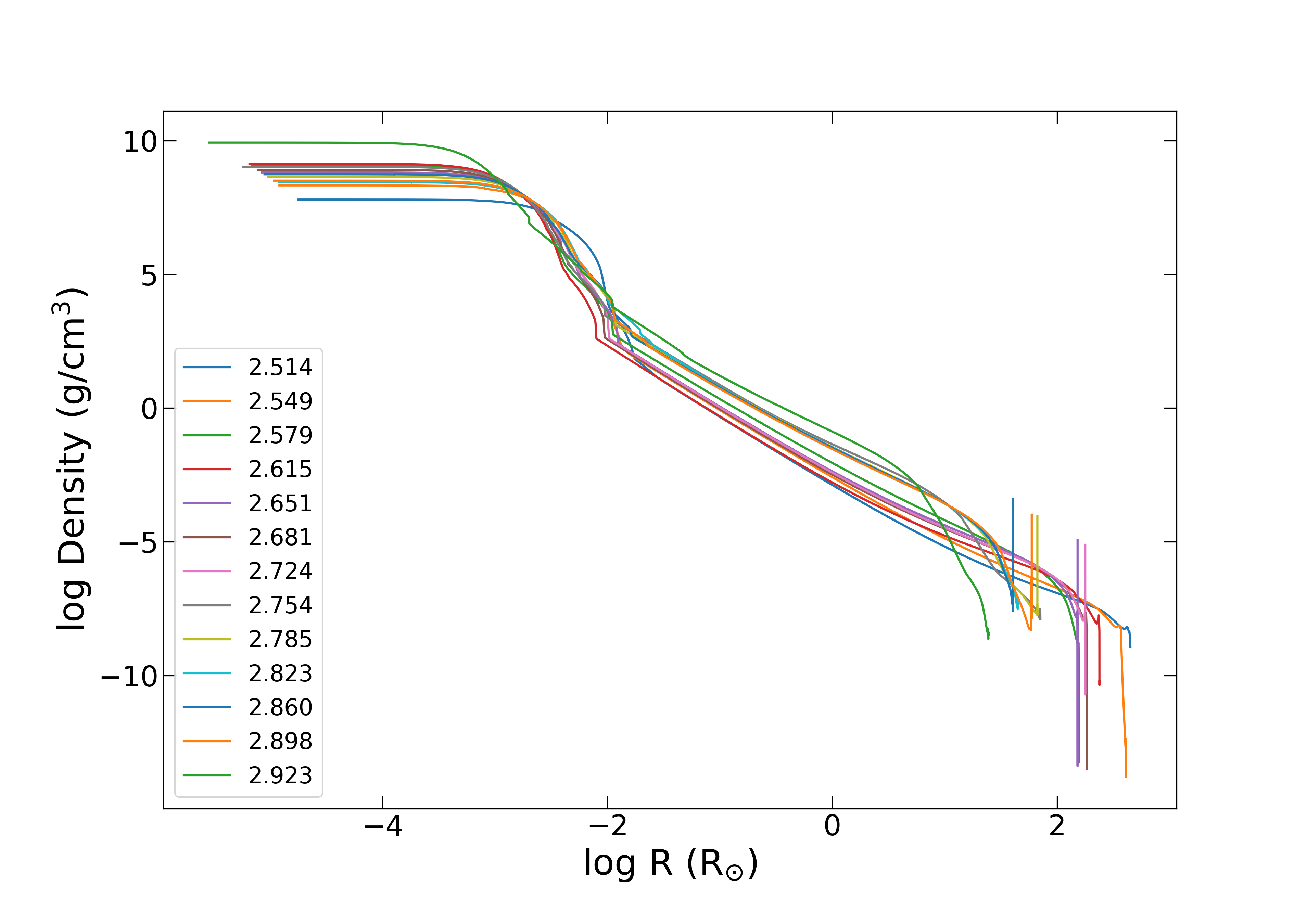}
    \caption{Density profile of the single He-stars of different masses before late-time mass transfer used in Section \ref{sec:bound1}.}
    \label{fig:ssdensity}
\end{figure*}

The density profiles of the single He-stars of different masses before late-time mass transfer used in Section \ref{sec:bound1} are shown in Figure \ref{fig:ssdensity}. The density profiles of the bound material of the stars after mass transfer can be found in Zenodo via\dataset[DOI: 10.5281/zenodo.11505429]{https://doi.org/10.5281/zenodo.11505429}.

\end{document}